\pdfoutput=1
\documentclass[manuscript,screen]{acmart}
\setcopyright{none}

\acmConference[FAccT '23]{2023 ACM Conference on Fairness, Accountability, and Transparency}{June 12--15, 2023}{Chicago, IL, USA}
\acmBooktitle{2023 ACM Conference on Fairness, Accountability, and Transparency (FAccT '23), June 12--15, 2023, Chicago, IL, USA}
\usepackage{cleveref}
\usepackage{breakurl}
\usepackage{subcaption}

\begin{document}

\title[Harms from Increasingly Agentic Algorithmic Systems]{Harms from Increasingly Agentic Algorithmic Systems} 

\author{Alan Chan}
\email{alan.chan@mila.quebec}
\authornote{Major contributions to the project direction and framing.}
\authornote{Major contributions to the writing. Authors with only this mark have had their order randomized. Authors without a mark also have their orders randomized.}
\authornote{Correspondence to alan.chan@mila.quebec.}
\affiliation{\institution{Mila, Université de Montréal}
\city{Montréal}
\country{Canada}}

\author{Rebecca Salganik}
\authornotemark[2]
\affiliation{\institution{Mila, Université de Montréal}
\city{Montréal}
\country{Canada}}

\author{Alva Markelius}
\authornotemark[2]
\affiliation{\institution{University of Cambridge}
\city{Cambridge}
\country{UK}}

\author{Chris Pang}
\authornotemark[2]
\affiliation{\institution{University of Cambridge}
\city{Cambridge}
\country{UK}}

\author{Nitarshan Rajkumar}
\authornotemark[2]
\affiliation{\institution{University of Cambridge}
\city{Cambridge}
\country{UK}}

\author{Dmitrii Krasheninnikov}
\authornotemark[2]
\affiliation{\institution{University of Cambridge}
\city{Cambridge}
\country{UK}}

\author{Lauro Langosco}
\authornotemark[2]
\affiliation{\institution{University of Cambridge}
\city{Cambridge}
\country{UK}}

\author{Zhonghao He}
\authornotemark[2]
\affiliation{\institution{University of Cambridge}
\city{Cambridge}
\country{UK}}

\author{Yawen Duan}
\authornotemark[2]
\affiliation{\institution{University of Cambridge}
\city{Cambridge}
\country{UK}}

\author{Micah Carroll}
\authornotemark[2]
\affiliation{\institution{University of California, Berkeley}
\city{Berkeley}
\country{USA}}

\author{Michelle Lin}
\affiliation{\institution{McGill University}
\city{Montréal}
\country{Canada}}

\author{Alex Mayhew}
\affiliation{\institution{University of Western Ontario}
\city{London}
\country{Canada}}

\author{Katherine Collins}
\affiliation{\institution{University of Cambridge}
\city{Cambridge}
\country{UK}}

\author{Maryam Molamohammadi}
\affiliation{\institution{Mila}
\city{Montréal}
\country{Canada}}

\author{John Burden}
\affiliation{\institution{Center for the Study of Existential Risk, University of Cambridge}
\city{Cambridge}
\country{UK}}

\author{Wanru Zhao}
\affiliation{\institution{University of Cambridge}
\city{Cambridge}
\country{UK}}

\author{Shalaleh Rismani}
\affiliation{\institution{McGill University, Mila}
\city{Montréal}
\country{Canada}}

\author{Konstantinos Voudouris}
\affiliation{\institution{University of Cambridge}
\city{Cambridge}
\country{UK}}

\author{Umang Bhatt}
\affiliation{\institution{University of Cambridge}
\city{Cambridge}
\country{UK}}

\author{Adrian Weller}
\affiliation{\institution{University of Cambridge}
\city{Cambridge}
\country{UK}}

\author{David Krueger}
\authornotemark[1]
\affiliation{\institution{University of Cambridge}
\city{Cambridge}
\country{UK}}

\author{Tegan Maharaj}
\authornotemark[1]
\authornotemark[2]
\affiliation{\institution{University of Toronto}
\city{Toronto}
\country{Canada}}

\authorsaddresses{}

\renewcommand{\shortauthors}{Chan et al.}
\newcommand\todo[1]{\textcolor{red}{#1}}

\begin{abstract}
Research in Fairness, Accountability, Transparency, and Ethics (FATE)\footnote{We use the term \textbf{FATE} as a shorthand, keeping in mind and valuing the ideological diversity of those who work on FATE and related disciplines not captured in this acronym.} has established many sources and forms of algorithmic harm, in domains as diverse as health care, finance, policing, and recommendations. Much work remains to be done to mitigate the serious harms of these systems, particularly those disproportionately affecting marginalized communities. Despite these ongoing harms, new systems are being developed and deployed, typically without strong regulatory barriers, threatening the perpetuation of the same harms and the creation of novel ones. In response, the FATE community has emphasized the importance of \textit{anticipating} harms, rather than just responding to them. Anticipation of harms is especially important given the rapid pace of developments in machine learning (ML). Our work focuses on the anticipation of harms from increasingly agentic systems. Rather than providing a definition of agency as a binary property, we identify 4 key characteristics which, particularly in combination, tend to increase the agency of a given algorithmic system: underspecification, directness of impact, goal-directedness, and long-term planning. We also discuss important harms which arise from increasing agency -- notably, these include systemic and/or long-range impacts, often on marginalized or unconsidered stakeholders. We emphasize that recognizing agency of algorithmic systems does not absolve or shift the human responsibility for algorithmic harms. Rather, we use the term agency to highlight the increasingly evident fact that ML systems are not fully under human control. Our work explores increasingly agentic algorithmic systems in three parts. First, we explain the notion of an increase in agency for algorithmic systems in the context of diverse perspectives on agency across disciplines. Second, we argue for the need to anticipate harms from increasingly agentic systems. Third, we discuss important harms from increasingly agentic systems and ways forward for addressing them. We conclude by reflecting on implications of our work for anticipating algorithmic harms from emerging systems. 
\end{abstract}



\maketitle

\section{Introduction}
The promised benefits of algorithmic systems have not always borne out, and benefits are often tempered by significant negative externalities. 
Although the deployment of algorithmic systems may result in increased safety or material improvements to human well-being \citep{li_applications_2017, abebe_mechanism_2018,jumper_highly_2021}, diverse lines of work in Fairness, Accountability, Transparency, and Ethics (FATE) have established the roles that algorithmic systems play in causing harm. Examples include the perpetuation of existing, unjust power relations \citep{buolamwini_gender_2018,barabas_interventions_2018,kasy_fairness_2021,steven_t_piantadosi_spiantado_yes_2022,wolfe_american_2022,ehsan_algorithmic_2022}, the generation of toxic language \citep{gehman_realtoxicityprompts_2020,abid_persistent_2021}, and informational harms \citep{jiang_degenerate_2019,weidinger_taxonomy_2022,li_fairness_2022, carroll_estimating_2022}.  


Despite the clear evidence of harms from existing systems, new types of algorithmic systems are continually being developed and deployed, often without strong regulatory barriers \citep{gesley_regulation_2019}. The pace of development has been particularly rapid in the machine learning (ML) community. Just in the last five years, we have witnessed large improvements in the capabilities of systems to perform a variety of real-world tasks, including search \cite{nayak_understanding_2019},  
drug discovery \cite{stokes_deep_2020, jumper_highly_2021}, and dialogue \cite{openai_chatgpt_2022}. 

Researchers in the FATE community have responded to the rapid pace of ML developments by emphasizing the need to \textit{anticipate} harms, rather than just react to them. In particular, many have identified the impact of computational modeling and development in social change \citep{abebe_roles_2020,selbst_fairness_2019,jacobs_measurement_2021} and scoped numerous taxonomies of risks, harms, and failures of algorithmic systems \citep{shelby_sociotechnical_2022,weidinger_taxonomy_2022,raji_fallacy_2022}. 
While it is crucial not to idealize or over-hype a model's performance by ignoring model failures~ \citep{bender_climbing_2020,blodgett_stereotyping_2021,vinsel_youre_2021,bender_dangers_2021,lin_truthfulqa_2022,collins_structured_2022, raji_fallacy_2022}
, it is also important not to understate (and thus fail to anticipate negative consequences of) what these models \textit{can do} and \textit{may be capable of doing} in the near future \citep{kaplan_scaling_2020,hoffmann_empirical_2022,bowman_dangers_2022}, especially given growing investments in the field \citep{giattino_artificial_2022}. 

In this paper, we continue the work of anticipating harms by drawing attention to increasingly agentic algorithmic systems. We use agency and agentic in a narrow sense for our work as applied to algorithmic systems, particularly ML systems. While recognizing the many meanings of \textbf{agency}, as well as the need not to absolve humans of responsibility pertaining to algorithmic harms \citep{nissenbaum_accountability_1996,wieringa_what_2020,cooper_accountability_2022}, we use the term agency consciously to counter the somewhat prevalent view that the developers of an algorithmic system have full control over its behaviour. E.g. \citet{johnson_reframing_2017} claim that ``the behaviour of computational artefacts is in the control of the humans that design them.'' 
And in a systematic review on algorithmic accountability, \citet{wieringa_what_2020} defines algorithms as ``basically instructions fed to a computer''. While this description is accurate for many purposes, we argue that, particularly for ML-based algorithmic systems, it elides autonomous, responsive, and interactive qualities of these systems which can so easily lead to unforeseen outcomes. \citet{nissenbaum_accountability_1996,cooper_accountability_2022} do identify bugs -- including faulty modeling premises and bad model performance -- as one way in which humans may not have total control of the operation of an algorithmic system. 
However, we view agency as distinct from mistakes or bugs and demonstrate the unique and important harms that can result. We note there are significant economic and military incentives to build increasingly agentic systems. Indeed, many in the ML community are explicitly building such systems as a research goal \citep{chen_decision_2021,sutton_alberta_2022,reed_generalist_2022}. \textbf{In summary, our contributions are}:
\vspace*{-2mm}
\begin{enumerate}
\item We identify characteristics that tend to increase agency of algorithmic systems, and situate our characterization in the context of diverse perspectives on agency across disciplines. We articulate that even when recognizing agency in algorithmic systems, we can and should emphasize the human responsibility to prevent harms.
\item We argue for the need to anticipate harms from increasingly agentic systems. Increasingly agentic systems are being developed and there exist strong incentives for this work to continue. 
\item We discuss some harms to be anticipated from increasingly agentic systems. In so doing, we connect to ongoing lines of work in the FATE community, including systemic and delayed effects, an impoverishment of collective decision-making power, and exacerbation of extreme concentrations of power in the hands of a few. We also discuss the role of increasing agency as a source of harms that are yet to be identified. 
\end{enumerate}
This paper is \textbf{not} about 
the moral agency or consciousness of algorithms or machines. Instead, we focus on identifying a property of emerging ML systems, argue for the need to anticipate harms from systems that increasingly satisfy this property, and discuss the harms to be anticipated.

\section{Agency}\label{sec:terminology}
In colloquial use, agency refers to the ability to take actions or affect outcomes. A difficulty of having concrete discussions on agency is the variety of perspectives through which such a concept can be defined, making confusion and disagreement common. In recognition of this variety of perspectives, we do not attempt to define agency in a binary manner, but instead present a set of characteristics we take to be associated with \textit{increasing} agency, i.e. the more of these characteristics a system has, particularly in combination, the more agency we can consider it to have. We first present our characterization, and follow by contextualizing it in some of the most relevant perspectives and related concepts to our work.

\subsection{Characteristics that are Associated with Increasing Agency in Algorithmic Systems}
 When we say that an algorithmic system has a degree of agency, we mean that it is to some extent an agent or agentic. \textbf{Agency} is the property, \textbf{agent} is the role, and \textbf{agentic} is the adjective.
Our characterization of agency is specific to algorithmic systems and is not meant to define agency for humans or arbitrary entities. We will sometimes use ``agentic system'' in place of ``agentic algorithmic system'' for brevity.

We identify 4 key characteristics associated with increasing agency in algorithmic systems, especially in combination: underspecification, directness of impact, goal-directedness, and long-term planning.
\begin{enumerate}
    \item \textbf{Underspecification:} the degree to which the algorithmic system can accomplish a goal provided by operators or designers, without a concrete specification of how the goal is to be accomplished \citep{damour_underspecification_2020}. 
    \item \textbf{Directness of impact:} the degree to which the algorithmic system's actions affect the world without mediation or intervention by a human, i.e. without a human in the loop. 
    \item \textbf{Goal-directedness:} the degree to which the system acts as if it is designed/trained to achieve a particular quantifiable objective.
    \item \textbf{Long-term planning:} the degree to which the algorithmic system is designed/trained to make decisions that are temporally dependent upon one another to achieve a goal and/or make predictions over a long time horizon. 
\end{enumerate}

To illustrate the notion of increasing agency, consider the task of compiling a literature review on a certain subject. With a search engine, the human user must type in queries, click on related works, read papers, look through bibliographies, record relevant information in a document, and edit the text. A system that was more agentic than the search engine, still for the same task, could simply be queried with the topic of the desired literature review, and would automatically look through related works on the internet without user intervention, like WebGPT can do to some extent \citep{nakano_webgpt_2022}. The user would not need (or be able to) to specify which papers were relevant nor have to compile papers manually into a document. 

\subsection{Prior Work on Agency}\label{sec:agency-other}
Agency is a central concept in many fields of academia \citep{schlosser_agency_2019}.
\citet{dennett_intentional_1981} provides one of the most popular analyses of when and how to attribute agency, focusing on the notion that agents behave intentionally.
\citet{orseau_agents_2018} and more recently \citet{kenton_discovering_2022} have attempted to formalize this notion of agency in the context of artificial intelligence.
In cognitive science and psychology, agency is conceptualized relatively similarly, as having intentions, plans, goals, communication, and reasoning \citep{spelke_core_2007,lake_building_2017} -- entities with agency can plan, act, memorize, exert self-control, and communicate with others. 
While these notions of agency focus on individuals making rational choices in pursuit of some goal, in sociology, agency is often thought of as contextualized within, constrained by, and/or contrasted with structure \citep{emirbayer_what_1998}. 

\textbf{Principal-agent theory} \citep{jensen_theory_1976,eisenhardt_agency_1989} provides more intuition for how we characterize agency. Principal-agent theory concerns itself with a \textit{principal} who delegates tasks to an \textit{agent} in order to achieve their goals. The agent acts (directness of impact) on behalf of the principal to achieve the principal's goals, which may be long-horizon (long-term planning). Crucially, the agent and principal have different incentives\footnote{It is coherent to talk about the incentives of algorithmic systems. See \citet{everitt_agent_2021}.} and information: the principal does not tell the agent how to complete the tasks (underspecification). In our context, we view the principal as humans and the agent as algorithmic systems, as done in prior work \cite{hadfield-menell_incomplete_2019}. It is in this sense that we consider algorithmic systems to have agency.

Our notion of increasing agency also takes inspiration from how the term agent is used in AI research. In the most popular introductory text on artificial intelligence, \citet[p.~58]{russell_artificial_2021} define a \textbf{rational agent} as follows: ``For each possible percept sequence, a rational agent should select an action that is expected to maximize its performance measure, given the evidence provided by the percept sequence and whatever built-in knowledge the agent has.'' \citet[p.~60]{russell_artificial_2021} further states that ``To the extent that an agent relies on the prior knowledge of its designer rather than on its own percepts and learning processes, we say that the agent lacks autonomy. A rational agent should be autonomous—it should learn what it can to compensate for partial or incorrect prior knowledge.'' While our characterization does not consider (ir)rationality, goal-directedness and underspecification are captured in this definition.

\textbf{Reinforcement learning} is a field that concentrates on the construction of agents. In the field's premier introductory text, \citet[p.~47-8]{sutton_reinforcement_2018} states that the ``learner and decision maker is called the agent. The thing it interacts with, comprising everything outside the agent, is called the environment. These interact continually, the agent selecting actions and the environment responding to these actions and presenting new situations to the agent. The environment also gives rise to rewards, special numerical values that the agent seeks to maximize over time through its choice of actions.'' 
Note that reinforcement learning is not the only way of constructing agents, however. 
For instance, recent work has shown that foundation models can perform planning tasks \citep{huang_inner_2022}. Even simple predictive algorithms, depending on their training procedure, can follow incentives to affect the world in unexpected ways -- for example by shifting user interests rather than improving at their predictive task \cite{krueger_hidden_2020}, thus increasing their agency.



One of the difficult discussions surrounding agency is interaction of agency and responsibility, for humans and for machines.  \citet{goetze_mind_2022} identifies a \textbf{responsibility gap} between engineers and the outcomes of their designed systems -- people  designing autonomous systems are far removed from the consequences of their deployment. The authors contend that regardless of the system's autonomy, human designers must be the ones held accountable. \cite{sullivan_moral_2022} investigate attitudes toward agency of fictional AI robots, and find survey respondents do not typically consider AI systems to be moral agents -- they tend to place moral responsibility on developers, not on AI systems as agents.   
Similarly, \citet{robinette_overtrust_2016} examines (over)trust of autonomous systems and find in emergency situations, people will follow robots into further danger, because they attribute the agency of the robot to the (assumed capable and responsible) designers. As people in these situations appear to, we distinguish agency from responsibility, and emphasize that attribution of agency to an autonomous system in no way shifts moral responsibility from humans onto that system. 

\citet{ai_myths_myth_nodate,leufer_why_2020} examine another aspect of this problem, describing AI \textbf{agency as a myth} which masks human agency (and therefore responsibility). The authors contend that anthropomorphization of AI systems contributes to mystification of the underlying technology and sociotechnical blindness \cite{johnson_reframing_2017}, wherein people ``believe AI systems got to be the way they are without human intervention'', and obscuring of the (often exploitative) human labour which enables AI systems to exist \citep{gray_ghost_2019,pasquinelli_nooscope_2021,perrigo_exclusive_2023}. While we strongly agree with all these points, we reach the opposite conclusion -- AI agency
(in the sense of our work) 
is not a myth, it is a reality of increasing sociotechnical importance. It is precisely because of the importance of problems like these (responsibility gap, mystification, sociotechnical blindness, masking human agency and labour, etc.) and their far-ranging implications that we need to carefully examine the agency of AI systems, not dismiss it out of hand. If we think it is categorically impossible for AI systems to have agency, we will never be able to accurately recognize when we are giving up our agency to them. 

A related concept we wish to distinguish from agency is autonomy. In our framework, autonomy corresponds most closely to directness of impact, with some overlap in the three other categories. While it is often an intuitive or useful description of a system, we find it combines distinct phenomena we wish to distinguish with our characteristics. \citet{bekey_autonomous_2005} defines \textbf{autonomy} as ``the ability to operate without a human operator for a protracted period of time.'' Many factory robots are highly autonomous, but they operate strictly within the confines of a factory, and the actions they take affect only the intended outcome (e.g. the product they're making) -- they are autonomous but do not have agency. \citet{welsh_regulating_2019} presents a series of protocols which can be used to govern the use of lethal autonomous weapons, emphasizing the need for human-in-the-loop decision making -- i.e. to ensure all agency rests with human controllers. 

In this vein, our focus on agency also shares many commonalities with work from the FATE community on establishing the harms of \textbf{automated decision-making (ADM)}. ADM involves the use of algorithms to make decisions or enact policies without human intervention. Given its applications in recommendations \citep{milano_recommender_2020,li_fairness_2022}, health-care systems \citep{obermeyer_dissecting_2019,sendak_human_2020,fogliato_who_2022}, the judicial sector \citep{barabas_studying_2020,green_false_2020,zilka_transparency_2022}, and public services \citep{loi_towards_2021,stapleton_imagining_2022,black_algorithmic_2022}, ADM can often exhibit similar kinds of diffuse and long-term harms to those we discuss coming from increased agency. Given the commonalities, many of the harms of ADM also apply to increasingly agentic systems, as we discuss in \Cref{sec:potential-harms}. But there are two key differences between the body of work on ADM and our work. First, with the term agency we emphasize lack of explicit or low-level instructions for behaviour - we might specify a task, but not \textit{how} to solve that task. 
Second, our work explicitly targets systems that are \textit{increasingly} agentic, such as reinforcement-learning systems that are capable of making decisions in an open-ended environment over long time horizons without human intervention. Such systems have not been the focus of work in ADM simply because they have not yet seen widespread public deployment. We thus consider our focus on agency to be a continuation of current work on ADM.

Some philosophical work on agency also focuses on mental states such as 
consciousness, emotions, and subjective experience \citep{schlosser_agency_2019}. Entities with these mental states have personalities, and feel things like pleasure, curiosity, pain, embarrassment, fear and joy. 
Our work does not address experience or consciousness, only agency.

\subsection{Potential Objections to our Use of Agency}

One objection against framing algorithmic systems as agents is that it distracts from the responsibility of humans. 
As noted above, we characterize agency as separate from responsiblity. As many authors suggest, we strongly agree that attention should be directed towards holding corporations, regulators, developers, etc. (\textit{actors} for short in this section) accountable \citep{nissenbaum_accountability_1996,johnson_reframing_2017,wieringa_what_2020,cooper_accountability_2022}. 
This claim is not in contention with the idea that algorithmic systems can be agentic in our narrow sense. Principals (actors) can be held responsible on behalf of their agents, such as when employers are held liable for negligent hiring when employees cause harm \citep{hickox_employer_2010}. 

We should also require more than just individual accountability. In addition to focusing on individual actors, we should also attend to structural factors that shape their behaviours. A developer is likely blameworthy at least to some extent when a system causes harm, but structural factors like economic incentives or company culture to push forward likely also play significant roles \citep{van_der_loeff_ai_2019,zwetsloot_thinking_2019}. 
As we will discuss in \Cref{sec:potential-harms}, viewing algorithmic systems as agents can in fact highlight harms and the collective responsibility we have to prevent them.

\section{The Need to Anticipate Harms from Increasingly Agentic Systems}\label{sec:ongoing}
We argue for the need to anticipate harms from increasingly agentic systems. Anticipation is about two things: (1) the development of systems with increasing agency and (2) the deployment of systems with more agency than those already deployed. We touch upon trends in ML development and deployment as well as some reasons to expect these trends to continue. In \Cref{sec:objections} we respond to some potential objections.

\subsection{Trends in Development and Deployment}

We aim to show two things in this section. First: development of increasingly agentic systems has proceeded by consistently overcoming technical challenges. Second: deployment of increasingly agentic systems has occurred because these systems have increasingly practical skills that are useful for real-world applications.


\subsubsection{Overcoming Technical Challenges to Build Increasingly Agentic Systems}


Reinforcement learning (RL), as one of the major paradigms of machine learning, has a major focus on the construction of agents \citep{sutton_reinforcement_2018,sutton_alberta_2022}. In particular, RL is about designing systems to learn, without human intervention, to achieve a goal encoded in a reward function. Prior to 2013, RL systems were developed mainly for a restricted set of simple domains \citep{coulom_reinforcement_2002}. The introduction of deep learning to RL systems produced superhuman performance on a wider variety of narrow tasks with limited to no human supervision, including but not limited to increasingly complex board games \citep{silver_mastering_2016,silver_mastering_2017,brown_superhuman_2019,perolat_mastering_2022} and video games \citep{mnih_playing_2013, schrittwieser_mastering_2020,ye_mastering_2021}. Subsequent work has greatly improved the performance of RL systems on more complex, open-ended environments. For instance, DreamerV3 \citep{hafner_mastering_2023} collected diamonds from scratch without human data or curricula in MineCraft, which has been a longstanding challenge because the task is extremely complex and open-ended.
Another striking example comes from Diplomacy, a complex, multiplayer board game involving tactical coordination and natural language negotiation. The recent Cicero \citep{bakhtin_human-level_2022}, integrating a language model with planning and RL algorithms, is the first AI to achieve human-level performance in Diplomacy. Such systems have demonstrated strong capabilities to interact with complex environments and humans to accomplish their goals that require long-horizon planning.

We emphasize that for all the systems we have mentioned in this section so far, designers do not specify how the tasks were to be completed. In the current scientific paradigm of large-scale deep-learning, one instead provides high-level learning algorithms that tend to be task-agnostic or adapt to new tasks efficiently \citep{beck_survey_2023,dong_survey_2022}. One particular example to highlight is AdA \citep{adaptive_agent_team_human-timescale_2023}, which adapts to open-ended, novel, embodied 3D problems as quickly as humans, without human specification of how to solve problems. 

\subsubsection{The Increasing Deployment of Increasingly Agentic Systems}\label{subsubsec:availability}
The practicality of systems has increased along with their agency. Increasing practicality means that increasingly agentic systems are more likely to be found making decisions in the real world. Major companies have been deploying increasingly agentic systems to control parts of their operation. For example, DeepMind and Google use RL for controlling commercial cooling systems and data centers
\citep{evans_deepmind_2016, kava_better_2014}.
Amazon has applied RL to supply chain optimization problems \citep{schmelzer_amazon_2019}.
Additionally, there has been an increasing amount of research in recommender systems to optimize long-term metrics such as engagement via reinforcement learning \citep{afsar_reinforcement_2022}. Major recommendation companies such as Meta \citep{gauci_horizon_2019}, YouTube \citep{association_for_computing_machinery_acm_reinforcement_2019}, and Spotify \citep{engineering_shifting_2021} have already deployed RL-based recommender systems on their live products.

Systems that can competently operate across different data modalities and tasks are plausibly more useful than more narrow systems, regardless of how agentic they are. Before the current era of large language models (LLMs) \citep{brown_language_2020, bommasani_opportunities_2022, rae_scaling_2022, srivastava_beyond_2022}, few systems competently performed out-of-the-box on a range of natural language tasks \citep{brown_language_2020}.
Recent models \citep{alayrac_flamingo_2022, zeng_socratic_2022, reed_generalist_2022} can even handle multiple data modalities simultaneously. GATO \citep{reed_generalist_2022} can complete tasks using the same model and weights in vastly different domains, such as Atari, image captioning, dialogue, and robotics. As systems become increasingly agentic, the systems that are increasingly domain general seem likely to see more practical application. 

Increasingly agentic systems are also becoming more available to the general public. Although language models are only trained on next-token prediction, they can be leveraged to interact with APIs and accomplish a wide variety of multi-step digital tasks with increasingly less explicit human intervention \citep{nakano_webgpt_2022,chase_langchain_2022, menick_teaching_2022}. Adept's ACT-1 \citep{adept_act-1_2022} is a system in development which purportedly can perform an arbitrary task on your computer, such as searching for and buying an item online, through a single text command. 
OpenAI's ChatGPT \citep{openai_chatgpt_2023} has plug-ins that can interface with a wideo variety of web-apps, including Gmail and a web browser. AutoGPT \citep{noauthor_auto-gpt_2023} chains together an arbitrary number of GPT-4 calls to accomplish a high-level task on one's computer without one's intervention. 

Despite the progress so far, systems still have limitations and there are still barriers to the deployment of more agentic systems. For example, the raw task performance of generalist systems \citep{reed_generalist_2022, zeng_socratic_2022} is still limited to tasks where expert data is available and has not achieved human level on all tasks. In the realm of language models, recent studies \citep{valmeekam_large_2022, ji_survey_2022} have shown that large language models can perform poorly on planning and reasoning tasks, and such systems are prone to hallucinate unintended text, which fails to meet users' intents on many real-world scenarios. However, we note the pace of development and deployment are still rapidly increasing, not decreasing. We anticipate that current limitations and barriers will be surpassed or ignored in the pressure to deploy.  
\subsection{Factors in the Continued Development and Deployment of Increasingly Agentic Systems}
For current AI models, there are strong incentives for continued investment and development despite uncertainty around how their future capabilities will emerge~\cite{ganguli_predictability_2022}.
Similarly, a number of reasons suggest the potential for development and deployment of increasingly agentic algorithmic systems. These factors are the economic and military advantages afforded by increasingly agentic systems, scientific curiosity and prestige, a lack of regulatory barriers, and emergent agency. The first three reasons are sociopolitical, while the last reason concerns potentially surprising technical properties of ML systems. Taken together, these increase our subjective likelihood that systems will become increasingly agentic.




\subsubsection{Economic Incentives}\label{sec:incentives}
Actors who deploy more agentic systems than their competitors would likely generate more profit because of increased automation.
First, more agentic systems might be able to perform tasks much more cheaply than a less agentic systems. A less agentic system by definition would require more human intervention, whether to make decisions or specify explicit procedures for task completion. 
%
Second, more agentic systems will often be more effective at performing tasks than less agentic systems. Part of an increase of agency is the degree to which a system achieves a goal without operators or designers to specify how. The upshot is that the search space of solutions to a problem is larger for a more agentic system, which could result in solutions that would be much more efficient than those a human could have found. That AlphaGo \citep{silver_mastering_2016} beat Lee Sedol, the world Go champion, with the apparently confusing move 37 is evidence of this possibility. 



\subsubsection{Military Incentives}
Militaries may perceive that increasingly agentic systems could provide capability advantages over adversaries that are constrained by human decision-making.
The introduction by any one military of a more agentic system could upset a balance of power and force others to pursue similar developments in an unsafe race to the bottom, mirroring other races for technologies such as nuclear weapons, and ballistic and hypersonic missiles \citep{dafoe_ai_2018}.
The UK's defence AI strategy \citep{noauthor_defence_2022} frames advances in AI as being an area of ``geostrategic competition'' and ``a battleground for competing ideologies'', but also imposes no governance or oversight mechanisms on increasingly agentic systems, focusing such efforts ``on effects rather than the nature of any particular technology.''
Total bans on developments for such highly autonomous systems may be difficult to introduce and maintain, and it may be easier to pursue nonproliferation of such advances beyond a small set of technologically advanced users \cite{trager_deliberating_2022}.


\subsubsection{Scientific Curiosity and Prestige}
Developing increasingly agentic systems is an object of scientific curiosity and also confers status, a motivation that contributes to a prestige race at varying levels between actors in the AI research system.
For individual researchers this emerges through standard metrics such as paper publications and grant awards that support climbing the academic career ladder, but for many leading figures the ambitions transcend these: Geoffrey Hinton -- a pioneer deep learning research -- has stated that ``the prospect of discovery is too sweet'' in spite of his beliefs that ``political systems will use [AI] to terrorize people'' \cite{khatchadourian_doomsday_2015}, and Rich Sutton -- a pioneer in reinforcement learning -- has stated that creating ``beings of far greater intelligence than current humans'' (that would necessarily be agentic) will be ``the greatest intellectual achievement of all time'' and ``a great and glorious goal'' \cite{sutton_notitle_2022}.
For companies, developing increasingly agentic systems could drive the most impactful research and development outputs, increasing attractiveness to the best scientific talent in a competitive hiring pool.
For nations, highly visible scientific demonstrations may act as demonstrations of broader state capacity, and prestige may be as motivating a force as security for competitive races with peers and adversaries~\cite{barnhart_emerging_2022}

\subsubsection{Lack of Regulatory Barriers}
Regulatory efforts for AI have focused largely on salient risks, rather than on anticipatory governance mechanisms that are proactive to future advances in AI capabilities \citep{gesley_regulation_2019}.
For example, the EU AI act currently proposes to target regulation according to tiers of risk determined by type of data use and deployment setting, and efforts in the UK take a sectoral focus on regulating only the applications of AI, but neither covers development of agentic AI systems that could both be intrinsically high-risk and could underlie progress and use across a variety of sectors and domains~\cite{edwards_eu_2022,noauthor_establishing_2022}.
Accordingly, development and deployment in this space is effectively unregulated and without any clear possibility of regulation in the near future.

\subsubsection{Emergent Agency}\label{sec:emergent-agency}
Even if designers do not explicitly build more agency into their systems, it may emerge from general capability improvements. Recent works discuss emergent behaviors of large language models. \citet{bommasani_opportunities_2022} introduce emergence as 
a ``behavior of a system [that] is implicitly induced rather than explicitly constructed; it is both the source of scientific excitement and anxiety about unintended consequences.''
For example, LLMs are trained to model a distribution of internet text; this training leads to \emph{emergent behavior} such as learning from very few examples \citep{olsson_-context_2022}, or arithmetic \citep{brown_language_2020}, or even the ability itself to perform sequential reasoning \citep{wei_chain--thought_2023}. Many of these abilities only emerge at a certain scale, or after a certain point in the training process \citep{wei_emergent_2022}.

When emergent behavior increases the agency of a system we can speak of \emph{emergent agency}. One particularly striking example is the ability of LLMs to simulate the \textbf{human agents} who are the sources of the training data. For example, \citet{maraoz_interviewing_2021} uses GPT-3 to write a transcript of a conversation between themselves and Albert Einstein, and others have used LLMs to retroactively simulate user studies from psychology and economics \cite{aher_using_2022}. The seeming fidelity of such texts has motivated some to argue that LLMs have a general ability to simulate human agents \citep{andreas_language_2022}. 

%

Some emergent capabilities relate directly to our characterization of agency. \citet{wei_chain--thought_2023} show that adding ``let's think step-by-step'' vastly improves sequential reasoning capabilities in LLMs, 
 a capability which is useful for  performing tasks over long time horizons.
\citet{adaptive_agent_team_human-timescale_2023} show that scaling up a particular approach leads to RL systems that capably adapt to open-ended, novel 3D problems as well as humans can, without human intervention on how to solve the problem.

\subsection{Potential Objections to our Characterization of ML Progress}\label{sec:objections}

\subsubsection{The Need for Anticipation of Increasingly Agentic Systems is Small}
Earlier, we distinguished between two things to anticipate:\textbf{ (1)} the increasing agency of developed systems and \textbf{(2) }the deployment of systems with more agency than those already deployed. We respond to objections against both points. 

One could accept the need for attention to \textbf{(1)}, but maintain that the need is small given that technical improvements to increase agency occur much more slowly than we have characterized. Indeed, past beliefs in rapid pace of artificial intelligence research have been overoptimistic \citep{dreyfus_alchemy_1965}. Barriers to increasing agency include acting capably over long time horizons \citep{valmeekam_large_2022} and with an accurate understanding of the world \citep{bender_climbing_2020}. These challenges are real and there is by no means any certainty that the ML research community will overcome them. Even the perceived agency of algorithmic systems depends heavily on (sometimes exploitative) human labor and data extraction \citep{gray_ghost_2019,pasquinelli_nooscope_2021}. Moreover, it can be difficult to measure the rate of progress towards agentic systems. \citet{dehghani_benchmark_2021} provide evidence that factors other than ``fundamental algorithmic superiority'' may lead to the perception that a particular method is superior. \citet{raji_ai_2021} discuss several issues with benchmarking, including construct invalidity and limitations in scope. 

We have no disagreements on the technical challenges of developing systems of increased agency. We are also not claiming that systems of significantly greater agency than those in development already (e.g., compared to ACT-1 \citep{adept_act-1_2022}, GATO \citep{reed_generalist_2022}) will be coming soon. Rather, our view is that even absent significant technical breakthroughs, continued work within the current scientific paradigm \cite{kuhn_structure_2012} of scaling deep-learning seems likely to generate systems that are appreciably more agentic than current systems. Scaling laws provide predictable relationships between the amount of compute and data used to train model of a given size, and the performance of a model on some metric. Of particular interest for increasing agency is that scaling laws have been derived for reinforcement learning \citep{gao_scaling_2022,hilton_scaling_2023,adaptive_agent_team_human-timescale_2023} and generative modeling \citep{kaplan_scaling_2020,hoffmann_empirical_2022}. There is also initial work into developing scaling laws for robotics \citep{caballero_broken_2023}. The upshot is that continued training of larger models with more compute and data seems likely to increase the ability of systems to act in environments of increasing scope, over longer time horizons, to achieve goals without significant designer/operator intervention. 

One could also object to the need for attention to (2). Even if a system that is more agentic than those currently deployed has been developed, there might still be strong reasons against deployment, despite the incentives in \Cref{sec:incentives}. \citet{raji_fallacy_2022} argues that deployed AI systems often simply do not work, suffering from issues such as robustness failures, missing safety features, or being set to perform impossible tasks (such as inferring criminality from appearance). Given that increasingly agentic systems would be more capable of achieving goals without human specification of how, the failures that \citet{raji_fallacy_2022} highlight could disincentivize adoption of increasingly agentic systems, even if they were developed.

The likely failures of a more agentic system (relative to what has been deployed already) are certainly a barrier to deployment -- given the disproportionate impact of these failures on already marginalized groups, we would hope and advocate for restrictions on deployment \citep{brundage_toward_2020}. However, we think that this barrier is unfortunately weak relative to countervailing forces.  Hype around the (claimed) functionalities of ML systems is strong \citep{broussard_artificial_2019,schulz_industry_2019,natale_imagining_2020}, which is unsurprising given massive financial investments \citep{giattino_artificial_2022}. Continued cycles of deployment and failure \citep{buolamwini_gender_2018,obermeyer_dissecting_2019,wallace_universal_2019,ribeiro_auditing_2020,barabas_studying_2020,wolfe_american_2022,piantadosi_meaning_2022} suggest that increasingly agentic systems will likely be deployed according to industry interests, and not the interest of those most likely to be harmed.

\subsubsection{Techno-Determinism}
An objection against our characterization of increasingly agentic systems is that it is techno-deterministic -- it assumes that AI development is inevitable and determines the direction of sociocultural development \citep{wyatt_technological_2008}. This objection comes in two parts. Firstly, the perceived inevitability of ML progress nullifies accountability of those developing the systems and removes reason to regulate or stop development. Secondly, techno-determinism neglects social and cultural structures and implies a reductionist view of the harms caused by ML systems. Related to this is the adoption of discourse around ML systems that their capabilities are both scientifically impossible to explain, and yet deterministic in their societal impact \citep{campolo_enchanted_2020}. 
Some who study the harms of more agentic systems have also been accused of techno-optimism -- optimism about the potential of technology to solve major social problems -- and techno-determinism \citep{cremer_democratising_2021}. The problems include a disproportionately high reliance on technological solutions and neglect of insights from structural aspects of risk-analysis.  

We do not dispute the dangers of techno-determinism or techno-optimism. However, careful work on identifying and mitigating harms of increasingly agentic systems need not rely on or contribute to either. For example, one can be engaged in activism to ban specific uses or developments of increasingly agentic AI, while concurrently pursuing sociotechnical research to mitigate those systems' potential harms. In this framing, the sociotechnical work can be seen as an attempt to reduce harm in the case that one's broader attempts to change the field's course of action are not successful. While it can be argued that working on such harm reduction contributes to perceptions of inevitability or deployment incentives, being thoughtful in the framing of one's work can significantly contribute to avoiding this issue.

\section{Anticipated Harms from Increasingly Agentic Systems}\label{sec:potential-harms}

The previous section argued for the need to anticipate the harms of increasingly agentic systems. We now delve into some of these harms and why they are of especial importance for the FATE community.

\subsection{Systemic, Delayed Harms}

A systemic harm is a harm that is pervasively embedded in society. A delayed harm is a harm whose cause has a non-immediate impact. Systemic, delayed harms from algorithmic systems 
negatively influence groups of people in non-immediate ways. While harder to analyze than immediate harms, systemic and delayed harms might also be more insidious, as they can be caused even by low-stakes decision making systems. Each action might not seem consequential on its own, but, in aggregate, the outcomes can be destructive, long-lasting, and hard to fix. For example, there has recently been evidence that a single rent-setting algorithm might have significantly contributed to an increase in housing rental costs across the US \cite{vogell_how_2022}.
 
The FATE community has studied systemic and delayed harms in the past, such as environmental risks \citep{bender_dangers_2021}, concentration of power \citep{pratyusha_world_2020, abdalla_grey_2021}, unfair algorithmic hiring decisions \citep{suhr_does_2021}, and privacy infringements \citep{ekstrand_all_2018}. Another line of work focusing on the long-term fairness implications of decisions \cite{bird_exploring_2016,joseph_fairness_2016,jabbari_fairness_2017,liu_delayed_2018, damour_fairness_2020, zhang_fairness_2020}. More broadly, many have identified the systemic nature of general classes of harms, such as financial risk \citep{armour_systemic_2014}, racism \citep{braveman_systemic_2022}, and misogyny \citep{manne_down_2017}.

Social media is speculated to be a contributing factor to many systemic and delayed harms, including mental health issues \citep{hou_social_2019, yoon_is_2019}, the amplification of political polarisation \citep{whittaker_recommender_2021}, and the spread of fake news \citep{allcott_social_2017}. There is evidence on both sides for many of these issues \citep{lewis-kraus_how_2022, boxell_greater_2017, keles_systematic_2020}, but caution seems warranted due to the sheer scale of these platforms (e.g., Facebook has almost three-billion users \cite{meta_meta_2023}). 

While many of these harms do not involve the use of algorithms that are trained to act over long time horizons \cite{jiang_degenerate_2019}, the application of reinforcement-learning based recommendation systems (RLRS) in today's social media platforms warrant additional reason for concern. In particular, \citet{krueger_hidden_2020, evans_user_2022,carroll_estimating_2022} show that long time-horizon systems, such as RLRS, will have incentives to change or manipulate users' internal states (e.g. preferences, beliefs, and psychology) for the purposes of increasing the metrics the RLRS systems are optimizing. While some work has also investigated potential solutions \citep{farquhar_path-specific_2022, carroll_estimating_2022}, how to practically measure and address these issues in real-world RLRS remains an open problem \citep{carroll_characterizing_2023}. Notably, such systems are not speculative: RLRS are now increasingly applied by major social media providers (such as YouTube or Facebook), as discussed in \Cref{subsubsec:availability}.


\subsection{Collective Disempowerment}\label{sec:disempowerment}
We take \textbf{collective self-governance} to be the capacity and ongoing act of deciding collectively how to govern one's community, whether it be a small, local community, a state, or human societies at large \citep{christiano_democracy_2022}. Collective self-governance is about power, which is a core theme in FATE work \citep{barabas_studying_2020,kasy_fairness_2021,abdalla_grey_2021,boag_tech_2022,young_confronting_2022,birhane_forgotten_2022}. Indeed, the FATE community has extensively studied the ways in which automated decision-making can disempower individuals, by impairing human decision-making \citep{green_disparate_2019,green_ai_2021} or subjecting individuals to oppressive institutions \citep{barabas_interventions_2018,barabas_studying_2020,green_false_2020,zilka_transparency_2022}. 
We extend this ongoing discussion by pointing to some ways in which increasingly agentic systems can result in collective disempowerment. A key underlying point will be that increasingly agentic systems will likely seem more capable of handling  more important societal functions without significant operator or designer intervention, as we discussed in \Cref{sec:incentives}. We discuss two possibilities: a situation in which power diffuses away from all humans, and a situation in which power concentrates in the hands of a few.

\subsubsection{Diffusion of Power Away from Humans}
As systems become increasingly agentic, they have increasing control of societal functions in many ways. At one end, humans in a particular social structure may decide to cede decision-making power to an particular system, such as one that decides taxation policy \citep{zheng_ai_2021}. 
At the other end, power may gradually be ceded, as separate groups are incentivized to delegate more central functions to increasingly agentic systems per \Cref{sec:incentives}. \citet{nissenbaum_accountability_1996,cooper_accountability_2022} examine the erosion of accountability that externalizes algorithmic harms. Even if collective disempowerment is a risk, it might not be a large enough risk for a single party to be concerned. 
In either case, ceding decision-making power to such systems is \textit{not} inevitable; it would be a result of collective human decisions.




Regardless of how power is ceded, any group might have increasing difficulty in controlling increasingly agentic systems. Specifying a correct objective function is quite difficult \citep{krakovna_specification_2020,skalse_defining_2022}. Even a system successfully trained under a correctly specified objective function may do something completely different in a different environment \citep{langosco_goal_2022,shah_goal_2022}. 
Additional problems remain in understanding how to manage the interests of multiple stakeholders \citep{dafoe_open_2020}. As well, it would likely be extremely difficult to understand the decisions of the controlling system(s). Analysis of a single decision is likely insufficient for understanding the reasons for a series of long-term decisions (i.e., the overall plan). Collective self-governance requires not just having decisions be made, but understanding why those decisions are made, which \citet{lazar_legitimacy_2022} terms the publicity requirement. \citet{lazar_legitimacy_2022} argues that failure to satisfy this requirement delegitimizes the exercise of political authority,  by nullifying the moral effectiveness of consent. 


\subsubsection{Exacerbating the Extreme Concentration of Power Amongst the ``Coding Elite''}
The FATE community has highlighted the concerning ways in which the deployment of algorithmic systems has concentrated power in the hands of designers and/or operators. \citet{kasy_fairness_2021} argues that common notions of fairness legitimize hierarchies that are the result of historical injustice. They also provide a framework to reason about the impact of algorithmic decisions on the distribution of power. \citet[p.~217]{burrell_society_2021} identifies the \textbf{coding elite} -- a nebula of software developers, tech CEOs, investors, and computer science and engineering professors, among
others, often circulating effortlessly between these influential roles -- as a main beneficiary of the concentration of power. According to \citet{burrell_society_2021}, the coding elite concentrates power by controlling the algorithms underlying the modern digital world, using that power to affect politics for their own gains. The amount of control exerted is already substantial with existing algorithmic systems, considering the centrality of the products of a handful of tech companies in our daily lives. 

Increasingly agentic systems threaten to exacerbate an already extreme concentration of power. First, \citet[p.~11]{ganguli_predictability_2022} show that the proportion of large-scale ML results from industry has dominated in the past few years. The importance of large-scale results for increasing agency is that, as we discussed in \Cref{sec:ongoing}, scaling up the compute, data, and parameters of a system provides a significant way to increase its agency, and is in some sense easier than deriving fundamental algorithmic insights. It therefore seems plausible that large industrial labs will continue to be the ones who deploy and profit the most from increasingly agentic systems. Second, increasingly agentic systems would likely enable the coding elite to integrate algorithms into more of society. There are many tasks now that are yet outside the reach of algorithmic systems, such as deciding national economic policy or running a business. Increasingly agentic systems seem more likely to be able to assume many of those tasks than current systems. 

\subsection{Harms Yet To Be Identified} 
In \Cref{sec:emergent-agency}, we identified emergent behaviours as a possible cause of increasingly agentic algorithmic systems. Here, we explain some emergent behaviours that could be the source of harms that have yet to be identified and the link of those behaviours with increasing agency. 


\subsubsection{Reward Hacking}
An RL system trained to maximize its score in the video game CoastRunners will drive off-track and keep turning in circles forever, thus achieving a high score despite not completing the race-track as intended by the programmers \citep{clark_faulty_2016}.
This kind of failure is called \textbf{reward hacking} \citep{krakovna_specification_2020,skalse_defining_2022}, which is when a system exploits a reward signal to achieve a goal in an unforeseen, perhaps undesirable way. As an instance of Goodhart's law \citep{goodhart_problems_1975}, reward hacking is a common problem in ML systems that involve elements associated with increasing agency, in particular goal-directedness.\footnote{A large number of examples of reward hacking are compiled in \href{https://docs.google.com/spreadsheets/d/e/2PACX-1vRPiprOaC3HsCf5Tuum8bRfzYUiKLRqJmbOoC-32JorNdfyTiRRsR7Ea5eWtvsWzuxo8bjOxCG84dAg/pubhtml}{this online spreadsheet}.}
Increased model size or training time can result in abrupt increases in reward hacking, because a more capable model is better able find unforeseen maxima of its reward function \citep{pan_effects_2022}. 

If increasingly agentic systems are deployed in consequential domains like finance, health care, and law, reward hacking could result in extremely negative outcomes. Even with knowledge of the possibility of reward hacking, designers might still deploy systems anyway if the harms from their systems are externalized, or if they judge the immediate likelihood of reward hacking to be low.

\subsubsection{Instrumental Goals}
An \textbf{instrumental goal} is a goal that is useful as a subobjective in pursuit of a specified goal.
A \textbf{convergent instrumental goal} is a goal that would be useful in pursuit of a wide range of possible goals.
For example, acquiring money is a convergent instrumental goal since money increases economic power and optionality.
Many convergent instrumental goals involve gaining some sort of power over the environment and other actors within it \citep{omohundro_basic_2008}. 
An algorithmic system that sought to gain power over other actors, such as through manipulation or threats \citep{kenton_alignment_2021}, would be concerning.
An additional concern would be if the same thing were to happen without explicit, malicious instructions from their designer(s) or operator(s) to perform such behaviour.
While this possibility remains uncertain, some initial evidence does not dismiss it.
\citet{perez_discovering_2022} show that increased training of a LLM with RL techniques can increase the proportion of the time that the LLM expresses the pursuit of convergent instrumental goals, such as gaining wealth and persuading the operator not to shut it off, without any apparent designer or operator instruction to do so.
To recall, RL is about the construction of agents, by training systems to act over long time horizons to achieve goals without explicit human intervention.
Training LLMs with RL techniques plausibly increases their agency; therefore, \citet{perez_discovering_2022} provides some evidence that increasing the agency of LLMs can be associated with an increase in the expression of convergent instrumental goals.
It is important not to overstate this early evidence; expressing a desire to pursue a goal is different from actually pursuing the goal in the world. 
Yet, this evidence should be taken as an additional reason for caution regarding increasingly agentic systems.

\section{Paths to Preventing Harms}\label{sec:future-work}


Much work remains in figuring out how to address the present need we have highlighted throughout our piece. We provide a preliminary discussion of some directions and tie them to existing work from the FATE community.

\subsection{Investigating the Sociotechnical Attributes of Increasingly Agentic Systems}
Several landmark works in the FATE community have involved audits \citep{raji_closing_2020} of algorithmic systems \citep{buolamwini_gender_2018,obermeyer_dissecting_2019,ribeiro_auditing_2020,black_algorithmic_2022}. Audits have motivated action from designers to reduce the harms of their systems \citep{raji_actionable_2019}. 

Since one typically audits a deployed system, it will be difficult to perform thorough audits of increasingly agentic systems before they are widely deployed. Nevertheless, there are a variety of ways to reason about the potential impacts of a system's deployment. Assuming either that we have the system in question or that we can simulate it faithfully \citep{elzayn_effects_2020,aher_using_2022,park_social_2022}, we can formulate and test hypotheses in simple experiments or simulation. 
Small-scale studies and simulations will almost certainly fail to capture perfectly what would happen if the system was actually deployed on a large. Nevertheless, such investigations can also highlight potentially concerning phenomena for further investigation. Failing to observe harm should not necessarily be taken to mean that a system is safe; on the other hand, observation of a potential harm in a pre-deployment study should motivate further research into understanding to what extent the harm would appear in practice.

More broadly, it might be possible to take inspiration from policymaking techniques such as scenario planning \cite{volkery_scenario_2009}, which involve thinking ahead about how to make effective policy decisions when uncertain about what the world will look like in the future. The emerging science of forecasting \citep{tetlock_superforecasting_2016} may also provide insights into anticipating the impacts of emerging systems.

Other tools from the FATE community may be helpful for characterizing the sociotechnical attributes of increasingly agentic systems, even before widespread deployment. For example, datasheets \citep{gebru_datasheets_2021} and model cards \citep{mitchell_model_2019} can highlight sources of harm, such as accountability gaps \citep{nissenbaum_accountability_1996}, in a way that does not depend upon a particular application. In the same vein, \citet{gilbert_reward_2022} introduces reward reports to document what it appears that systems are optimizing for, which may help to reduce the likelihood of unintended negative consequences from system operation. Interpretability work may also help us understand how a system is achieving a goal \citep{olsson_-context_2022,adebayo_post_2022}.

Another promising line of work is to propose both quantitative and qualitative metrics that build upon our characterization of agency. For instance, we could have metrics for measuring the degree to which an AI system is capable of accomplishing tasks in the real world. Having metrics for agency would facilitate the study of when agency causes or is correlated with observed negative impacts. Some existing work already measures aspects of agency, such as long-term planning \citep{valmeekam_large_2022} and goal-directedness \citep{pan_rewards_2023}.

\subsection{Regulatory and Institutional Interventions}
Stronger regulations could prevent some harms of increasingly agentic systems from occurring. Compute limits enforced by compute usage tracking \citep{brundage_toward_2020,shavit_what_2023}, while a source of serious privacy risks, could help to control the pace at which systems become increasingly agentic and permit more time to develop mitigations. Along this line, it might be collectively beneficial to decide upon a threshold of agency as a deployment bar. If an AI system surpassed this level of agency, it could be forbidden from application in certain consequential sectors, like energy, the military, finance, health care, and criminal justice. The FATE community has previously rallied around deeming certain applications off-limits for particular technologies, such as the use of deep learning to predict criminality \citep{coalition_for_critical_technology_abolish_2020}. 

Efforts to improve democratic control and oversight over AI development could help to address the incentives for the development of increasingly agentic systems that we highlighted in \Cref{sec:incentives}. \citet{tutt_fda_2017} proposes an ``FDA for algorithms'', which would scrutinize each algorithmic system before permitting its deployment, just as drugs are regulated in the United States. \citet{jernite_data_2022,huang_generative_2023} highlight the imbalance of power between AI developers and the rest of society, and propose frameworks and vehicles for collective data governance. Given the importance of data for training state-of-the-art systems \citep{hoffmann_empirical_2022}, democratic control over data usage could be an important check on the development of increasingly agentic systems.

\section{Conclusion}

Our work focused on the increasing prevalence of agency in machine-learning systems and associated harms. 
We situated our characterization of increasing agency in the context of diverse work on the meaning of agency. We argued that there is a need to anticipate the harms from increasingly agentic systems, given a strong track record of, and incentives for, technical developments and increasing deployment. We described some anticipated harms from increasingly agentic systems, namely that they could cause systemic and delayed harms, disempower human decision-making, exacerbate extreme concentrations of power, and be a source of additional unknown threats through emergent capabilities. 

Addressing the harms of increasingly agentic systems shares commonalities with central lines of work in the FATE community on anticipating the harms of algorithmic decision-making systems. Future work, such as investigations into the sociotechnical attributes of increasingly agentic systems and interventions upon the structural factors underlying their harms, readily follows from ongoing efforts. Immense pressure to develop and deploy emerging technologies should be met with similarly strong attempts to guide and constrain their impact.

\begin{acks}
In no particular order, we would like to thank Robert Harling, Max Kaufmann, Fernando Diaz, Seth Lazar, Janarthanan Rajendran, Nicolas Le Roux, Shahar Avin, Kayla Matteucci, Thomas Gilbert, Herbie Bradley, and Usman Anwar for insightful comments about the direction of the project and parts of the paper. All mistakes remain our own.

AC acknowledges funding from Open Philanthropy and the Effective Ventures Foundation. AW acknowledges support from a Turing AI Fellowship under EPSRC grant EP/V025279/1, and the Leverhulme Trust via CFI. KC acknowledges funding from the Marshall Scholarship and Cambridge Trust. YD acknowledges funding from Longview Philanthropy.
\end{acks}

\bibliographystyle{ACM-Reference-Format}
\bibliography{ARXIV}


\begin{thebibliography}{207}


\ifx \showCODEN    \undefined \def \showCODEN     #1{\unskip}     \fi
\ifx \showDOI      \undefined \def \showDOI       #1{#1}\fi
\ifx \showISBNx    \undefined \def \showISBNx     #1{\unskip}     \fi
\ifx \showISBNxiii \undefined \def \showISBNxiii  #1{\unskip}     \fi
\ifx \showISSN     \undefined \def \showISSN      #1{\unskip}     \fi
\ifx \showLCCN     \undefined \def \showLCCN      #1{\unskip}     \fi
\ifx \shownote     \undefined \def \shownote      #1{#1}          \fi
\ifx \showarticletitle \undefined \def \showarticletitle #1{#1}   \fi
\ifx \showURL      \undefined \def \showURL       {\relax}        \fi
\providecommand\bibfield[2]{#2}
\providecommand\bibinfo[2]{#2}
\providecommand\natexlab[1]{#1}
\providecommand\showeprint[2][]{arXiv:#2}

\bibitem[noa(2022a)]%
        {noauthor_defence_2022}
 \bibinfo{year}{2022}\natexlab{a}.
\newblock \bibinfo{booktitle}{\emph{Defence {Artificial} {Intelligence}
  {Strategy}}}.
\newblock \bibinfo{type}{{T}echnical {R}eport}. \bibinfo{institution}{Ministry
  of Defence}.
\newblock
\urldef\tempurl%
\url{https://www.gov.uk/government/publications/defence-artificial-intelligence-strategy/defence-artificial-intelligence-strategy}
\showURL{%
\tempurl}


\bibitem[noa(2022b)]%
        {noauthor_establishing_2022}
 \bibinfo{year}{2022}\natexlab{b}.
\newblock \bibinfo{booktitle}{\emph{Establishing a pro-innovation approach to
  regulating {AI}}}.
\newblock \bibinfo{type}{{T}echnical {R}eport}. \bibinfo{institution}{Office
  for Artificial Intelligence}.
\newblock
\urldef\tempurl%
\url{https://www.gov.uk/government/publications/establishing-a-pro-innovation-approach-to-regulating-ai/establishing-a-pro-innovation-approach-to-regulating-ai-policy-statement}
\showURL{%
\tempurl}


\bibitem[noa(2023)]%
        {noauthor_auto-gpt_2023}
 \bibinfo{year}{2023}\natexlab{}.
\newblock \bibinfo{title}{Auto-{GPT}: {An} {Autonomous} {GPT}-4 {Experiment}}.
\newblock
\newblock
\urldef\tempurl%
\url{https://github.com/Significant-Gravitas/Auto-GPT}
\showURL{%
\tempurl}
\newblock
\shownote{original-date: 2023-03-16T09:21:07Z}.


\bibitem[Abdalla and Abdalla(2021)]%
        {abdalla_grey_2021}
\bibfield{author}{\bibinfo{person}{Mohamed Abdalla} {and}
  \bibinfo{person}{Moustafa Abdalla}.} \bibinfo{year}{2021}\natexlab{}.
\newblock \showarticletitle{The {Grey} {Hoodie} {Project}: {Big} {Tobacco},
  {Big} {Tech}, and the {Threat} on {Academic} {Integrity}}. In
  \bibinfo{booktitle}{\emph{Proceedings of the 2021 {AAAI}/{ACM} {Conference}
  on {AI}, {Ethics}, and {Society}}}. \bibinfo{publisher}{ACM}.
\newblock
\urldef\tempurl%
\url{https://doi.org/10.1145/3461702.3462563}
\showDOI{\tempurl}


\bibitem[Abebe et~al\mbox{.}(2020)]%
        {abebe_roles_2020}
\bibfield{author}{\bibinfo{person}{Rediet Abebe}, \bibinfo{person}{Solon
  Barocas}, \bibinfo{person}{Jon Kleinberg}, \bibinfo{person}{Karen Levy},
  \bibinfo{person}{Manish Raghavan}, {and} \bibinfo{person}{David~G.
  Robinson}.} \bibinfo{year}{2020}\natexlab{}.
\newblock \showarticletitle{Roles for computing in social change}. In
  \bibinfo{booktitle}{\emph{Proceedings of the 2020 {Conference} on {Fairness},
  {Accountability}, and {Transparency}}}. \bibinfo{publisher}{ACM}.
\newblock
\urldef\tempurl%
\url{https://doi.org/10.1145/3351095.3372871}
\showDOI{\tempurl}


\bibitem[Abebe and Goldner(2018)]%
        {abebe_mechanism_2018}
\bibfield{author}{\bibinfo{person}{Rediet Abebe} {and} \bibinfo{person}{Kira
  Goldner}.} \bibinfo{year}{2018}\natexlab{}.
\newblock \showarticletitle{Mechanism design for social good}.
\newblock \bibinfo{journal}{\emph{AI Matters}} \bibinfo{volume}{4},
  \bibinfo{number}{3} (\bibinfo{date}{Oct.} \bibinfo{year}{2018}),
  \bibinfo{pages}{27--34}.
\newblock
\urldef\tempurl%
\url{https://doi.org/10.1145/3284751.3284761}
\showDOI{\tempurl}


\bibitem[Abid et~al\mbox{.}(2021)]%
        {abid_persistent_2021}
\bibfield{author}{\bibinfo{person}{Abubakar Abid}, \bibinfo{person}{Maheen
  Farooqi}, {and} \bibinfo{person}{James Zou}.}
  \bibinfo{year}{2021}\natexlab{}.
\newblock \showarticletitle{Persistent {Anti}-{Muslim} {Bias} in {Large}
  {Language} {Models}}. In \bibinfo{booktitle}{\emph{Proceedings of the 2021
  {AAAI}/{ACM} {Conference} on {AI}, {Ethics}, and {Society}}}.
  \bibinfo{publisher}{ACM}.
\newblock
\urldef\tempurl%
\url{https://doi.org/10.1145/3461702.3462624}
\showDOI{\tempurl}


\bibitem[Adebayo et~al\mbox{.}(2022)]%
        {adebayo_post_2022}
\bibfield{author}{\bibinfo{person}{Julius Adebayo}, \bibinfo{person}{Michael
  Muelly}, \bibinfo{person}{Harold Abelson}, {and} \bibinfo{person}{Been Kim}.}
  \bibinfo{year}{2022}\natexlab{}.
\newblock \showarticletitle{Post hoc {Explanations} may be {Ineffective} for
  {Detecting} {Unknown} {Spurious} {Correlation}}. In
  \bibinfo{booktitle}{\emph{International {Conference} on {Learning}
  {Representations}}}.
\newblock
\urldef\tempurl%
\url{https://openreview.net/forum?id=xNOVfCCvDpM}
\showURL{%
\tempurl}


\bibitem[Adept(2022)]%
        {adept_act-1_2022}
\bibfield{author}{\bibinfo{person}{Adept}.} \bibinfo{year}{2022}\natexlab{}.
\newblock \bibinfo{title}{{ACT}-1: {Transformer} for {Actions}}.
\newblock
\newblock
\urldef\tempurl%
\url{https://www.adept.ai/act}
\showURL{%
\tempurl}


\bibitem[Afsar et~al\mbox{.}(2022)]%
        {afsar_reinforcement_2022}
\bibfield{author}{\bibinfo{person}{M.~Mehdi Afsar}, \bibinfo{person}{Trafford
  Crump}, {and} \bibinfo{person}{Behrouz Far}.}
  \bibinfo{year}{2022}\natexlab{}.
\newblock \bibinfo{title}{Reinforcement learning based recommender systems: {A}
  survey}.
\newblock
\newblock
\urldef\tempurl%
\url{https://doi.org/10.48550/arXiv.2101.06286}
\showDOI{\tempurl}
\newblock
\shownote{arXiv:2101.06286 [cs]}.


\bibitem[Aher et~al\mbox{.}(2022)]%
        {aher_using_2022}
\bibfield{author}{\bibinfo{person}{Gati Aher}, \bibinfo{person}{Rosa~I
  Arriaga}, {and} \bibinfo{person}{Adam~Tauman Kalai}.}
  \bibinfo{year}{2022}\natexlab{}.
\newblock \showarticletitle{Using {Large} {Language} {Models} to {Simulate}
  {Multiple} {Humans}}.
\newblock \bibinfo{journal}{\emph{arXiv preprint arXiv:2208.10264}}
  (\bibinfo{year}{2022}).
\newblock


\bibitem[Alayrac et~al\mbox{.}(2022)]%
        {alayrac_flamingo_2022}
\bibfield{author}{\bibinfo{person}{Jean-Baptiste Alayrac},
  \bibinfo{person}{Jeff Donahue}, \bibinfo{person}{Pauline Luc},
  \bibinfo{person}{Antoine Miech}, \bibinfo{person}{Iain Barr},
  \bibinfo{person}{Yana Hasson}, \bibinfo{person}{Karel Lenc},
  \bibinfo{person}{Arthur Mensch}, \bibinfo{person}{Katie Millican},
  \bibinfo{person}{Malcolm Reynolds}, \bibinfo{person}{Roman Ring},
  \bibinfo{person}{Eliza Rutherford}, \bibinfo{person}{Serkan Cabi},
  \bibinfo{person}{Tengda Han}, \bibinfo{person}{Zhitao Gong},
  \bibinfo{person}{Sina Samangooei}, \bibinfo{person}{Marianne Monteiro},
  \bibinfo{person}{Jacob Menick}, \bibinfo{person}{Sebastian Borgeaud},
  \bibinfo{person}{Andrew Brock}, \bibinfo{person}{Aida Nematzadeh},
  \bibinfo{person}{Sahand Sharifzadeh}, \bibinfo{person}{Mikolaj Binkowski},
  \bibinfo{person}{Ricardo Barreira}, \bibinfo{person}{Oriol Vinyals},
  \bibinfo{person}{Andrew Zisserman}, {and} \bibinfo{person}{Karen Simonyan}.}
  \bibinfo{year}{2022}\natexlab{}.
\newblock \bibinfo{title}{Flamingo: a {Visual} {Language} {Model} for
  {Few}-{Shot} {Learning}}.
\newblock
\newblock
\urldef\tempurl%
\url{https://doi.org/10.48550/arXiv.2204.14198}
\showDOI{\tempurl}
\newblock
\shownote{arXiv:2204.14198 [cs]}.


\bibitem[Allcott and Gentzkow(2017)]%
        {allcott_social_2017}
\bibfield{author}{\bibinfo{person}{Hunt Allcott} {and} \bibinfo{person}{Matthew
  Gentzkow}.} \bibinfo{year}{2017}\natexlab{}.
\newblock \showarticletitle{Social {Media} and {Fake} {News} in the 2016
  {Election}}.
\newblock \bibinfo{journal}{\emph{Journal of Economic Perspectives}}
  \bibinfo{volume}{31}, \bibinfo{number}{2} (\bibinfo{date}{May}
  \bibinfo{year}{2017}), \bibinfo{pages}{211--236}.
\newblock
\showISSN{0895-3309}
\urldef\tempurl%
\url{https://doi.org/10.1257/jep.31.2.211}
\showDOI{\tempurl}


\bibitem[Andreas(2022)]%
        {andreas_language_2022}
\bibfield{author}{\bibinfo{person}{Jacob Andreas}.}
  \bibinfo{year}{2022}\natexlab{}.
\newblock \bibinfo{title}{Language {Models} as {Agent} {Models}}.
\newblock
\newblock
\urldef\tempurl%
\url{https://doi.org/10.48550/arXiv.2212.01681}
\showDOI{\tempurl}
\newblock
\shownote{arXiv:2212.01681 [cs]}.


\bibitem[Armour and Gordon(2014)]%
        {armour_systemic_2014}
\bibfield{author}{\bibinfo{person}{John Armour} {and}
  \bibinfo{person}{Jeffrey~N. Gordon}.} \bibinfo{year}{2014}\natexlab{}.
\newblock \showarticletitle{Systemic {Harms} and {Shareholder} {Value}}.
\newblock \bibinfo{journal}{\emph{Journal of Legal Analysis}}
  \bibinfo{volume}{6}, \bibinfo{number}{1} (\bibinfo{date}{Oct.}
  \bibinfo{year}{2014}), \bibinfo{pages}{35--85}.
\newblock
\showISSN{2161-7201}
\urldef\tempurl%
\url{https://doi.org/10.1093/jla/lau004}
\showDOI{\tempurl}


\bibitem[{Association for Computing Machinery (ACM)}(2019)]%
        {association_for_computing_machinery_acm_reinforcement_2019}
\bibfield{author}{\bibinfo{person}{{Association for Computing Machinery
  (ACM)}}.} \bibinfo{year}{2019}\natexlab{}.
\newblock \bibinfo{title}{"{Reinforcement} {Learning} for {Recommender}
  {Systems}: {A} {Case} {Study} on {Youtube}," by {Minmin} {Chen}}.
\newblock
\newblock
\urldef\tempurl%
\url{https://www.youtube.com/watch?v=HEqQ2_1XRTs}
\showURL{%
\tempurl}


\bibitem[Bakhtin et~al\mbox{.}(2022)]%
        {bakhtin_human-level_2022}
\bibfield{author}{\bibinfo{person}{Anton Bakhtin}, \bibinfo{person}{Noam
  Brown}, \bibinfo{person}{Emily Dinan}, \bibinfo{person}{Gabriele Farina},
  \bibinfo{person}{Colin Flaherty}, \bibinfo{person}{Daniel Fried},
  \bibinfo{person}{Andrew Goff}, \bibinfo{person}{Jonathan Gray},
  \bibinfo{person}{Hengyuan Hu}, \bibinfo{person}{Athul~Paul Jacob},
  \bibinfo{person}{Mojtaba Komeili}, \bibinfo{person}{Karthik Konath},
  \bibinfo{person}{Minae Kwon}, \bibinfo{person}{Adam Lerer},
  \bibinfo{person}{Mike Lewis}, \bibinfo{person}{Alexander~H. Miller},
  \bibinfo{person}{Sasha Mitts}, \bibinfo{person}{Adithya Renduchintala},
  \bibinfo{person}{Stephen Roller}, \bibinfo{person}{Dirk Rowe},
  \bibinfo{person}{Weiyan Shi}, \bibinfo{person}{Joe Spisak},
  \bibinfo{person}{Alexander Wei}, \bibinfo{person}{David Wu},
  \bibinfo{person}{Hugh Zhang}, {and} \bibinfo{person}{Markus Zijlstra}.}
  \bibinfo{year}{2022}\natexlab{}.
\newblock \showarticletitle{Human-level play in the game of {Diplomacy} by
  combining language models with strategic reasoning}.
\newblock \bibinfo{journal}{\emph{Science}} \bibinfo{volume}{378},
  \bibinfo{number}{6624} (\bibinfo{date}{Dec.} \bibinfo{year}{2022}),
  \bibinfo{pages}{1067--1074}.
\newblock
\urldef\tempurl%
\url{https://doi.org/10.1126/science.ade9097}
\showDOI{\tempurl}
\newblock
\shownote{Publisher: American Association for the Advancement of Science}.


\bibitem[Barabas et~al\mbox{.}(2020)]%
        {barabas_studying_2020}
\bibfield{author}{\bibinfo{person}{Chelsea Barabas}, \bibinfo{person}{Colin
  Doyle}, \bibinfo{person}{J.~B. Rubinovitz}, {and} \bibinfo{person}{Karthik
  Dinakar}.} \bibinfo{year}{2020}\natexlab{}.
\newblock \showarticletitle{Studying up}. In
  \bibinfo{booktitle}{\emph{Proceedings of the 2020 {Conference} on {Fairness},
  {Accountability}, and {Transparency}}}. \bibinfo{publisher}{ACM}.
\newblock
\urldef\tempurl%
\url{https://doi.org/10.1145/3351095.3372859}
\showDOI{\tempurl}


\bibitem[Barabas et~al\mbox{.}(2018)]%
        {barabas_interventions_2018}
\bibfield{author}{\bibinfo{person}{Chelsea Barabas}, \bibinfo{person}{Madars
  Virza}, \bibinfo{person}{Karthik Dinakar}, \bibinfo{person}{Joichi Ito},
  {and} \bibinfo{person}{Jonathan Zittrain}.} \bibinfo{year}{2018}\natexlab{}.
\newblock \showarticletitle{Interventions over {Predictions}: {Reframing} the
  {Ethical} {Debate} for {Actuarial} {Risk} {Assessment}}. In
  \bibinfo{booktitle}{\emph{Proceedings of the 1st {Conference} on {Fairness},
  {Accountability} and {Transparency}}} \emph{(\bibinfo{series}{Proceedings of
  {Machine} {Learning} {Research}}, Vol.~\bibinfo{volume}{81})},
  \bibfield{editor}{\bibinfo{person}{Sorelle~A. Friedler} {and}
  \bibinfo{person}{Christo Wilson}} (Eds.). \bibinfo{publisher}{PMLR},
  \bibinfo{pages}{62--76}.
\newblock
\urldef\tempurl%
\url{https://proceedings.mlr.press/v81/barabas18a.html}
\showURL{%
\tempurl}


\bibitem[Barnhart(2022)]%
        {barnhart_emerging_2022}
\bibfield{author}{\bibinfo{person}{Joslyn Barnhart}.}
  \bibinfo{year}{2022}\natexlab{}.
\newblock \bibinfo{title}{Emerging {Technologies}, {Prestige} {Motivations} and
  the {Dynamics} of {International} {Competition}}.  (\bibinfo{date}{Jan.}
  \bibinfo{year}{2022}).
\newblock
\urldef\tempurl%
\url{https://www.governance.ai/research-paper/emerging-technologies-prestige-motivations-and-the-dynamics-of-international-competition}
\showURL{%
\tempurl}


\bibitem[Beck et~al\mbox{.}(2023)]%
        {beck_survey_2023}
\bibfield{author}{\bibinfo{person}{Jacob Beck}, \bibinfo{person}{Risto Vuorio},
  \bibinfo{person}{Evan~Zheran Liu}, \bibinfo{person}{Zheng Xiong},
  \bibinfo{person}{Luisa Zintgraf}, \bibinfo{person}{Chelsea Finn}, {and}
  \bibinfo{person}{Shimon Whiteson}.} \bibinfo{year}{2023}\natexlab{}.
\newblock \bibinfo{title}{A {Survey} of {Meta}-{Reinforcement} {Learning}}.
\newblock
\newblock
\urldef\tempurl%
\url{https://doi.org/10.48550/arXiv.2301.08028}
\showDOI{\tempurl}
\newblock
\shownote{arXiv:2301.08028 [cs]}.


\bibitem[Bekey(2005)]%
        {bekey_autonomous_2005}
\bibfield{author}{\bibinfo{person}{G.A. Bekey}.}
  \bibinfo{year}{2005}\natexlab{}.
\newblock \bibinfo{booktitle}{\emph{Autonomous {Robots}: {From} {Biological}
  {Inspiration} to {Implementation} and {Control}}}.
\newblock \bibinfo{publisher}{MIT Press}.
\newblock
\showISBNx{978-0-262-02578-2}
\showLCCN{20040668}
\urldef\tempurl%
\url{https://books.google.ca/books?id=3xwfia2DpmoC}
\showURL{%
\tempurl}


\bibitem[Bender et~al\mbox{.}(2021)]%
        {bender_dangers_2021}
\bibfield{author}{\bibinfo{person}{Emily~M. Bender}, \bibinfo{person}{Timnit
  Gebru}, \bibinfo{person}{Angelina McMillan-Major}, {and}
  \bibinfo{person}{Shmargaret Shmitchell}.} \bibinfo{year}{2021}\natexlab{}.
\newblock \showarticletitle{On the {Dangers} of {Stochastic} {Parrots}: {Can}
  {Language} {Models} {Be} {Too} {Big}?}. In
  \bibinfo{booktitle}{\emph{Proceedings of the 2021 {ACM} {Conference} on
  {Fairness}, {Accountability}, and {Transparency}}}
  \emph{(\bibinfo{series}{{FAccT} '21})}. \bibinfo{publisher}{Association for
  Computing Machinery}, \bibinfo{address}{New York, NY, USA},
  \bibinfo{pages}{610--623}.
\newblock
\showISBNx{978-1-4503-8309-7}
\urldef\tempurl%
\url{https://doi.org/10.1145/3442188.3445922}
\showDOI{\tempurl}
\newblock
\shownote{event-place: Virtual Event, Canada}.


\bibitem[Bender and Koller(2020)]%
        {bender_climbing_2020}
\bibfield{author}{\bibinfo{person}{Emily~M. Bender} {and}
  \bibinfo{person}{Alexander Koller}.} \bibinfo{year}{2020}\natexlab{}.
\newblock \showarticletitle{Climbing towards {NLU}: {On} {Meaning}, {Form}, and
  {Understanding} in the {Age} of {Data}}. In
  \bibinfo{booktitle}{\emph{Proceedings of the 58th {Annual} {Meeting} of the
  {Association} for {Computational} {Linguistics}}}.
  \bibinfo{publisher}{Association for Computational Linguistics},
  \bibinfo{address}{Online}, \bibinfo{pages}{5185--5198}.
\newblock
\urldef\tempurl%
\url{https://doi.org/10.18653/v1/2020.acl-main.463}
\showDOI{\tempurl}


\bibitem[Bird et~al\mbox{.}(2016)]%
        {bird_exploring_2016}
\bibfield{author}{\bibinfo{person}{Sarah Bird}, \bibinfo{person}{Solon
  Barocas}, \bibinfo{person}{Kate Crawford}, \bibinfo{person}{Fernando Diaz},
  {and} \bibinfo{person}{Hanna Wallach}.} \bibinfo{year}{2016}\natexlab{}.
\newblock \bibinfo{title}{Exploring or {Exploiting}? {Social} and {Ethical}
  {Implications} of {Autonomous} {Experimentation} in {AI}}.
\newblock
\newblock
\urldef\tempurl%
\url{https://papers.ssrn.com/abstract=2846909}
\showURL{%
\tempurl}


\bibitem[Birhane et~al\mbox{.}(2022)]%
        {birhane_forgotten_2022}
\bibfield{author}{\bibinfo{person}{Abeba Birhane}, \bibinfo{person}{Elayne
  Ruane}, \bibinfo{person}{Thomas Laurent}, \bibinfo{person}{Matthew~S. Brown},
  \bibinfo{person}{Johnathan Flowers}, \bibinfo{person}{Anthony Ventresque},
  {and} \bibinfo{person}{Christopher~L. Dancy}.}
  \bibinfo{year}{2022}\natexlab{}.
\newblock \showarticletitle{The {Forgotten} {Margins} of {AI} {Ethics}}. In
  \bibinfo{booktitle}{\emph{2022 {ACM} {Conference} on {Fairness},
  {Accountability}, and {Transparency}}}. \bibinfo{publisher}{ACM}.
\newblock
\urldef\tempurl%
\url{https://doi.org/10.1145/3531146.3533157}
\showDOI{\tempurl}


\bibitem[Black et~al\mbox{.}(2022)]%
        {black_algorithmic_2022}
\bibfield{author}{\bibinfo{person}{Emily Black}, \bibinfo{person}{Hadi Elzayn},
  \bibinfo{person}{Alexandra Chouldechova}, \bibinfo{person}{Jacob Goldin},
  {and} \bibinfo{person}{Daniel Ho}.} \bibinfo{year}{2022}\natexlab{}.
\newblock \showarticletitle{Algorithmic {Fairness} and {Vertical} {Equity}:
  {Income} {Fairness} with {IRS} {Tax} {Audit} {Models}}. In
  \bibinfo{booktitle}{\emph{2022 {ACM} {Conference} on {Fairness},
  {Accountability}, and {Transparency}}}. \bibinfo{publisher}{ACM}.
\newblock
\urldef\tempurl%
\url{https://doi.org/10.1145/3531146.3533204}
\showDOI{\tempurl}


\bibitem[Blodgett et~al\mbox{.}(2021)]%
        {blodgett_stereotyping_2021}
\bibfield{author}{\bibinfo{person}{Su~Lin Blodgett}, \bibinfo{person}{Gilsinia
  Lopez}, \bibinfo{person}{Alexandra Olteanu}, \bibinfo{person}{Robert Sim},
  {and} \bibinfo{person}{Hanna Wallach}.} \bibinfo{year}{2021}\natexlab{}.
\newblock \showarticletitle{Stereotyping {Norwegian} {Salmon}: {An} {Inventory}
  of {Pitfalls} in {Fairness} {Benchmark} {Datasets}}. In
  \bibinfo{booktitle}{\emph{Proceedings of the 59th {Annual} {Meeting} of the
  {Association} for {Computational} {Linguistics} and the 11th {International}
  {Joint} {Conference} on {Natural} {Language} {Processing} ({Volume} 1: {Long}
  {Papers})}}. \bibinfo{publisher}{Association for Computational Linguistics},
  \bibinfo{address}{Online}, \bibinfo{pages}{1004--1015}.
\newblock
\urldef\tempurl%
\url{https://doi.org/10.18653/v1/2021.acl-long.81}
\showDOI{\tempurl}


\bibitem[Boag et~al\mbox{.}(2022)]%
        {boag_tech_2022}
\bibfield{author}{\bibinfo{person}{William Boag}, \bibinfo{person}{Harini
  Suresh}, \bibinfo{person}{Bianca Lepe}, {and} \bibinfo{person}{Catherine
  D'Ignazio}.} \bibinfo{year}{2022}\natexlab{}.
\newblock \showarticletitle{Tech {Worker} {Organizing} for {Power} and
  {Accountability}}. In \bibinfo{booktitle}{\emph{2022 {ACM} {Conference} on
  {Fairness}, {Accountability}, and {Transparency}}}. \bibinfo{publisher}{ACM}.
\newblock
\urldef\tempurl%
\url{https://doi.org/10.1145/3531146.3533111}
\showDOI{\tempurl}


\bibitem[Bommasani et~al\mbox{.}(2022)]%
        {bommasani_opportunities_2022}
\bibfield{author}{\bibinfo{person}{Rishi Bommasani}, \bibinfo{person}{Drew~A.
  Hudson}, \bibinfo{person}{Ehsan Adeli}, \bibinfo{person}{Russ Altman},
  \bibinfo{person}{Simran Arora}, \bibinfo{person}{Sydney von Arx},
  \bibinfo{person}{Michael~S. Bernstein}, \bibinfo{person}{Jeannette Bohg},
  \bibinfo{person}{Antoine Bosselut}, \bibinfo{person}{Emma Brunskill},
  \bibinfo{person}{Erik Brynjolfsson}, \bibinfo{person}{Shyamal Buch},
  \bibinfo{person}{Dallas Card}, \bibinfo{person}{Rodrigo Castellon},
  \bibinfo{person}{Niladri Chatterji}, \bibinfo{person}{Annie Chen},
  \bibinfo{person}{Kathleen Creel}, \bibinfo{person}{Jared~Quincy Davis},
  \bibinfo{person}{Dora Demszky}, \bibinfo{person}{Chris Donahue},
  \bibinfo{person}{Moussa Doumbouya}, \bibinfo{person}{Esin Durmus},
  \bibinfo{person}{Stefano Ermon}, \bibinfo{person}{John Etchemendy},
  \bibinfo{person}{Kawin Ethayarajh}, \bibinfo{person}{Li Fei-Fei},
  \bibinfo{person}{Chelsea Finn}, \bibinfo{person}{Trevor Gale},
  \bibinfo{person}{Lauren Gillespie}, \bibinfo{person}{Karan Goel},
  \bibinfo{person}{Noah Goodman}, \bibinfo{person}{Shelby Grossman},
  \bibinfo{person}{Neel Guha}, \bibinfo{person}{Tatsunori Hashimoto},
  \bibinfo{person}{Peter Henderson}, \bibinfo{person}{John Hewitt},
  \bibinfo{person}{Daniel~E. Ho}, \bibinfo{person}{Jenny Hong},
  \bibinfo{person}{Kyle Hsu}, \bibinfo{person}{Jing Huang},
  \bibinfo{person}{Thomas Icard}, \bibinfo{person}{Saahil Jain},
  \bibinfo{person}{Dan Jurafsky}, \bibinfo{person}{Pratyusha Kalluri},
  \bibinfo{person}{Siddharth Karamcheti}, \bibinfo{person}{Geoff Keeling},
  \bibinfo{person}{Fereshte Khani}, \bibinfo{person}{Omar Khattab},
  \bibinfo{person}{Pang~Wei Koh}, \bibinfo{person}{Mark Krass},
  \bibinfo{person}{Ranjay Krishna}, \bibinfo{person}{Rohith Kuditipudi},
  \bibinfo{person}{Ananya Kumar}, \bibinfo{person}{Faisal Ladhak},
  \bibinfo{person}{Mina Lee}, \bibinfo{person}{Tony Lee}, \bibinfo{person}{Jure
  Leskovec}, \bibinfo{person}{Isabelle Levent}, \bibinfo{person}{Xiang~Lisa
  Li}, \bibinfo{person}{Xuechen Li}, \bibinfo{person}{Tengyu Ma},
  \bibinfo{person}{Ali Malik}, \bibinfo{person}{Christopher~D. Manning},
  \bibinfo{person}{Suvir Mirchandani}, \bibinfo{person}{Eric Mitchell},
  \bibinfo{person}{Zanele Munyikwa}, \bibinfo{person}{Suraj Nair},
  \bibinfo{person}{Avanika Narayan}, \bibinfo{person}{Deepak Narayanan},
  \bibinfo{person}{Ben Newman}, \bibinfo{person}{Allen Nie},
  \bibinfo{person}{Juan~Carlos Niebles}, \bibinfo{person}{Hamed Nilforoshan},
  \bibinfo{person}{Julian Nyarko}, \bibinfo{person}{Giray Ogut},
  \bibinfo{person}{Laurel Orr}, \bibinfo{person}{Isabel Papadimitriou},
  \bibinfo{person}{Joon~Sung Park}, \bibinfo{person}{Chris Piech},
  \bibinfo{person}{Eva Portelance}, \bibinfo{person}{Christopher Potts},
  \bibinfo{person}{Aditi Raghunathan}, \bibinfo{person}{Rob Reich},
  \bibinfo{person}{Hongyu Ren}, \bibinfo{person}{Frieda Rong},
  \bibinfo{person}{Yusuf Roohani}, \bibinfo{person}{Camilo Ruiz},
  \bibinfo{person}{Jack Ryan}, \bibinfo{person}{Christopher Ré},
  \bibinfo{person}{Dorsa Sadigh}, \bibinfo{person}{Shiori Sagawa},
  \bibinfo{person}{Keshav Santhanam}, \bibinfo{person}{Andy Shih},
  \bibinfo{person}{Krishnan Srinivasan}, \bibinfo{person}{Alex Tamkin},
  \bibinfo{person}{Rohan Taori}, \bibinfo{person}{Armin~W. Thomas},
  \bibinfo{person}{Florian Tramèr}, \bibinfo{person}{Rose~E. Wang},
  \bibinfo{person}{William Wang}, \bibinfo{person}{Bohan Wu},
  \bibinfo{person}{Jiajun Wu}, \bibinfo{person}{Yuhuai Wu},
  \bibinfo{person}{Sang~Michael Xie}, \bibinfo{person}{Michihiro Yasunaga},
  \bibinfo{person}{Jiaxuan You}, \bibinfo{person}{Matei Zaharia},
  \bibinfo{person}{Michael Zhang}, \bibinfo{person}{Tianyi Zhang},
  \bibinfo{person}{Xikun Zhang}, \bibinfo{person}{Yuhui Zhang},
  \bibinfo{person}{Lucia Zheng}, \bibinfo{person}{Kaitlyn Zhou}, {and}
  \bibinfo{person}{Percy Liang}.} \bibinfo{year}{2022}\natexlab{}.
\newblock \bibinfo{title}{On the {Opportunities} and {Risks} of {Foundation}
  {Models}}.
\newblock
\newblock
\urldef\tempurl%
\url{https://doi.org/10.48550/arXiv.2108.07258}
\showDOI{\tempurl}
\newblock
\shownote{arXiv:2108.07258 [cs]}.


\bibitem[Bowman(2022)]%
        {bowman_dangers_2022}
\bibfield{author}{\bibinfo{person}{Samuel Bowman}.}
  \bibinfo{year}{2022}\natexlab{}.
\newblock \showarticletitle{The {Dangers} of {Underclaiming}: {Reasons} for
  {Caution} {When} {Reporting} {How} {NLP} {Systems} {Fail}}. In
  \bibinfo{booktitle}{\emph{Proceedings of the 60th {Annual} {Meeting} of the
  {Association} for {Computational} {Linguistics} ({Volume} 1: {Long}
  {Papers})}}. \bibinfo{publisher}{Association for Computational Linguistics},
  \bibinfo{address}{Dublin, Ireland}, \bibinfo{pages}{7484--7499}.
\newblock
\urldef\tempurl%
\url{https://doi.org/10.18653/v1/2022.acl-long.516}
\showDOI{\tempurl}


\bibitem[Boxell et~al\mbox{.}(2017)]%
        {boxell_greater_2017}
\bibfield{author}{\bibinfo{person}{Levi Boxell}, \bibinfo{person}{Matthew
  Gentzkow}, {and} \bibinfo{person}{Jesse~M Shapiro}.}
  \bibinfo{year}{2017}\natexlab{}.
\newblock \showarticletitle{Greater {Internet} use is not associated with
  faster growth in political polarization among {US} demographic groups}.
\newblock \bibinfo{journal}{\emph{Proceedings of the National Academy of
  Sciences}} \bibinfo{volume}{114}, \bibinfo{number}{40}
  (\bibinfo{year}{2017}), \bibinfo{pages}{10612--10617}.
\newblock
\newblock
\shownote{Publisher: National Acad Sciences}.


\bibitem[Braveman et~al\mbox{.}(2022)]%
        {braveman_systemic_2022}
\bibfield{author}{\bibinfo{person}{Paula~A. Braveman}, \bibinfo{person}{Elaine
  Arkin}, \bibinfo{person}{Dwayne Proctor}, \bibinfo{person}{Tina Kauh}, {and}
  \bibinfo{person}{Nicole Holm}.} \bibinfo{year}{2022}\natexlab{}.
\newblock \showarticletitle{Systemic {And} {Structural} {Racism}:
  {Definitions}, {Examples}, {Health} {Damages}, {And} {Approaches} {To}
  {Dismantling}}.
\newblock \bibinfo{journal}{\emph{Health Affairs}} \bibinfo{volume}{41},
  \bibinfo{number}{2} (\bibinfo{date}{Feb.} \bibinfo{year}{2022}),
  \bibinfo{pages}{171--178}.
\newblock
\showISSN{0278-2715}
\urldef\tempurl%
\url{https://doi.org/10.1377/hlthaff.2021.01394}
\showDOI{\tempurl}
\newblock
\shownote{Publisher: Health Affairs}.


\bibitem[Broussard et~al\mbox{.}(2019)]%
        {broussard_artificial_2019}
\bibfield{author}{\bibinfo{person}{Meredith Broussard},
  \bibinfo{person}{Nicholas Diakopoulos}, \bibinfo{person}{Andrea~L Guzman},
  \bibinfo{person}{Rediet Abebe}, \bibinfo{person}{Michel Dupagne}, {and}
  \bibinfo{person}{Ching-Hua Chuan}.} \bibinfo{year}{2019}\natexlab{}.
\newblock \showarticletitle{Artificial intelligence and journalism}.
\newblock \bibinfo{journal}{\emph{Journalism \& Mass Communication Quarterly}}
  \bibinfo{volume}{96}, \bibinfo{number}{3} (\bibinfo{year}{2019}),
  \bibinfo{pages}{673--695}.
\newblock
\newblock
\shownote{Publisher: SAGE Publications Sage CA: Los Angeles, CA}.


\bibitem[Brown and Sandholm(2019)]%
        {brown_superhuman_2019}
\bibfield{author}{\bibinfo{person}{Noam Brown} {and} \bibinfo{person}{Tuomas
  Sandholm}.} \bibinfo{year}{2019}\natexlab{}.
\newblock \showarticletitle{Superhuman {AI} for multiplayer poker}.
\newblock \bibinfo{journal}{\emph{Science}} \bibinfo{volume}{365},
  \bibinfo{number}{6456} (\bibinfo{date}{Aug.} \bibinfo{year}{2019}),
  \bibinfo{pages}{885--890}.
\newblock
\urldef\tempurl%
\url{https://doi.org/10.1126/science.aay2400}
\showDOI{\tempurl}
\newblock
\shownote{Publisher: American Association for the Advancement of Science}.


\bibitem[Brown et~al\mbox{.}(2020)]%
        {brown_language_2020}
\bibfield{author}{\bibinfo{person}{Tom Brown}, \bibinfo{person}{Benjamin Mann},
  \bibinfo{person}{Nick Ryder}, \bibinfo{person}{Melanie Subbiah},
  \bibinfo{person}{Jared~D Kaplan}, \bibinfo{person}{Prafulla Dhariwal},
  \bibinfo{person}{Arvind Neelakantan}, \bibinfo{person}{Pranav Shyam},
  \bibinfo{person}{Girish Sastry}, \bibinfo{person}{Amanda Askell},
  \bibinfo{person}{Sandhini Agarwal}, \bibinfo{person}{Ariel Herbert-Voss},
  \bibinfo{person}{Gretchen Krueger}, \bibinfo{person}{Tom Henighan},
  \bibinfo{person}{Rewon Child}, \bibinfo{person}{Aditya Ramesh},
  \bibinfo{person}{Daniel Ziegler}, \bibinfo{person}{Jeffrey Wu},
  \bibinfo{person}{Clemens Winter}, \bibinfo{person}{Chris Hesse},
  \bibinfo{person}{Mark Chen}, \bibinfo{person}{Eric Sigler},
  \bibinfo{person}{Mateusz Litwin}, \bibinfo{person}{Scott Gray},
  \bibinfo{person}{Benjamin Chess}, \bibinfo{person}{Jack Clark},
  \bibinfo{person}{Christopher Berner}, \bibinfo{person}{Sam McCandlish},
  \bibinfo{person}{Alec Radford}, \bibinfo{person}{Ilya Sutskever}, {and}
  \bibinfo{person}{Dario Amodei}.} \bibinfo{year}{2020}\natexlab{}.
\newblock \showarticletitle{Language {Models} are {Few}-{Shot} {Learners}}. In
  \bibinfo{booktitle}{\emph{Advances in {Neural} {Information} {Processing}
  {Systems}}}, Vol.~\bibinfo{volume}{33}. \bibinfo{publisher}{Curran
  Associates, Inc.}, \bibinfo{pages}{1877--1901}.
\newblock
\urldef\tempurl%
\url{https://proceedings.neurips.cc/paper/2020/hash/1457c0d6bfcb4967418bfb8ac142f64a-Abstract.html}
\showURL{%
\tempurl}


\bibitem[Brundage et~al\mbox{.}(2020)]%
        {brundage_toward_2020}
\bibfield{author}{\bibinfo{person}{Miles Brundage}, \bibinfo{person}{Shahar
  Avin}, \bibinfo{person}{Jasmine Wang}, \bibinfo{person}{Haydn Belfield},
  \bibinfo{person}{Gretchen Krueger}, \bibinfo{person}{Gillian Hadfield},
  \bibinfo{person}{Heidy Khlaaf}, \bibinfo{person}{Jingying Yang},
  \bibinfo{person}{Helen Toner}, \bibinfo{person}{Ruth Fong},
  \bibinfo{person}{Tegan Maharaj}, \bibinfo{person}{Pang~Wei Koh},
  \bibinfo{person}{Sara Hooker}, \bibinfo{person}{Jade Leung},
  \bibinfo{person}{Andrew Trask}, \bibinfo{person}{Emma Bluemke},
  \bibinfo{person}{Jonathan Lebensold}, \bibinfo{person}{Cullen O'Keefe},
  \bibinfo{person}{Mark Koren}, \bibinfo{person}{Théo Ryffel},
  \bibinfo{person}{J.~B. Rubinovitz}, \bibinfo{person}{Tamay Besiroglu},
  \bibinfo{person}{Federica Carugati}, \bibinfo{person}{Jack Clark},
  \bibinfo{person}{Peter Eckersley}, \bibinfo{person}{Sarah de Haas},
  \bibinfo{person}{Maritza Johnson}, \bibinfo{person}{Ben Laurie},
  \bibinfo{person}{Alex Ingerman}, \bibinfo{person}{Igor Krawczuk},
  \bibinfo{person}{Amanda Askell}, \bibinfo{person}{Rosario Cammarota},
  \bibinfo{person}{Andrew Lohn}, \bibinfo{person}{David Krueger},
  \bibinfo{person}{Charlotte Stix}, \bibinfo{person}{Peter Henderson},
  \bibinfo{person}{Logan Graham}, \bibinfo{person}{Carina Prunkl},
  \bibinfo{person}{Bianca Martin}, \bibinfo{person}{Elizabeth Seger},
  \bibinfo{person}{Noa Zilberman}, \bibinfo{person}{Seán~Ó hÉigeartaigh},
  \bibinfo{person}{Frens Kroeger}, \bibinfo{person}{Girish Sastry},
  \bibinfo{person}{Rebecca Kagan}, \bibinfo{person}{Adrian Weller},
  \bibinfo{person}{Brian Tse}, \bibinfo{person}{Elizabeth Barnes},
  \bibinfo{person}{Allan Dafoe}, \bibinfo{person}{Paul Scharre},
  \bibinfo{person}{Ariel Herbert-Voss}, \bibinfo{person}{Martijn Rasser},
  \bibinfo{person}{Shagun Sodhani}, \bibinfo{person}{Carrick Flynn},
  \bibinfo{person}{Thomas~Krendl Gilbert}, \bibinfo{person}{Lisa Dyer},
  \bibinfo{person}{Saif Khan}, \bibinfo{person}{Yoshua Bengio}, {and}
  \bibinfo{person}{Markus Anderljung}.} \bibinfo{year}{2020}\natexlab{}.
\newblock \bibinfo{title}{Toward {Trustworthy} {AI} {Development}: {Mechanisms}
  for {Supporting} {Verifiable} {Claims}}.
\newblock
\newblock
\urldef\tempurl%
\url{https://doi.org/10.48550/arXiv.2004.07213}
\showDOI{\tempurl}
\newblock
\shownote{arXiv:2004.07213 [cs]}.


\bibitem[Buolamwini and Gebru(2018)]%
        {buolamwini_gender_2018}
\bibfield{author}{\bibinfo{person}{Joy Buolamwini} {and}
  \bibinfo{person}{Timnit Gebru}.} \bibinfo{year}{2018}\natexlab{}.
\newblock \showarticletitle{Gender {Shades}: {Intersectional} {Accuracy}
  {Disparities} in {Commercial} {Gender} {Classification}}. In
  \bibinfo{booktitle}{\emph{Proceedings of the 1st {Conference} on {Fairness},
  {Accountability} and {Transparency}}} \emph{(\bibinfo{series}{Proceedings of
  {Machine} {Learning} {Research}}, Vol.~\bibinfo{volume}{81})},
  \bibfield{editor}{\bibinfo{person}{Sorelle~A. Friedler} {and}
  \bibinfo{person}{Christo Wilson}} (Eds.). \bibinfo{publisher}{PMLR},
  \bibinfo{pages}{77--91}.
\newblock
\urldef\tempurl%
\url{https://proceedings.mlr.press/v81/buolamwini18a.html}
\showURL{%
\tempurl}


\bibitem[Burrell and Fourcade(2021)]%
        {burrell_society_2021}
\bibfield{author}{\bibinfo{person}{Jenna Burrell} {and} \bibinfo{person}{Marion
  Fourcade}.} \bibinfo{year}{2021}\natexlab{}.
\newblock \showarticletitle{The {Society} of {Algorithms}}.
\newblock \bibinfo{journal}{\emph{Annual Review of Sociology}}
  \bibinfo{volume}{47}, \bibinfo{number}{1} (\bibinfo{year}{2021}),
  \bibinfo{pages}{213--237}.
\newblock
\urldef\tempurl%
\url{https://doi.org/10.1146/annurev-soc-090820-020800}
\showDOI{\tempurl}
\newblock
\shownote{\_eprint: https://doi.org/10.1146/annurev-soc-090820-020800}.


\bibitem[Caballero et~al\mbox{.}(2023)]%
        {caballero_broken_2023}
\bibfield{author}{\bibinfo{person}{Ethan Caballero}, \bibinfo{person}{Kshitij
  Gupta}, \bibinfo{person}{Irina Rish}, {and} \bibinfo{person}{David Krueger}.}
  \bibinfo{year}{2023}\natexlab{}.
\newblock \bibinfo{title}{Broken {Neural} {Scaling} {Laws}}.
\newblock
\newblock
\urldef\tempurl%
\url{https://doi.org/10.48550/arXiv.2210.14891}
\showDOI{\tempurl}
\newblock
\shownote{arXiv:2210.14891 [cs]}.


\bibitem[Campolo and Crawford(2020)]%
        {campolo_enchanted_2020}
\bibfield{author}{\bibinfo{person}{Alexander Campolo} {and}
  \bibinfo{person}{Kate Crawford}.} \bibinfo{year}{2020}\natexlab{}.
\newblock \showarticletitle{Enchanted {Determinism}: {Power} without
  {Responsibility} in {Artificial} {Intelligence}}.
\newblock \bibinfo{journal}{\emph{Engaging Science, Technology, and Society}}
  \bibinfo{volume}{6} (\bibinfo{date}{Jan.} \bibinfo{year}{2020}),
  \bibinfo{pages}{1--19}.
\newblock
\showISSN{2413-8053}
\urldef\tempurl%
\url{https://doi.org/10.17351/ests2020.277}
\showDOI{\tempurl}


\bibitem[Carroll et~al\mbox{.}(2023)]%
        {carroll_characterizing_2023}
\bibfield{author}{\bibinfo{person}{Micah Carroll}, \bibinfo{person}{Alan Chan},
  \bibinfo{person}{Henry Ashton}, {and} \bibinfo{person}{David Krueger}.}
  \bibinfo{year}{2023}\natexlab{}.
\newblock \bibinfo{title}{Characterizing {Manipulation} from {AI} {Systems}}.
\newblock
\newblock
\urldef\tempurl%
\url{https://doi.org/10.48550/arXiv.2303.09387}
\showDOI{\tempurl}
\newblock
\shownote{arXiv:2303.09387 [cs]}.


\bibitem[Carroll et~al\mbox{.}(2022)]%
        {carroll_estimating_2022}
\bibfield{author}{\bibinfo{person}{Micah~D. Carroll}, \bibinfo{person}{Anca
  Dragan}, \bibinfo{person}{Stuart Russell}, {and} \bibinfo{person}{Dylan
  Hadfield-Menell}.} \bibinfo{year}{2022}\natexlab{}.
\newblock \showarticletitle{Estimating and {Penalizing} {Induced} {Preference}
  {Shifts} in {Recommender} {Systems}}. In
  \bibinfo{booktitle}{\emph{Proceedings of the 39th {International}
  {Conference} on {Machine} {Learning}}}. \bibinfo{publisher}{PMLR},
  \bibinfo{pages}{2686--2708}.
\newblock
\urldef\tempurl%
\url{https://proceedings.mlr.press/v162/carroll22a.html}
\showURL{%
\tempurl}
\newblock
\shownote{ISSN: 2640-3498}.


\bibitem[Chase(2022)]%
        {chase_langchain_2022}
\bibfield{author}{\bibinfo{person}{Harrison Chase}.}
  \bibinfo{year}{2022}\natexlab{}.
\newblock \bibinfo{title}{{LangChain} 0.0.77 {Docs}}.
\newblock
\newblock
\urldef\tempurl%
\url{https://langchain.readthedocs.io/en/latest/modules/agents/getting_started.html}
\showURL{%
\tempurl}


\bibitem[Chen et~al\mbox{.}(2021)]%
        {chen_decision_2021}
\bibfield{author}{\bibinfo{person}{Lili Chen}, \bibinfo{person}{Kevin Lu},
  \bibinfo{person}{Aravind Rajeswaran}, \bibinfo{person}{Kimin Lee},
  \bibinfo{person}{Aditya Grover}, \bibinfo{person}{Misha Laskin},
  \bibinfo{person}{Pieter Abbeel}, \bibinfo{person}{Aravind Srinivas}, {and}
  \bibinfo{person}{Igor Mordatch}.} \bibinfo{year}{2021}\natexlab{}.
\newblock \showarticletitle{Decision {Transformer}: {Reinforcement} {Learning}
  via {Sequence} {Modeling}}. In \bibinfo{booktitle}{\emph{Advances in {Neural}
  {Information} {Processing} {Systems}}}, Vol.~\bibinfo{volume}{34}.
  \bibinfo{publisher}{Curran Associates, Inc.}, \bibinfo{pages}{15084--15097}.
\newblock
\urldef\tempurl%
\url{https://proceedings.neurips.cc/paper/2021/hash/7f489f642a0ddb10272b5c31057f0663-Abstract.html}
\showURL{%
\tempurl}


\bibitem[Christiano and Bajaj(2022)]%
        {christiano_democracy_2022}
\bibfield{author}{\bibinfo{person}{Tom Christiano} {and}
  \bibinfo{person}{Sameer Bajaj}.} \bibinfo{year}{2022}\natexlab{}.
\newblock \showarticletitle{Democracy}.
\newblock In \bibinfo{booktitle}{\emph{The {Stanford} {Encyclopedia} of
  {Philosophy}} (\bibinfo{edition}{spring 2022} ed.)},
  \bibfield{editor}{\bibinfo{person}{Edward~N. Zalta}} (Ed.).
  \bibinfo{publisher}{Metaphysics Research Lab, Stanford University}.
\newblock
\urldef\tempurl%
\url{https://plato.stanford.edu/archives/spr2022/entries/democracy/}
\showURL{%
\tempurl}


\bibitem[Clark and Amodei(2016)]%
        {clark_faulty_2016}
\bibfield{author}{\bibinfo{person}{Jack Clark} {and} \bibinfo{person}{Dario
  Amodei}.} \bibinfo{year}{2016}\natexlab{}.
\newblock \bibinfo{title}{Faulty {Reward} {Functions} in the {Wild}}.
\newblock
\newblock
\urldef\tempurl%
\url{https://openai.com/blog/faulty-reward-functions/}
\showURL{%
\tempurl}


\bibitem[Collins et~al\mbox{.}(2022)]%
        {collins_structured_2022}
\bibfield{author}{\bibinfo{person}{Katherine~M. Collins},
  \bibinfo{person}{Catherine Wong}, \bibinfo{person}{Jiahai Feng},
  \bibinfo{person}{Megan Wei}, {and} \bibinfo{person}{Joshua~B. Tenenbaum}.}
  \bibinfo{year}{2022}\natexlab{}.
\newblock \showarticletitle{Structured, flexible, and robust: benchmarking and
  improving large language models towards more human-like behavior in
  out-of-distribution reasoning tasks}.
\newblock  (\bibinfo{date}{May} \bibinfo{year}{2022}).
\newblock
\urldef\tempurl%
\url{https://doi.org/10.48550/arXiv.2205.05718}
\showDOI{\tempurl}


\bibitem[Cooper et~al\mbox{.}(2022)]%
        {cooper_accountability_2022}
\bibfield{author}{\bibinfo{person}{A.~Feder Cooper}, \bibinfo{person}{Emanuel
  Moss}, \bibinfo{person}{Benjamin Laufer}, {and} \bibinfo{person}{Helen
  Nissenbaum}.} \bibinfo{year}{2022}\natexlab{}.
\newblock \showarticletitle{Accountability in an {Algorithmic} {Society}:
  {Relationality}, {Responsibility}, and {Robustness} in {Machine} {Learning}}.
  In \bibinfo{booktitle}{\emph{2022 {ACM} {Conference} on {Fairness},
  {Accountability}, and {Transparency}}}. \bibinfo{publisher}{ACM}.
\newblock
\urldef\tempurl%
\url{https://doi.org/10.1145/3531146.3533150}
\showDOI{\tempurl}


\bibitem[Coulom(2002)]%
        {coulom_reinforcement_2002}
\bibfield{author}{\bibinfo{person}{Rémi Coulom}.}
  \bibinfo{year}{2002}\natexlab{}.
\newblock \showarticletitle{Reinforcement {Learning} {Using} {Neural}
  {Networks}, with {Applications} to {Motor} {Control}}.
\newblock  (\bibinfo{date}{June} \bibinfo{year}{2002}).
\newblock


\bibitem[Cremer and Kemp(2021)]%
        {cremer_democratising_2021}
\bibfield{author}{\bibinfo{person}{Carla~Zoe Cremer} {and}
  \bibinfo{person}{Luke Kemp}.} \bibinfo{year}{2021}\natexlab{}.
\newblock \bibinfo{title}{Democratising {Risk}: {In} {Search} of a
  {Methodology} to {Study} {Existential} {Risk}}.
\newblock
\newblock
\urldef\tempurl%
\url{https://papers.ssrn.com/abstract=3995225}
\showURL{%
\tempurl}


\bibitem[Dafoe(2018)]%
        {dafoe_ai_2018}
\bibfield{author}{\bibinfo{person}{Allan Dafoe}.}
  \bibinfo{year}{2018}\natexlab{}.
\newblock \bibinfo{title}{{AI} {Governance}: {A} {Research} {Agenda}}.
  (\bibinfo{date}{Aug.} \bibinfo{year}{2018}).
\newblock
\urldef\tempurl%
\url{https://www.fhi.ox.ac.uk/wp-content/uploads/GovAI-Agenda.pdf}
\showURL{%
\tempurl}


\bibitem[Dafoe et~al\mbox{.}(2020)]%
        {dafoe_open_2020}
\bibfield{author}{\bibinfo{person}{Allan Dafoe}, \bibinfo{person}{Edward
  Hughes}, \bibinfo{person}{Yoram Bachrach}, \bibinfo{person}{Tantum Collins},
  \bibinfo{person}{Kevin~R. McKee}, \bibinfo{person}{Joel~Z. Leibo},
  \bibinfo{person}{Kate Larson}, {and} \bibinfo{person}{Thore Graepel}.}
  \bibinfo{year}{2020}\natexlab{}.
\newblock \bibinfo{title}{Open {Problems} in {Cooperative} {AI}}.
\newblock
\newblock
\urldef\tempurl%
\url{https://doi.org/10.48550/arXiv.2012.08630}
\showDOI{\tempurl}
\newblock
\shownote{arXiv:2012.08630 [cs]}.


\bibitem[D'Amour et~al\mbox{.}(2020a)]%
        {damour_underspecification_2020}
\bibfield{author}{\bibinfo{person}{Alexander D'Amour},
  \bibinfo{person}{Katherine Heller}, \bibinfo{person}{Dan Moldovan},
  \bibinfo{person}{Ben Adlam}, \bibinfo{person}{Babak Alipanahi},
  \bibinfo{person}{Alex Beutel}, \bibinfo{person}{Christina Chen},
  \bibinfo{person}{Jonathan Deaton}, \bibinfo{person}{Jacob Eisenstein},
  \bibinfo{person}{Matthew~D. Hoffman}, \bibinfo{person}{Farhad Hormozdiari},
  \bibinfo{person}{Neil Houlsby}, \bibinfo{person}{Shaobo Hou},
  \bibinfo{person}{Ghassen Jerfel}, \bibinfo{person}{Alan Karthikesalingam},
  \bibinfo{person}{Mario Lucic}, \bibinfo{person}{Yian Ma},
  \bibinfo{person}{Cory McLean}, \bibinfo{person}{Diana Mincu},
  \bibinfo{person}{Akinori Mitani}, \bibinfo{person}{Andrea Montanari},
  \bibinfo{person}{Zachary Nado}, \bibinfo{person}{Vivek Natarajan},
  \bibinfo{person}{Christopher Nielson}, \bibinfo{person}{Thomas~F. Osborne},
  \bibinfo{person}{Rajiv Raman}, \bibinfo{person}{Kim Ramasamy},
  \bibinfo{person}{Rory Sayres}, \bibinfo{person}{Jessica Schrouff},
  \bibinfo{person}{Martin Seneviratne}, \bibinfo{person}{Shannon Sequeira},
  \bibinfo{person}{Harini Suresh}, \bibinfo{person}{Victor Veitch},
  \bibinfo{person}{Max Vladymyrov}, \bibinfo{person}{Xuezhi Wang},
  \bibinfo{person}{Kellie Webster}, \bibinfo{person}{Steve Yadlowsky},
  \bibinfo{person}{Taedong Yun}, \bibinfo{person}{Xiaohua Zhai}, {and}
  \bibinfo{person}{D. Sculley}.} \bibinfo{year}{2020}\natexlab{a}.
\newblock \bibinfo{title}{Underspecification {Presents} {Challenges} for
  {Credibility} in {Modern} {Machine} {Learning}}.
\newblock
\newblock
\urldef\tempurl%
\url{https://doi.org/10.48550/arXiv.2011.03395}
\showDOI{\tempurl}
\newblock
\shownote{arXiv:2011.03395 [cs, stat]}.


\bibitem[D'Amour et~al\mbox{.}(2020b)]%
        {damour_fairness_2020}
\bibfield{author}{\bibinfo{person}{Alexander D'Amour}, \bibinfo{person}{Hansa
  Srinivasan}, \bibinfo{person}{James Atwood}, \bibinfo{person}{Pallavi
  Baljekar}, \bibinfo{person}{D. Sculley}, {and} \bibinfo{person}{Yoni
  Halpern}.} \bibinfo{year}{2020}\natexlab{b}.
\newblock \showarticletitle{Fairness is not static}. In
  \bibinfo{booktitle}{\emph{Proceedings of the 2020 {Conference} on {Fairness},
  {Accountability}, and {Transparency}}}. \bibinfo{publisher}{ACM}.
\newblock
\urldef\tempurl%
\url{https://doi.org/10.1145/3351095.3372878}
\showDOI{\tempurl}


\bibitem[Dehghani et~al\mbox{.}(2021)]%
        {dehghani_benchmark_2021}
\bibfield{author}{\bibinfo{person}{Mostafa Dehghani}, \bibinfo{person}{Yi Tay},
  \bibinfo{person}{Alexey~A. Gritsenko}, \bibinfo{person}{Zhe Zhao},
  \bibinfo{person}{Neil Houlsby}, \bibinfo{person}{Fernando Diaz},
  \bibinfo{person}{Donald Metzler}, {and} \bibinfo{person}{Oriol Vinyals}.}
  \bibinfo{year}{2021}\natexlab{}.
\newblock \bibinfo{title}{The {Benchmark} {Lottery}}.
\newblock
\newblock
\urldef\tempurl%
\url{https://doi.org/10.48550/arXiv.2107.07002}
\showDOI{\tempurl}
\newblock
\shownote{arXiv:2107.07002 [cs]}.


\bibitem[Dennett(1981)]%
        {dennett_intentional_1981}
\bibfield{author}{\bibinfo{person}{Daniel~Clement Dennett}.}
  \bibinfo{year}{1981}\natexlab{}.
\newblock \bibinfo{booktitle}{\emph{The {Intentional} {Stance}}}.
\newblock \bibinfo{publisher}{MIT Press}.
\newblock


\bibitem[Dong et~al\mbox{.}(2022)]%
        {dong_survey_2022}
\bibfield{author}{\bibinfo{person}{Qingxiu Dong}, \bibinfo{person}{Lei Li},
  \bibinfo{person}{Damai Dai}, \bibinfo{person}{Ce Zheng},
  \bibinfo{person}{Zhiyong Wu}, \bibinfo{person}{Baobao Chang},
  \bibinfo{person}{Xu Sun}, \bibinfo{person}{Jingjing Xu}, \bibinfo{person}{Lei
  Li}, {and} \bibinfo{person}{Zhifang Sui}.} \bibinfo{year}{2022}\natexlab{}.
\newblock \bibinfo{title}{A {Survey} for {In}-context {Learning}}.
\newblock
\newblock
\urldef\tempurl%
\url{https://doi.org/10.48550/arXiv.2301.00234}
\showDOI{\tempurl}
\newblock
\shownote{arXiv:2301.00234 [cs]}.


\bibitem[Dreyfus(1965)]%
        {dreyfus_alchemy_1965}
\bibfield{author}{\bibinfo{person}{Hubert~L Dreyfus}.}
  \bibinfo{year}{1965}\natexlab{}.
\newblock \bibinfo{booktitle}{\emph{Alchemy and artificial intelligence}}.
\newblock \bibinfo{type}{{T}echnical {R}eport}. \bibinfo{institution}{RAND CORP
  SANTA MONICA CA}.
\newblock


\bibitem[Edwards(2022)]%
        {edwards_eu_2022}
\bibfield{author}{\bibinfo{person}{Lilian Edwards}.}
  \bibinfo{year}{2022}\natexlab{}.
\newblock \bibinfo{booktitle}{\emph{The {EU} {AI} {Act}: a summary of its
  significance and scope}}.
\newblock \bibinfo{type}{{T}echnical {R}eport}.
\newblock
\urldef\tempurl%
\url{https://www.adalovelaceinstitute.org/wp-content/uploads/2022/04/Expert-explainer-The-EU-AI-Act-11-April-2022.pdf}
\showURL{%
\tempurl}


\bibitem[Ehsan et~al\mbox{.}(2022)]%
        {ehsan_algorithmic_2022}
\bibfield{author}{\bibinfo{person}{Upol Ehsan}, \bibinfo{person}{Ranjit Singh},
  \bibinfo{person}{Jacob Metcalf}, {and} \bibinfo{person}{Mark Riedl}.}
  \bibinfo{year}{2022}\natexlab{}.
\newblock \showarticletitle{The {Algorithmic} {Imprint}}. In
  \bibinfo{booktitle}{\emph{2022 {ACM} {Conference} on {Fairness},
  {Accountability}, and {Transparency}}}. \bibinfo{publisher}{ACM}.
\newblock
\urldef\tempurl%
\url{https://doi.org/10.1145/3531146.3533186}
\showDOI{\tempurl}


\bibitem[Eisenhardt(1989)]%
        {eisenhardt_agency_1989}
\bibfield{author}{\bibinfo{person}{Kathleen~M. Eisenhardt}.}
  \bibinfo{year}{1989}\natexlab{}.
\newblock \showarticletitle{Agency {Theory}: {An} {Assessment} and {Review}}.
\newblock \bibinfo{journal}{\emph{The Academy of Management Review}}
  \bibinfo{volume}{14}, \bibinfo{number}{1} (\bibinfo{year}{1989}),
  \bibinfo{pages}{57--74}.
\newblock
\showISSN{0363-7425}
\urldef\tempurl%
\url{https://doi.org/10.2307/258191}
\showDOI{\tempurl}
\newblock
\shownote{Publisher: Academy of Management}.


\bibitem[Ekstrand et~al\mbox{.}(2018)]%
        {ekstrand_all_2018}
\bibfield{author}{\bibinfo{person}{Michael~D. Ekstrand}, \bibinfo{person}{Mucun
  Tian}, \bibinfo{person}{Ion~Madrazo Azpiazu}, \bibinfo{person}{Jennifer~D.
  Ekstrand}, \bibinfo{person}{Oghenemaro Anuyah}, \bibinfo{person}{David
  McNeill}, {and} \bibinfo{person}{Maria~Soledad Pera}.}
  \bibinfo{year}{2018}\natexlab{}.
\newblock \showarticletitle{All {The} {Cool} {Kids}, {How} {Do} {They} {Fit}
  {In}?: {Popularity} and {Demographic} {Biases} in {Recommender} {Evaluation}
  and {Effectiveness}}. In \bibinfo{booktitle}{\emph{Proceedings of the 1st
  {Conference} on {Fairness}, {Accountability} and {Transparency}}}
  \emph{(\bibinfo{series}{Proceedings of {Machine} {Learning} {Research}},
  Vol.~\bibinfo{volume}{81})}, \bibfield{editor}{\bibinfo{person}{Sorelle~A.
  Friedler} {and} \bibinfo{person}{Christo Wilson}} (Eds.).
  \bibinfo{publisher}{PMLR}, \bibinfo{pages}{172--186}.
\newblock
\urldef\tempurl%
\url{https://proceedings.mlr.press/v81/ekstrand18b.html}
\showURL{%
\tempurl}


\bibitem[Elzayn and Fish(2020)]%
        {elzayn_effects_2020}
\bibfield{author}{\bibinfo{person}{Hadi Elzayn} {and} \bibinfo{person}{Benjamin
  Fish}.} \bibinfo{year}{2020}\natexlab{}.
\newblock \showarticletitle{The effects of competition and regulation on error
  inequality in data-driven markets}. In \bibinfo{booktitle}{\emph{Proceedings
  of the 2020 {Conference} on {Fairness}, {Accountability}, and
  {Transparency}}}. \bibinfo{publisher}{ACM}.
\newblock
\urldef\tempurl%
\url{https://doi.org/10.1145/3351095.3372842}
\showDOI{\tempurl}


\bibitem[Emirbayer and Mische(1998)]%
        {emirbayer_what_1998}
\bibfield{author}{\bibinfo{person}{Mustafa Emirbayer} {and}
  \bibinfo{person}{Ann Mische}.} \bibinfo{year}{1998}\natexlab{}.
\newblock \showarticletitle{What is {Agency}?}
\newblock \bibinfo{journal}{\emph{American journal of sociology}}
  \bibinfo{volume}{103}, \bibinfo{number}{4} (\bibinfo{year}{1998}),
  \bibinfo{pages}{962--1023}.
\newblock
\newblock
\shownote{Publisher: The University of Chicago Press}.


\bibitem[Engineering(2021)]%
        {engineering_shifting_2021}
\bibfield{author}{\bibinfo{person}{Spotify Engineering}.}
  \bibinfo{year}{2021}\natexlab{}.
\newblock \bibinfo{title}{Shifting {Consumption} towards {Diverse} content via
  {Reinforcement} {Learning}}.
\newblock
\newblock
\urldef\tempurl%
\url{https://research.atspotify.com/2021/03/shifting-consumption-towards-diverse-content-via-reinforcement-learning/}
\showURL{%
\tempurl}
\newblock
\shownote{Section: Algorithmic Responsibility}.


\bibitem[Evans and Kasirzadeh(2022)]%
        {evans_user_2022}
\bibfield{author}{\bibinfo{person}{Charles Evans} {and} \bibinfo{person}{Atoosa
  Kasirzadeh}.} \bibinfo{year}{2022}\natexlab{}.
\newblock \bibinfo{title}{User {Tampering} in {Reinforcement} {Learning}
  {Recommender} {Systems}}.
\newblock
\newblock
\urldef\tempurl%
\url{https://doi.org/10.48550/arXiv.2109.04083}
\showDOI{\tempurl}
\newblock
\shownote{arXiv:2109.04083 [cs]}.


\bibitem[Evans and Gao(2016)]%
        {evans_deepmind_2016}
\bibfield{author}{\bibinfo{person}{Richard Evans} {and} \bibinfo{person}{Jim
  Gao}.} \bibinfo{year}{2016}\natexlab{}.
\newblock \bibinfo{title}{{DeepMind} {AI} {Reduces} {Google} {Data} {Centre}
  {Cooling} {Bill} by 40\%}.
\newblock
\newblock
\urldef\tempurl%
\url{https://www.deepmind.com/blog/deepmind-ai-reduces-google-data-centre-cooling-bill-by-40}
\showURL{%
\tempurl}


\bibitem[Everitt et~al\mbox{.}(2021)]%
        {everitt_agent_2021}
\bibfield{author}{\bibinfo{person}{Tom Everitt}, \bibinfo{person}{Ryan Carey},
  \bibinfo{person}{Eric~D. Langlois}, \bibinfo{person}{Pedro~A. Ortega}, {and}
  \bibinfo{person}{Shane Legg}.} \bibinfo{year}{2021}\natexlab{}.
\newblock \showarticletitle{Agent {Incentives}: {A} {Causal} {Perspective}}.
\newblock \bibinfo{journal}{\emph{Proceedings of the AAAI Conference on
  Artificial Intelligence}} \bibinfo{volume}{35}, \bibinfo{number}{13}
  (\bibinfo{date}{May} \bibinfo{year}{2021}), \bibinfo{pages}{11487--11495}.
\newblock
\showISSN{2374-3468}
\urldef\tempurl%
\url{https://doi.org/10.1609/aaai.v35i13.17368}
\showDOI{\tempurl}
\newblock
\shownote{Number: 13}.


\bibitem[Farquhar et~al\mbox{.}(2022)]%
        {farquhar_path-specific_2022}
\bibfield{author}{\bibinfo{person}{Sebastian Farquhar}, \bibinfo{person}{Ryan
  Carey}, {and} \bibinfo{person}{Tom Everitt}.}
  \bibinfo{year}{2022}\natexlab{}.
\newblock \bibinfo{title}{Path-{Specific} {Objectives} for {Safer} {Agent}
  {Incentives}}.
\newblock
\newblock
\urldef\tempurl%
\url{https://doi.org/10.48550/arXiv.2204.10018}
\showDOI{\tempurl}
\newblock
\shownote{arXiv:2204.10018 [cs, stat]}.


\bibitem[Fogliato et~al\mbox{.}(2022)]%
        {fogliato_who_2022}
\bibfield{author}{\bibinfo{person}{Riccardo Fogliato}, \bibinfo{person}{Shreya
  Chappidi}, \bibinfo{person}{Matthew Lungren}, \bibinfo{person}{Paul Fisher},
  \bibinfo{person}{Diane Wilson}, \bibinfo{person}{Michael Fitzke},
  \bibinfo{person}{Mark Parkinson}, \bibinfo{person}{Eric Horvitz},
  \bibinfo{person}{Kori Inkpen}, {and} \bibinfo{person}{Besmira Nushi}.}
  \bibinfo{year}{2022}\natexlab{}.
\newblock \showarticletitle{Who {Goes} {First}? {Influences} of {Human}-{AI}
  {Workflow} on {Decision} {Making} in {Clinical} {Imaging}}. In
  \bibinfo{booktitle}{\emph{2022 {ACM} {Conference} on {Fairness},
  {Accountability}, and {Transparency}}}. \bibinfo{publisher}{ACM}.
\newblock
\urldef\tempurl%
\url{https://doi.org/10.1145/3531146.3533193}
\showDOI{\tempurl}


\bibitem[for Critical~Technology(2020)]%
        {coalition_for_critical_technology_abolish_2020}
\bibfield{author}{\bibinfo{person}{Coalition for Critical~Technology}.}
  \bibinfo{year}{2020}\natexlab{}.
\newblock \bibinfo{title}{Abolish the \#{TechToPrisonPipeline}}.
\newblock
\newblock
\urldef\tempurl%
\url{https://medium.com/@CoalitionForCriticalTechnology/abolish-the-techtoprisonpipeline-9b5b14366b16}
\showURL{%
\tempurl}


\bibitem[Ganguli et~al\mbox{.}(2022)]%
        {ganguli_predictability_2022}
\bibfield{author}{\bibinfo{person}{Deep Ganguli}, \bibinfo{person}{Danny
  Hernandez}, \bibinfo{person}{Liane Lovitt}, \bibinfo{person}{Amanda Askell},
  \bibinfo{person}{Yuntao Bai}, \bibinfo{person}{Anna Chen},
  \bibinfo{person}{Tom Conerly}, \bibinfo{person}{Nova Dassarma},
  \bibinfo{person}{Dawn Drain}, \bibinfo{person}{Nelson Elhage},
  \bibinfo{person}{Sheer~El Showk}, \bibinfo{person}{Stanislav Fort},
  \bibinfo{person}{Zac Hatfield-Dodds}, \bibinfo{person}{Tom Henighan},
  \bibinfo{person}{Scott Johnston}, \bibinfo{person}{Andy Jones},
  \bibinfo{person}{Nicholas Joseph}, \bibinfo{person}{Jackson Kernian},
  \bibinfo{person}{Shauna Kravec}, \bibinfo{person}{Ben Mann},
  \bibinfo{person}{Neel Nanda}, \bibinfo{person}{Kamal Ndousse},
  \bibinfo{person}{Catherine Olsson}, \bibinfo{person}{Daniela Amodei},
  \bibinfo{person}{Tom Brown}, \bibinfo{person}{Jared Kaplan},
  \bibinfo{person}{Sam McCandlish}, \bibinfo{person}{Christopher Olah},
  \bibinfo{person}{Dario Amodei}, {and} \bibinfo{person}{Jack Clark}.}
  \bibinfo{year}{2022}\natexlab{}.
\newblock \showarticletitle{Predictability and {Surprise} in {Large}
  {Generative} {Models}}. In \bibinfo{booktitle}{\emph{2022 {ACM} {Conference}
  on {Fairness}, {Accountability}, and {Transparency}}}.
  \bibinfo{publisher}{ACM}.
\newblock
\urldef\tempurl%
\url{https://doi.org/10.1145/3531146.3533229}
\showDOI{\tempurl}


\bibitem[Gao et~al\mbox{.}(2022)]%
        {gao_scaling_2022}
\bibfield{author}{\bibinfo{person}{Leo Gao}, \bibinfo{person}{John Schulman},
  {and} \bibinfo{person}{Jacob Hilton}.} \bibinfo{year}{2022}\natexlab{}.
\newblock \bibinfo{title}{Scaling {Laws} for {Reward} {Model}
  {Overoptimization}}.
\newblock
\newblock
\urldef\tempurl%
\url{https://doi.org/10.48550/arXiv.2210.10760}
\showDOI{\tempurl}
\newblock
\shownote{arXiv:2210.10760 [cs, stat]}.


\bibitem[Gauci et~al\mbox{.}(2019)]%
        {gauci_horizon_2019}
\bibfield{author}{\bibinfo{person}{Jason Gauci}, \bibinfo{person}{Edoardo
  Conti}, \bibinfo{person}{Yitao Liang}, \bibinfo{person}{Kittipat Virochsiri},
  \bibinfo{person}{Yuchen He}, \bibinfo{person}{Zachary Kaden},
  \bibinfo{person}{Vivek Narayanan}, \bibinfo{person}{Xiaohui Ye},
  \bibinfo{person}{Zhengxing Chen}, {and} \bibinfo{person}{Scott Fujimoto}.}
  \bibinfo{year}{2019}\natexlab{}.
\newblock \bibinfo{title}{Horizon: {Facebook}'s {Open} {Source} {Applied}
  {Reinforcement} {Learning} {Platform}}.
\newblock
\newblock
\urldef\tempurl%
\url{https://doi.org/10.48550/arXiv.1811.00260}
\showDOI{\tempurl}
\newblock
\shownote{arXiv:1811.00260 [cs, stat]}.


\bibitem[Gebru et~al\mbox{.}(2021)]%
        {gebru_datasheets_2021}
\bibfield{author}{\bibinfo{person}{Timnit Gebru}, \bibinfo{person}{Jamie
  Morgenstern}, \bibinfo{person}{Briana Vecchione},
  \bibinfo{person}{Jennifer~Wortman Vaughan}, \bibinfo{person}{Hanna Wallach},
  \bibinfo{person}{Hal~Daumé Iii}, {and} \bibinfo{person}{Kate Crawford}.}
  \bibinfo{year}{2021}\natexlab{}.
\newblock \showarticletitle{Datasheets for datasets}.
\newblock \bibinfo{journal}{\emph{Commun. ACM}} \bibinfo{volume}{64},
  \bibinfo{number}{12} (\bibinfo{date}{Dec.} \bibinfo{year}{2021}),
  \bibinfo{pages}{86--92}.
\newblock
\showISSN{0001-0782, 1557-7317}
\urldef\tempurl%
\url{https://doi.org/10.1145/3458723}
\showDOI{\tempurl}


\bibitem[Gehman et~al\mbox{.}(2020)]%
        {gehman_realtoxicityprompts_2020}
\bibfield{author}{\bibinfo{person}{Samuel Gehman}, \bibinfo{person}{Suchin
  Gururangan}, \bibinfo{person}{Maarten Sap}, \bibinfo{person}{Yejin Choi},
  {and} \bibinfo{person}{Noah~A. Smith}.} \bibinfo{year}{2020}\natexlab{}.
\newblock \showarticletitle{{RealToxicityPrompts}: {Evaluating} {Neural}
  {Toxic} {Degeneration} in {Language} {Models}}. In
  \bibinfo{booktitle}{\emph{Findings of the {Association} for {Computational}
  {Linguistics}: {EMNLP} 2020}}. \bibinfo{publisher}{Association for
  Computational Linguistics}, \bibinfo{address}{Online},
  \bibinfo{pages}{3356--3369}.
\newblock
\urldef\tempurl%
\url{https://doi.org/10.18653/v1/2020.findings-emnlp.301}
\showDOI{\tempurl}


\bibitem[Gesley et~al\mbox{.}(2019)]%
        {gesley_regulation_2019}
\bibfield{author}{\bibinfo{person}{Jenny Gesley}, \bibinfo{person}{Tariq
  Ahmad}, \bibinfo{person}{Edouardo Soares}, \bibinfo{person}{Ruth Levush},
  \bibinfo{person}{Gustavo Guerra}, \bibinfo{person}{James Martin},
  \bibinfo{person}{Kelly Buchanan}, \bibinfo{person}{Laney Zhang},
  \bibinfo{person}{Sayuri Umeda}, \bibinfo{person}{Astghik Grigoryan},
  \bibinfo{person}{Nicolas Boring}, \bibinfo{person}{Elin Hofverberg},
  \bibinfo{person}{Clare Feikhert-Ahalt}, \bibinfo{person}{Graciela
  Rodriguez-Ferrand}, \bibinfo{person}{George Sadek}, {and}
  \bibinfo{person}{Hanibal Goitom}.} \bibinfo{year}{2019}\natexlab{}.
\newblock \showarticletitle{Regulation of {Artificial} {Intelligence} in
  {Selected} {Jurisdictions}}.
\newblock \bibinfo{journal}{\emph{Copyright, Fair Use, Scholarly Communication,
  etc.}} (\bibinfo{date}{Jan.} \bibinfo{year}{2019}).
\newblock
\urldef\tempurl%
\url{https://digitalcommons.unl.edu/scholcom/177}
\showURL{%
\tempurl}


\bibitem[Giattino et~al\mbox{.}(2022)]%
        {giattino_artificial_2022}
\bibfield{author}{\bibinfo{person}{Charlie Giattino}, \bibinfo{person}{Edouard
  Mathieu}, \bibinfo{person}{Julia Broden}, {and} \bibinfo{person}{Max Roser}.}
  \bibinfo{year}{2022}\natexlab{}.
\newblock \showarticletitle{Artificial {Intelligence}}.
\newblock \bibinfo{journal}{\emph{Our World in Data}} (\bibinfo{year}{2022}).
\newblock


\bibitem[Gilbert et~al\mbox{.}(2022)]%
        {gilbert_reward_2022}
\bibfield{author}{\bibinfo{person}{Thomas~Krendl Gilbert},
  \bibinfo{person}{Sarah Dean}, \bibinfo{person}{Nathan Lambert},
  \bibinfo{person}{Tom Zick}, {and} \bibinfo{person}{Aaron Snoswell}.}
  \bibinfo{year}{2022}\natexlab{}.
\newblock \bibinfo{title}{Reward {Reports} for {Reinforcement} {Learning}}.
\newblock
\newblock
\urldef\tempurl%
\url{https://doi.org/10.48550/arXiv.2204.10817}
\showDOI{\tempurl}
\newblock
\shownote{arXiv:2204.10817 [cs]}.


\bibitem[Goetze(2022)]%
        {goetze_mind_2022}
\bibfield{author}{\bibinfo{person}{Trystan~S. Goetze}.}
  \bibinfo{year}{2022}\natexlab{}.
\newblock \showarticletitle{Mind the {Gap}: {Autonomous} {Systems}, the
  {Responsibility} {Gap}, and {Moral} {Entanglement}}. In
  \bibinfo{booktitle}{\emph{2022 {ACM} {Conference} on {Fairness},
  {Accountability}, and {Transparency}}}. \bibinfo{publisher}{ACM}.
\newblock
\urldef\tempurl%
\url{https://doi.org/10.1145/3531146.3533106}
\showDOI{\tempurl}


\bibitem[Goodhart(1975)]%
        {goodhart_problems_1975}
\bibfield{author}{\bibinfo{person}{Charles Goodhart}.}
  \bibinfo{year}{1975}\natexlab{}.
\newblock \showarticletitle{Problems of monetary management: the {UK}
  experience in papers in monetary economics}.
\newblock \bibinfo{journal}{\emph{Monetary Economics}}  \bibinfo{volume}{1}
  (\bibinfo{year}{1975}).
\newblock


\bibitem[Gray and Suri(2019)]%
        {gray_ghost_2019}
\bibfield{author}{\bibinfo{person}{Mary~L. Gray} {and}
  \bibinfo{person}{Siddharth Suri}.} \bibinfo{year}{2019}\natexlab{}.
\newblock \bibinfo{booktitle}{\emph{Ghost work: how to stop {Silicon} {Valley}
  from building a new global underclass}}.
\newblock \bibinfo{publisher}{Houghton Mifflin Harcourt},
  \bibinfo{address}{Boston}.
\newblock
\showISBNx{9781328566287}


\bibitem[Green(2020)]%
        {green_false_2020}
\bibfield{author}{\bibinfo{person}{Ben Green}.}
  \bibinfo{year}{2020}\natexlab{}.
\newblock \showarticletitle{The false promise of risk assessments}. In
  \bibinfo{booktitle}{\emph{Proceedings of the 2020 {Conference} on {Fairness},
  {Accountability}, and {Transparency}}}. \bibinfo{publisher}{ACM}.
\newblock
\urldef\tempurl%
\url{https://doi.org/10.1145/3351095.3372869}
\showDOI{\tempurl}


\bibitem[Green and Chen(2019)]%
        {green_disparate_2019}
\bibfield{author}{\bibinfo{person}{Ben Green} {and} \bibinfo{person}{Yiling
  Chen}.} \bibinfo{year}{2019}\natexlab{}.
\newblock \showarticletitle{Disparate {Interactions}}. In
  \bibinfo{booktitle}{\emph{Proceedings of the {Conference} on {Fairness},
  {Accountability}, and {Transparency}}}. \bibinfo{publisher}{ACM}.
\newblock
\urldef\tempurl%
\url{https://doi.org/10.1145/3287560.3287563}
\showDOI{\tempurl}


\bibitem[Green(2021)]%
        {green_ai_2021}
\bibfield{author}{\bibinfo{person}{Nancy Green}.}
  \bibinfo{year}{2021}\natexlab{}.
\newblock \showarticletitle{An {AI} {Ethics} {Course} {Highlighting} {Explicit}
  {Ethical} {Agents}}. In \bibinfo{booktitle}{\emph{Proceedings of the 2021
  {AAAI}/{ACM} {Conference} on {AI}, {Ethics}, and {Society}}}.
  \bibinfo{publisher}{ACM}.
\newblock
\urldef\tempurl%
\url{https://doi.org/10.1145/3461702.3462552}
\showDOI{\tempurl}


\bibitem[Hadfield-Menell and Hadfield(2019)]%
        {hadfield-menell_incomplete_2019}
\bibfield{author}{\bibinfo{person}{Dylan Hadfield-Menell} {and}
  \bibinfo{person}{Gillian~K. Hadfield}.} \bibinfo{year}{2019}\natexlab{}.
\newblock \showarticletitle{Incomplete {Contracting} and {AI} {Alignment}}. In
  \bibinfo{booktitle}{\emph{Proceedings of the 2019 {AAAI}/{ACM} {Conference}
  on {AI}, {Ethics}, and {Society}}}. \bibinfo{publisher}{ACM}.
\newblock
\urldef\tempurl%
\url{https://doi.org/10.1145/3306618.3314250}
\showDOI{\tempurl}


\bibitem[Hafner et~al\mbox{.}(2023)]%
        {hafner_mastering_2023}
\bibfield{author}{\bibinfo{person}{Danijar Hafner}, \bibinfo{person}{Jurgis
  Pasukonis}, \bibinfo{person}{Jimmy Ba}, {and} \bibinfo{person}{Timothy
  Lillicrap}.} \bibinfo{year}{2023}\natexlab{}.
\newblock \bibinfo{title}{Mastering {Diverse} {Domains} through {World}
  {Models}}.
\newblock
\newblock
\urldef\tempurl%
\url{https://doi.org/10.48550/arXiv.2301.04104}
\showDOI{\tempurl}
\newblock
\shownote{arXiv:2301.04104 [cs, stat]}.


\bibitem[Hickox(2010)]%
        {hickox_employer_2010}
\bibfield{author}{\bibinfo{person}{Stacy~A. Hickox}.}
  \bibinfo{year}{2010}\natexlab{}.
\newblock \showarticletitle{Employer {Liability} of {Negligent} {Hiring} of
  {Ex}-{Offenders}}.
\newblock \bibinfo{journal}{\emph{Saint Louis University Law Journal}}
  \bibinfo{volume}{55}, \bibinfo{number}{3} (\bibinfo{year}{2010}),
  \bibinfo{pages}{1001--1046}.
\newblock
\urldef\tempurl%
\url{https://heinonline.org/HOL/P?h=hein.journals/stlulj55&i=1029}
\showURL{%
\tempurl}


\bibitem[Hilton et~al\mbox{.}(2023)]%
        {hilton_scaling_2023}
\bibfield{author}{\bibinfo{person}{Jacob Hilton}, \bibinfo{person}{Jie Tang},
  {and} \bibinfo{person}{John Schulman}.} \bibinfo{year}{2023}\natexlab{}.
\newblock \bibinfo{title}{Scaling laws for single-agent reinforcement
  learning}.
\newblock
\newblock
\urldef\tempurl%
\url{https://doi.org/10.48550/arXiv.2301.13442}
\showDOI{\tempurl}
\newblock
\shownote{arXiv:2301.13442 [cs, stat]}.


\bibitem[Hoffmann et~al\mbox{.}(2022)]%
        {hoffmann_empirical_2022}
\bibfield{author}{\bibinfo{person}{Jordan Hoffmann}, \bibinfo{person}{Sebastian
  Borgeaud}, \bibinfo{person}{Arthur Mensch}, \bibinfo{person}{Elena
  Buchatskaya}, \bibinfo{person}{Trevor Cai}, \bibinfo{person}{Eliza
  Rutherford}, \bibinfo{person}{Diego de~las Casas}, \bibinfo{person}{Lisa~Anne
  Hendricks}, \bibinfo{person}{Johannes Welbl}, \bibinfo{person}{Aidan Clark},
  \bibinfo{person}{Tom Hennigan}, \bibinfo{person}{Eric Noland},
  \bibinfo{person}{Katherine Millican}, \bibinfo{person}{George van~den
  Driessche}, \bibinfo{person}{Bogdan Damoc}, \bibinfo{person}{Aurelia Guy},
  \bibinfo{person}{Simon Osindero}, \bibinfo{person}{Karen Simonyan},
  \bibinfo{person}{Erich Elsen}, \bibinfo{person}{Oriol Vinyals},
  \bibinfo{person}{Jack~William Rae}, {and} \bibinfo{person}{Laurent Sifre}.}
  \bibinfo{year}{2022}\natexlab{}.
\newblock \showarticletitle{An empirical analysis of compute-optimal large
  language model training}. In \bibinfo{booktitle}{\emph{Advances in {Neural}
  {Information} {Processing} {Systems}}},
  \bibfield{editor}{\bibinfo{person}{Alice~H. Oh}, \bibinfo{person}{Alekh
  Agarwal}, \bibinfo{person}{Danielle Belgrave}, {and}
  \bibinfo{person}{Kyunghyun Cho}} (Eds.).
\newblock
\urldef\tempurl%
\url{https://openreview.net/forum?id=iBBcRUlOAPR}
\showURL{%
\tempurl}


\bibitem[Hou et~al\mbox{.}(2019)]%
        {hou_social_2019}
\bibfield{author}{\bibinfo{person}{Yubo Hou}, \bibinfo{person}{Dan Xiong},
  \bibinfo{person}{Tonglin Jiang}, \bibinfo{person}{Lily Song}, {and}
  \bibinfo{person}{Qi Wang}.} \bibinfo{year}{2019}\natexlab{}.
\newblock \showarticletitle{Social media addiction: {Its} impact, mediation,
  and intervention}.
\newblock \bibinfo{journal}{\emph{Cyberpsychology: Journal of Psychosocial
  Research on Cyberspace}} \bibinfo{volume}{13}, \bibinfo{number}{1}
  (\bibinfo{date}{Feb.} \bibinfo{year}{2019}).
\newblock
\showISSN{1802-7962}
\urldef\tempurl%
\url{https://doi.org/10.5817/CP2019-1-4}
\showDOI{\tempurl}
\newblock
\shownote{Number: 1}.


\bibitem[Huang and Siddarth(2023)]%
        {huang_generative_2023}
\bibfield{author}{\bibinfo{person}{Saffron Huang} {and} \bibinfo{person}{Divya
  Siddarth}.} \bibinfo{year}{2023}\natexlab{}.
\newblock \bibinfo{title}{Generative {AI} and the {Digital} {Commons}}.
\newblock
\newblock
\urldef\tempurl%
\url{https://cip.org/research/generative-ai-digital-commons}
\showURL{%
\tempurl}


\bibitem[Huang et~al\mbox{.}(2022)]%
        {huang_inner_2022}
\bibfield{author}{\bibinfo{person}{Wenlong Huang}, \bibinfo{person}{Fei Xia},
  \bibinfo{person}{Ted Xiao}, \bibinfo{person}{Harris Chan},
  \bibinfo{person}{Jacky Liang}, \bibinfo{person}{Pete Florence},
  \bibinfo{person}{Andy Zeng}, \bibinfo{person}{Jonathan Tompson},
  \bibinfo{person}{Igor Mordatch}, \bibinfo{person}{Yevgen Chebotar},
  \bibinfo{person}{Pierre Sermanet}, \bibinfo{person}{Noah Brown},
  \bibinfo{person}{Tomas Jackson}, \bibinfo{person}{Linda Luu},
  \bibinfo{person}{Sergey Levine}, \bibinfo{person}{Karol Hausman}, {and}
  \bibinfo{person}{Brian Ichter}.} \bibinfo{year}{2022}\natexlab{}.
\newblock \bibinfo{title}{Inner {Monologue}: {Embodied} {Reasoning} through
  {Planning} with {Language} {Models}}.
\newblock
\newblock
\urldef\tempurl%
\url{https://doi.org/10.48550/arXiv.2207.05608}
\showDOI{\tempurl}
\newblock
\shownote{arXiv:2207.05608 [cs]}.


\bibitem[Jabbari et~al\mbox{.}(2017)]%
        {jabbari_fairness_2017}
\bibfield{author}{\bibinfo{person}{Shahin Jabbari}, \bibinfo{person}{Matthew
  Joseph}, \bibinfo{person}{Michael Kearns}, \bibinfo{person}{Jamie
  Morgenstern}, {and} \bibinfo{person}{Aaron Roth}.}
  \bibinfo{year}{2017}\natexlab{}.
\newblock \showarticletitle{Fairness in {Reinforcement} {Learning}}. In
  \bibinfo{booktitle}{\emph{Proceedings of the 34th {International}
  {Conference} on {Machine} {Learning}}}. \bibinfo{publisher}{PMLR},
  \bibinfo{pages}{1617--1626}.
\newblock
\urldef\tempurl%
\url{https://proceedings.mlr.press/v70/jabbari17a.html}
\showURL{%
\tempurl}
\newblock
\shownote{ISSN: 2640-3498}.


\bibitem[Jacobs and Wallach(2021)]%
        {jacobs_measurement_2021}
\bibfield{author}{\bibinfo{person}{Abigail~Z. Jacobs} {and}
  \bibinfo{person}{Hanna Wallach}.} \bibinfo{year}{2021}\natexlab{}.
\newblock \showarticletitle{Measurement and {Fairness}}. In
  \bibinfo{booktitle}{\emph{Proceedings of the 2021 {ACM} {Conference} on
  {Fairness}, {Accountability}, and {Transparency}}}
  \emph{(\bibinfo{series}{{FAccT} '21})}. \bibinfo{publisher}{Association for
  Computing Machinery}, \bibinfo{address}{New York, NY, USA},
  \bibinfo{pages}{375--385}.
\newblock
\showISBNx{978-1-4503-8309-7}
\urldef\tempurl%
\url{https://doi.org/10.1145/3442188.3445901}
\showDOI{\tempurl}
\newblock
\shownote{event-place: Virtual Event, Canada}.


\bibitem[Jensen and Meckling(1976)]%
        {jensen_theory_1976}
\bibfield{author}{\bibinfo{person}{Michael~C. Jensen} {and}
  \bibinfo{person}{William~H. Meckling}.} \bibinfo{year}{1976}\natexlab{}.
\newblock \showarticletitle{Theory of the firm: {Managerial} behavior, agency
  costs and ownership structure}.
\newblock \bibinfo{journal}{\emph{Journal of Financial Economics}}
  \bibinfo{volume}{3}, \bibinfo{number}{4} (\bibinfo{date}{Oct.}
  \bibinfo{year}{1976}), \bibinfo{pages}{305--360}.
\newblock
\showISSN{0304-405X}
\urldef\tempurl%
\url{https://doi.org/10.1016/0304-405X(76)90026-X}
\showDOI{\tempurl}


\bibitem[Jernite et~al\mbox{.}(2022)]%
        {jernite_data_2022}
\bibfield{author}{\bibinfo{person}{Yacine Jernite}, \bibinfo{person}{Huu
  Nguyen}, \bibinfo{person}{Stella Biderman}, \bibinfo{person}{Anna Rogers},
  \bibinfo{person}{Maraim Masoud}, \bibinfo{person}{Valentin Danchev},
  \bibinfo{person}{Samson Tan}, \bibinfo{person}{Alexandra~Sasha Luccioni},
  \bibinfo{person}{Nishant Subramani}, \bibinfo{person}{Gérard Dupont},
  \bibinfo{person}{Jesse Dodge}, \bibinfo{person}{Kyle Lo},
  \bibinfo{person}{Zeerak Talat}, \bibinfo{person}{Isaac Johnson},
  \bibinfo{person}{Dragomir Radev}, \bibinfo{person}{Somaieh Nikpoor},
  \bibinfo{person}{Jörg Frohberg}, \bibinfo{person}{Aaron Gokaslan},
  \bibinfo{person}{Peter Henderson}, \bibinfo{person}{Rishi Bommasani}, {and}
  \bibinfo{person}{Margaret Mitchell}.} \bibinfo{year}{2022}\natexlab{}.
\newblock \showarticletitle{Data {Governance} in the {Age} of {Large}-{Scale}
  {Data}-{Driven} {Language} {Technology}}. In \bibinfo{booktitle}{\emph{2022
  {ACM} {Conference} on {Fairness}, {Accountability}, and {Transparency}}}.
  \bibinfo{pages}{2206--2222}.
\newblock
\urldef\tempurl%
\url{https://doi.org/10.1145/3531146.3534637}
\showDOI{\tempurl}
\newblock
\shownote{arXiv:2206.03216 [cs]}.


\bibitem[Ji et~al\mbox{.}(2022)]%
        {ji_survey_2022}
\bibfield{author}{\bibinfo{person}{Ziwei Ji}, \bibinfo{person}{Nayeon Lee},
  \bibinfo{person}{Rita Frieske}, \bibinfo{person}{Tiezheng Yu},
  \bibinfo{person}{Dan Su}, \bibinfo{person}{Yan Xu}, \bibinfo{person}{Etsuko
  Ishii}, \bibinfo{person}{Yejin Bang}, \bibinfo{person}{Wenliang Dai},
  \bibinfo{person}{Andrea Madotto}, {and} \bibinfo{person}{Pascale Fung}.}
  \bibinfo{year}{2022}\natexlab{}.
\newblock \showarticletitle{Survey of {Hallucination} in {Natural} {Language}
  {Generation}}.
\newblock \bibinfo{journal}{\emph{Comput. Surveys}} (\bibinfo{date}{Nov.}
  \bibinfo{year}{2022}), \bibinfo{pages}{3571730}.
\newblock
\showISSN{0360-0300, 1557-7341}
\urldef\tempurl%
\url{https://doi.org/10.1145/3571730}
\showDOI{\tempurl}
\newblock
\shownote{arXiv:2202.03629 [cs]}.


\bibitem[Jiang et~al\mbox{.}(2019)]%
        {jiang_degenerate_2019}
\bibfield{author}{\bibinfo{person}{Ray Jiang}, \bibinfo{person}{Silvia
  Chiappa}, \bibinfo{person}{Tor Lattimore}, \bibinfo{person}{András György},
  {and} \bibinfo{person}{Pushmeet Kohli}.} \bibinfo{year}{2019}\natexlab{}.
\newblock \showarticletitle{Degenerate {Feedback} {Loops} in {Recommender}
  {Systems}}. In \bibinfo{booktitle}{\emph{Proceedings of the 2019 {AAAI}/{ACM}
  {Conference} on {AI}, {Ethics}, and {Society}}}. \bibinfo{publisher}{ACM}.
\newblock
\urldef\tempurl%
\url{https://doi.org/10.1145/3306618.3314288}
\showDOI{\tempurl}


\bibitem[Johnson and Verdicchio(2017)]%
        {johnson_reframing_2017}
\bibfield{author}{\bibinfo{person}{Deborah~G. Johnson} {and}
  \bibinfo{person}{Mario Verdicchio}.} \bibinfo{year}{2017}\natexlab{}.
\newblock \showarticletitle{Reframing {AI} {Discourse}}.
\newblock \bibinfo{journal}{\emph{Minds and Machines}} \bibinfo{volume}{27},
  \bibinfo{number}{4} (\bibinfo{date}{Dec.} \bibinfo{year}{2017}),
  \bibinfo{pages}{575--590}.
\newblock
\showISSN{1572-8641}
\urldef\tempurl%
\url{https://doi.org/10.1007/s11023-017-9417-6}
\showDOI{\tempurl}


\bibitem[Joseph et~al\mbox{.}(2016)]%
        {joseph_fairness_2016}
\bibfield{author}{\bibinfo{person}{Matthew Joseph}, \bibinfo{person}{Michael
  Kearns}, \bibinfo{person}{Jamie~H Morgenstern}, {and} \bibinfo{person}{Aaron
  Roth}.} \bibinfo{year}{2016}\natexlab{}.
\newblock \showarticletitle{Fairness in {Learning}: {Classic} and {Contextual}
  {Bandits}}. In \bibinfo{booktitle}{\emph{Advances in {Neural} {Information}
  {Processing} {Systems}}}, Vol.~\bibinfo{volume}{29}.
  \bibinfo{publisher}{Curran Associates, Inc.}
\newblock
\urldef\tempurl%
\url{https://papers.nips.cc/paper/2016/hash/eb163727917cbba1eea208541a643e74-Abstract.html}
\showURL{%
\tempurl}


\bibitem[Jumper et~al\mbox{.}(2021)]%
        {jumper_highly_2021}
\bibfield{author}{\bibinfo{person}{John Jumper}, \bibinfo{person}{Richard
  Evans}, \bibinfo{person}{Alexander Pritzel}, \bibinfo{person}{Tim Green},
  \bibinfo{person}{Michael Figurnov}, \bibinfo{person}{Olaf Ronneberger},
  \bibinfo{person}{Kathryn Tunyasuvunakool}, \bibinfo{person}{Russ Bates},
  \bibinfo{person}{Augustin Žídek}, \bibinfo{person}{Anna Potapenko},
  \bibinfo{person}{Alex Bridgland}, \bibinfo{person}{Clemens Meyer},
  \bibinfo{person}{Simon A.~A. Kohl}, \bibinfo{person}{Andrew~J. Ballard},
  \bibinfo{person}{Andrew Cowie}, \bibinfo{person}{Bernardino Romera-Paredes},
  \bibinfo{person}{Stanislav Nikolov}, \bibinfo{person}{Rishub Jain},
  \bibinfo{person}{Jonas Adler}, \bibinfo{person}{Trevor Back},
  \bibinfo{person}{Stig Petersen}, \bibinfo{person}{David Reiman},
  \bibinfo{person}{Ellen Clancy}, \bibinfo{person}{Michal Zielinski},
  \bibinfo{person}{Martin Steinegger}, \bibinfo{person}{Michalina Pacholska},
  \bibinfo{person}{Tamas Berghammer}, \bibinfo{person}{Sebastian Bodenstein},
  \bibinfo{person}{David Silver}, \bibinfo{person}{Oriol Vinyals},
  \bibinfo{person}{Andrew~W. Senior}, \bibinfo{person}{Koray Kavukcuoglu},
  \bibinfo{person}{Pushmeet Kohli}, {and} \bibinfo{person}{Demis Hassabis}.}
  \bibinfo{year}{2021}\natexlab{}.
\newblock \showarticletitle{Highly accurate protein structure prediction with
  {AlphaFold}}.
\newblock \bibinfo{journal}{\emph{Nature}} \bibinfo{volume}{596},
  \bibinfo{number}{7873} (\bibinfo{date}{Aug.} \bibinfo{year}{2021}),
  \bibinfo{pages}{583--589}.
\newblock
\showISSN{1476-4687}
\urldef\tempurl%
\url{https://doi.org/10.1038/s41586-021-03819-2}
\showDOI{\tempurl}
\newblock
\shownote{Number: 7873 Publisher: Nature Publishing Group}.


\bibitem[Kaplan et~al\mbox{.}(2020)]%
        {kaplan_scaling_2020}
\bibfield{author}{\bibinfo{person}{Jared Kaplan}, \bibinfo{person}{Sam
  McCandlish}, \bibinfo{person}{Tom Henighan}, \bibinfo{person}{Tom~B. Brown},
  \bibinfo{person}{Benjamin Chess}, \bibinfo{person}{Rewon Child},
  \bibinfo{person}{Scott Gray}, \bibinfo{person}{Alec Radford},
  \bibinfo{person}{Jeffrey Wu}, {and} \bibinfo{person}{Dario Amodei}.}
  \bibinfo{year}{2020}\natexlab{}.
\newblock \bibinfo{title}{Scaling {Laws} for {Neural} {Language} {Models}}.
\newblock
\newblock
\urldef\tempurl%
\url{https://doi.org/10.48550/arXiv.2001.08361}
\showDOI{\tempurl}
\newblock
\shownote{arXiv:2001.08361 [cs, stat]}.


\bibitem[Kasy and Abebe(2021)]%
        {kasy_fairness_2021}
\bibfield{author}{\bibinfo{person}{Maximilian Kasy} {and}
  \bibinfo{person}{Rediet Abebe}.} \bibinfo{year}{2021}\natexlab{}.
\newblock \showarticletitle{Fairness, {Equality}, and {Power} in {Algorithmic}
  {Decision}-{Making}}. In \bibinfo{booktitle}{\emph{Proceedings of the 2021
  {ACM} {Conference} on {Fairness}, {Accountability}, and {Transparency}}}
  \emph{(\bibinfo{series}{{FAccT} '21})}. \bibinfo{publisher}{Association for
  Computing Machinery}, \bibinfo{address}{New York, NY, USA},
  \bibinfo{pages}{576--586}.
\newblock
\showISBNx{978-1-4503-8309-7}
\urldef\tempurl%
\url{https://doi.org/10.1145/3442188.3445919}
\showDOI{\tempurl}
\newblock
\shownote{event-place: Virtual Event, Canada}.


\bibitem[Kava(2014)]%
        {kava_better_2014}
\bibfield{author}{\bibinfo{person}{Joe Kava}.} \bibinfo{year}{2014}\natexlab{}.
\newblock \bibinfo{title}{Better data centers through machine learning}.
\newblock
\newblock
\urldef\tempurl%
\url{https://blog.google/inside-google/infrastructure/better-data-centers-through-machine/}
\showURL{%
\tempurl}


\bibitem[Keles et~al\mbox{.}(2020)]%
        {keles_systematic_2020}
\bibfield{author}{\bibinfo{person}{Betul Keles}, \bibinfo{person}{Niall
  McCrae}, {and} \bibinfo{person}{Annmarie Grealish}.}
  \bibinfo{year}{2020}\natexlab{}.
\newblock \showarticletitle{A systematic review: the influence of social media
  on depression, anxiety and psychological distress in adolescents}.
\newblock \bibinfo{journal}{\emph{International Journal of Adolescence and
  Youth}} \bibinfo{volume}{25}, \bibinfo{number}{1} (\bibinfo{date}{Dec.}
  \bibinfo{year}{2020}), \bibinfo{pages}{79--93}.
\newblock
\showISSN{0267-3843}
\urldef\tempurl%
\url{https://doi.org/10.1080/02673843.2019.1590851}
\showDOI{\tempurl}
\newblock
\shownote{Publisher: Routledge \_eprint:
  https://doi.org/10.1080/02673843.2019.1590851}.


\bibitem[Kenton et~al\mbox{.}(2021)]%
        {kenton_alignment_2021}
\bibfield{author}{\bibinfo{person}{Zachary Kenton}, \bibinfo{person}{Tom
  Everitt}, \bibinfo{person}{Laura Weidinger}, \bibinfo{person}{Iason Gabriel},
  \bibinfo{person}{Vladimir Mikulik}, {and} \bibinfo{person}{Geoffrey Irving}.}
  \bibinfo{year}{2021}\natexlab{}.
\newblock \showarticletitle{Alignment of {Language} {Agents}}.
\newblock \bibinfo{journal}{\emph{arXiv:2103.14659 [cs]}}
  (\bibinfo{date}{March} \bibinfo{year}{2021}).
\newblock
\urldef\tempurl%
\url{http://arxiv.org/abs/2103.14659}
\showURL{%
\tempurl}
\newblock
\shownote{arXiv: 2103.14659}.


\bibitem[Kenton et~al\mbox{.}(2022)]%
        {kenton_discovering_2022}
\bibfield{author}{\bibinfo{person}{Zachary Kenton}, \bibinfo{person}{Ramana
  Kumar}, \bibinfo{person}{Sebastian Farquhar}, \bibinfo{person}{Jonathan
  Richens}, \bibinfo{person}{Matt MacDermott}, {and} \bibinfo{person}{Tom
  Everitt}.} \bibinfo{year}{2022}\natexlab{}.
\newblock \bibinfo{title}{Discovering {Agents}}.
\newblock
\newblock
\urldef\tempurl%
\url{https://doi.org/10.48550/arXiv.2208.08345}
\showDOI{\tempurl}
\newblock
\shownote{arXiv:2208.08345 [cs]}.


\bibitem[Khatchadourian(2015)]%
        {khatchadourian_doomsday_2015}
\bibfield{author}{\bibinfo{person}{Raffi Khatchadourian}.}
  \bibinfo{year}{2015}\natexlab{}.
\newblock \showarticletitle{The {Doomsday} {Invention}}.
\newblock \bibinfo{journal}{\emph{The New Yorker}} (\bibinfo{date}{Nov.}
  \bibinfo{year}{2015}).
\newblock
\urldef\tempurl%
\url{https://www.newyorker.com/magazine/2015/11/23/doomsday-invention-artificial-intelligence-nick-bostrom}
\showURL{%
\tempurl}


\bibitem[Krakovna et~al\mbox{.}(2020)]%
        {krakovna_specification_2020}
\bibfield{author}{\bibinfo{person}{Victoria Krakovna},
  \bibinfo{person}{Jonathan Uesato}, \bibinfo{person}{Vladimir Mikulik},
  \bibinfo{person}{Matthew Rahtz}, \bibinfo{person}{Tom Everitt},
  \bibinfo{person}{Ramana Kumar}, \bibinfo{person}{Zac Kenton},
  \bibinfo{person}{Jan Leike}, {and} \bibinfo{person}{Shane Legg}.}
  \bibinfo{year}{2020}\natexlab{}.
\newblock \showarticletitle{Specification gaming: the flip side of {AI}
  ingenuity}.
\newblock \bibinfo{journal}{\emph{DeepMind Blog}} (\bibinfo{year}{2020}).
\newblock


\bibitem[Krueger et~al\mbox{.}(2020)]%
        {krueger_hidden_2020}
\bibfield{author}{\bibinfo{person}{David Krueger}, \bibinfo{person}{Tegan
  Maharaj}, {and} \bibinfo{person}{Jan Leike}.}
  \bibinfo{year}{2020}\natexlab{}.
\newblock \bibinfo{title}{Hidden {Incentives} for {Auto}-{Induced}
  {Distributional} {Shift}}.
\newblock
\newblock
\urldef\tempurl%
\url{https://doi.org/10.48550/arXiv.2009.09153}
\showDOI{\tempurl}
\newblock
\shownote{arXiv:2009.09153 [cs, stat]}.


\bibitem[Kuhn and Hacking(2012)]%
        {kuhn_structure_2012}
\bibfield{author}{\bibinfo{person}{T.S. Kuhn} {and} \bibinfo{person}{I.
  Hacking}.} \bibinfo{year}{2012}\natexlab{}.
\newblock \bibinfo{booktitle}{\emph{The {Structure} of {Scientific}
  {Revolutions}}}.
\newblock \bibinfo{publisher}{University of Chicago Press}.
\newblock
\showISBNx{978-0-226-45814-4}
\showLCCN{2011042476}
\urldef\tempurl%
\url{https://books.google.co.uk/books?id=3eP5Y\_OOuzwC}
\showURL{%
\tempurl}


\bibitem[Lake et~al\mbox{.}(2017)]%
        {lake_building_2017}
\bibfield{author}{\bibinfo{person}{Brenden~M. Lake}, \bibinfo{person}{Tomer~D.
  Ullman}, \bibinfo{person}{Joshua~B. Tenenbaum}, {and}
  \bibinfo{person}{Samuel~J. Gershman}.} \bibinfo{year}{2017}\natexlab{}.
\newblock \showarticletitle{Building machines that learn and think like
  people}.
\newblock \bibinfo{journal}{\emph{Behavioral and Brain Sciences}}
  \bibinfo{volume}{40} (\bibinfo{year}{2017}), \bibinfo{pages}{e253}.
\newblock
\showISSN{0140-525X, 1469-1825}
\urldef\tempurl%
\url{https://doi.org/10.1017/S0140525X16001837}
\showDOI{\tempurl}
\newblock
\shownote{Publisher: Cambridge University Press}.


\bibitem[Langosco et~al\mbox{.}(2022)]%
        {langosco_goal_2022}
\bibfield{author}{\bibinfo{person}{Lauro Langosco~Di Langosco},
  \bibinfo{person}{Jack Koch}, \bibinfo{person}{Lee~D Sharkey},
  \bibinfo{person}{Jacob Pfau}, {and} \bibinfo{person}{David Krueger}.}
  \bibinfo{year}{2022}\natexlab{}.
\newblock \showarticletitle{Goal {Misgeneralization} in {Deep} {Reinforcement}
  {Learning}}. In \bibinfo{booktitle}{\emph{Proceedings of the 39th
  {International} {Conference} on {Machine} {Learning}}}
  \emph{(\bibinfo{series}{Proceedings of {Machine} {Learning} {Research}},
  Vol.~\bibinfo{volume}{162})}, \bibfield{editor}{\bibinfo{person}{Kamalika
  Chaudhuri}, \bibinfo{person}{Stefanie Jegelka}, \bibinfo{person}{Le~Song},
  \bibinfo{person}{Csaba Szepesvari}, \bibinfo{person}{Gang Niu}, {and}
  \bibinfo{person}{Sivan Sabato}} (Eds.). \bibinfo{publisher}{PMLR},
  \bibinfo{pages}{12004--12019}.
\newblock
\urldef\tempurl%
\url{https://proceedings.mlr.press/v162/langosco22a.html}
\showURL{%
\tempurl}


\bibitem[Lazar(2022)]%
        {lazar_legitimacy_2022}
\bibfield{author}{\bibinfo{person}{Seth Lazar}.}
  \bibinfo{year}{2022}\natexlab{}.
\newblock \bibinfo{title}{Legitimacy, {Authority}, and the {Political} {Value}
  of {Explanations}}.
\newblock
\newblock
\urldef\tempurl%
\url{https://doi.org/10.48550/arXiv.2208.08628}
\showDOI{\tempurl}
\newblock
\shownote{arXiv:2208.08628 [cs]}.


\bibitem[Leufer(2020)]%
        {leufer_why_2020}
\bibfield{author}{\bibinfo{person}{Daniel Leufer}.}
  \bibinfo{year}{2020}\natexlab{}.
\newblock \showarticletitle{Why {We} {Need} to {Bust} {Some} {Myths} about
  {AI}}.
\newblock \bibinfo{journal}{\emph{Patterns}} \bibinfo{volume}{1},
  \bibinfo{number}{7} (\bibinfo{date}{Oct.} \bibinfo{year}{2020}),
  \bibinfo{pages}{100124}.
\newblock
\showISSN{2666-3899}
\urldef\tempurl%
\url{https://doi.org/10.1016/j.patter.2020.100124}
\showDOI{\tempurl}


\bibitem[Lewis-Kraus(2022)]%
        {lewis-kraus_how_2022}
\bibfield{author}{\bibinfo{person}{Gideon Lewis-Kraus}.}
  \bibinfo{year}{2022}\natexlab{}.
\newblock \bibinfo{title}{How harmful is social media?}
\newblock
\newblock
\urldef\tempurl%
\url{https://www.newyorker.com/culture/annals-of-inquiry/we-know-less-about-social-media-than-we-think}
\showURL{%
\tempurl}
\newblock
\shownote{Publication Title: The New Yorker}.


\bibitem[Li et~al\mbox{.}(2017)]%
        {li_applications_2017}
\bibfield{author}{\bibinfo{person}{Bo-hu Li}, \bibinfo{person}{Bao-cun Hou},
  \bibinfo{person}{Wen-tao Yu}, \bibinfo{person}{Xiao-bing Lu}, {and}
  \bibinfo{person}{Chun-wei Yang}.} \bibinfo{year}{2017}\natexlab{}.
\newblock \showarticletitle{Applications of artificial intelligence in
  intelligent manufacturing: a review}.
\newblock \bibinfo{journal}{\emph{Frontiers of Information Technology \&
  Electronic Engineering}} \bibinfo{volume}{18}, \bibinfo{number}{1}
  (\bibinfo{date}{Jan.} \bibinfo{year}{2017}), \bibinfo{pages}{86--96}.
\newblock
\showISSN{2095-9230}
\urldef\tempurl%
\url{https://doi.org/10.1631/FITEE.1601885}
\showDOI{\tempurl}


\bibitem[Li et~al\mbox{.}(2022)]%
        {li_fairness_2022}
\bibfield{author}{\bibinfo{person}{Yunqi Li}, \bibinfo{person}{Hanxiong Chen},
  \bibinfo{person}{Shuyuan Xu}, \bibinfo{person}{Yingqiang Ge},
  \bibinfo{person}{Juntao Tan}, \bibinfo{person}{Shuchang Liu}, {and}
  \bibinfo{person}{Yongfeng Zhang}.} \bibinfo{year}{2022}\natexlab{}.
\newblock \bibinfo{title}{Fairness in {Recommendation}: {A} {Survey}}.
\newblock
\newblock
\urldef\tempurl%
\url{https://doi.org/10.48550/arXiv.2205.13619}
\showDOI{\tempurl}
\newblock
\shownote{arXiv:2205.13619 [cs]}.


\bibitem[Lin et~al\mbox{.}(2022)]%
        {lin_truthfulqa_2022}
\bibfield{author}{\bibinfo{person}{Stephanie Lin}, \bibinfo{person}{Jacob
  Hilton}, {and} \bibinfo{person}{Owain Evans}.}
  \bibinfo{year}{2022}\natexlab{}.
\newblock \showarticletitle{{TruthfulQA}: {Measuring} {How} {Models} {Mimic}
  {Human} {Falsehoods}}. In \bibinfo{booktitle}{\emph{Proceedings of the 60th
  {Annual} {Meeting} of the {Association} for {Computational} {Linguistics}
  ({Volume} 1: {Long} {Papers})}}. \bibinfo{publisher}{Association for
  Computational Linguistics}, \bibinfo{address}{Dublin, Ireland},
  \bibinfo{pages}{3214--3252}.
\newblock
\urldef\tempurl%
\url{https://doi.org/10.18653/v1/2022.acl-long.229}
\showDOI{\tempurl}


\bibitem[Liu et~al\mbox{.}(2018)]%
        {liu_delayed_2018}
\bibfield{author}{\bibinfo{person}{Lydia~T Liu}, \bibinfo{person}{Sarah Dean},
  \bibinfo{person}{Esther Rolf}, \bibinfo{person}{Max Simchowitz}, {and}
  \bibinfo{person}{Moritz Hardt}.} \bibinfo{year}{2018}\natexlab{}.
\newblock \showarticletitle{Delayed impact of fair machine learning}. In
  \bibinfo{booktitle}{\emph{International {Conference} on {Machine}
  {Learning}}}. \bibinfo{publisher}{PMLR}, \bibinfo{pages}{3150--3158}.
\newblock


\bibitem[Loi and Spielkamp(2021)]%
        {loi_towards_2021}
\bibfield{author}{\bibinfo{person}{Michele Loi} {and} \bibinfo{person}{Matthias
  Spielkamp}.} \bibinfo{year}{2021}\natexlab{}.
\newblock \showarticletitle{Towards {Accountability} in the {Use} of
  {Artificial} {Intelligence} for {Public} {Administrations}}. In
  \bibinfo{booktitle}{\emph{Proceedings of the 2021 {AAAI}/{ACM} {Conference}
  on {AI}, {Ethics}, and {Society}}}. \bibinfo{publisher}{ACM}.
\newblock
\urldef\tempurl%
\url{https://doi.org/10.1145/3461702.3462631}
\showDOI{\tempurl}


\bibitem[Manne(2017)]%
        {manne_down_2017}
\bibfield{author}{\bibinfo{person}{Kate Manne}.}
  \bibinfo{year}{2017}\natexlab{}.
\newblock \bibinfo{booktitle}{\emph{Down {Girl}: {The} {Logic} of {Misogyny}}}.
\newblock \bibinfo{publisher}{Oxford University Press}.
\newblock


\bibitem[maraoz(2021)]%
        {maraoz_interviewing_2021}
\bibfield{author}{\bibinfo{person}{maraoz}.} \bibinfo{year}{2021}\natexlab{}.
\newblock \bibinfo{title}{Interviewing {Albert} {Einstein} via {GPT}-3}.
\newblock
\newblock
\urldef\tempurl%
\url{https://maraoz.substack.com/embed}
\showURL{%
\tempurl}


\bibitem[Menick et~al\mbox{.}(2022)]%
        {menick_teaching_2022}
\bibfield{author}{\bibinfo{person}{Jacob Menick}, \bibinfo{person}{Maja
  Trebacz}, \bibinfo{person}{Vladimir Mikulik}, \bibinfo{person}{John
  Aslanides}, \bibinfo{person}{Francis Song}, \bibinfo{person}{Martin
  Chadwick}, \bibinfo{person}{Mia Glaese}, \bibinfo{person}{Susannah Young},
  \bibinfo{person}{Lucy Campbell-Gillingham}, \bibinfo{person}{Geoffrey
  Irving}, {and} \bibinfo{person}{Nat McAleese}.}
  \bibinfo{year}{2022}\natexlab{}.
\newblock \bibinfo{title}{Teaching language models to support answers with
  verified quotes}.
\newblock
\newblock
\urldef\tempurl%
\url{https://doi.org/10.48550/arXiv.2203.11147}
\showDOI{\tempurl}
\newblock
\shownote{arXiv:2203.11147 [cs]}.


\bibitem[Meta(2023)]%
        {meta_meta_2023}
\bibfield{author}{\bibinfo{person}{Meta}.} \bibinfo{year}{2023}\natexlab{}.
\newblock \bibinfo{title}{Meta {Reports} {Fourth} {Quarter} and {Full} {Year}
  2022 {Results}}.
\newblock
\newblock
\urldef\tempurl%
\url{https://investor.fb.com/investor-news/press-release-details/2023/Meta-Reports-Fourth-Quarter-and-Full-Year-2022-Results/default.aspx}
\showURL{%
\tempurl}


\bibitem[Milano et~al\mbox{.}(2020)]%
        {milano_recommender_2020}
\bibfield{author}{\bibinfo{person}{Silvia Milano},
  \bibinfo{person}{Mariarosaria Taddeo}, {and} \bibinfo{person}{Luciano
  Floridi}.} \bibinfo{year}{2020}\natexlab{}.
\newblock \showarticletitle{Recommender systems and their ethical challenges}.
\newblock \bibinfo{journal}{\emph{AI \& SOCIETY}} \bibinfo{volume}{35},
  \bibinfo{number}{4} (\bibinfo{date}{Dec.} \bibinfo{year}{2020}),
  \bibinfo{pages}{957--967}.
\newblock
\showISSN{1435-5655}
\urldef\tempurl%
\url{https://doi.org/10.1007/s00146-020-00950-y}
\showDOI{\tempurl}


\bibitem[Mitchell et~al\mbox{.}(2019)]%
        {mitchell_model_2019}
\bibfield{author}{\bibinfo{person}{Margaret Mitchell}, \bibinfo{person}{Simone
  Wu}, \bibinfo{person}{Andrew Zaldivar}, \bibinfo{person}{Parker Barnes},
  \bibinfo{person}{Lucy Vasserman}, \bibinfo{person}{Ben Hutchinson},
  \bibinfo{person}{Elena Spitzer}, \bibinfo{person}{Inioluwa~Deborah Raji},
  {and} \bibinfo{person}{Timnit Gebru}.} \bibinfo{year}{2019}\natexlab{}.
\newblock \showarticletitle{Model {Cards} for {Model} {Reporting}}. In
  \bibinfo{booktitle}{\emph{Proceedings of the {Conference} on {Fairness},
  {Accountability}, and {Transparency}}}. \bibinfo{publisher}{ACM}.
\newblock
\urldef\tempurl%
\url{https://doi.org/10.1145/3287560.3287596}
\showDOI{\tempurl}


\bibitem[Mnih et~al\mbox{.}(2013)]%
        {mnih_playing_2013}
\bibfield{author}{\bibinfo{person}{Volodymyr Mnih}, \bibinfo{person}{Koray
  Kavukcuoglu}, \bibinfo{person}{David Silver}, \bibinfo{person}{Alex Graves},
  \bibinfo{person}{Ioannis Antonoglou}, \bibinfo{person}{Daan Wierstra}, {and}
  \bibinfo{person}{Martin Riedmiller}.} \bibinfo{year}{2013}\natexlab{}.
\newblock \bibinfo{title}{Playing {Atari} with {Deep} {Reinforcement}
  {Learning}}.
\newblock
\newblock
\urldef\tempurl%
\url{https://doi.org/10.48550/arXiv.1312.5602}
\showDOI{\tempurl}
\newblock
\shownote{arXiv:1312.5602 [cs]}.


\bibitem[Myths({[n.\,d.]})]%
        {ai_myths_myth_nodate}
\bibfield{author}{\bibinfo{person}{AI Myths}.}
  \bibinfo{year}{[n.\,d.]}\natexlab{}.
\newblock \bibinfo{title}{Myth: {AI} has agency}.
\newblock
\newblock
\urldef\tempurl%
\url{https://www.aimyths.org/ai-has-agency}
\showURL{%
\tempurl}


\bibitem[Nakano et~al\mbox{.}(2022)]%
        {nakano_webgpt_2022}
\bibfield{author}{\bibinfo{person}{Reiichiro Nakano}, \bibinfo{person}{Jacob
  Hilton}, \bibinfo{person}{Suchir Balaji}, \bibinfo{person}{Jeff Wu},
  \bibinfo{person}{Long Ouyang}, \bibinfo{person}{Christina Kim},
  \bibinfo{person}{Christopher Hesse}, \bibinfo{person}{Shantanu Jain},
  \bibinfo{person}{Vineet Kosaraju}, \bibinfo{person}{William Saunders},
  \bibinfo{person}{Xu Jiang}, \bibinfo{person}{Karl Cobbe},
  \bibinfo{person}{Tyna Eloundou}, \bibinfo{person}{Gretchen Krueger},
  \bibinfo{person}{Kevin Button}, \bibinfo{person}{Matthew Knight},
  \bibinfo{person}{Benjamin Chess}, {and} \bibinfo{person}{John Schulman}.}
  \bibinfo{year}{2022}\natexlab{}.
\newblock \bibinfo{title}{{WebGPT}: {Browser}-assisted question-answering with
  human feedback}.
\newblock
\newblock
\urldef\tempurl%
\url{https://doi.org/10.48550/arXiv.2112.09332}
\showDOI{\tempurl}
\newblock
\shownote{arXiv:2112.09332 [cs]}.


\bibitem[Natale and Ballatore(2020)]%
        {natale_imagining_2020}
\bibfield{author}{\bibinfo{person}{Simone Natale} {and} \bibinfo{person}{Andrea
  Ballatore}.} \bibinfo{year}{2020}\natexlab{}.
\newblock \showarticletitle{Imagining the thinking machine: {Technological}
  myths and the rise of artificial intelligence}.
\newblock \bibinfo{journal}{\emph{Convergence}} \bibinfo{volume}{26},
  \bibinfo{number}{1} (\bibinfo{year}{2020}), \bibinfo{pages}{3--18}.
\newblock
\newblock
\shownote{Publisher: SAGE Publications Sage UK: London, England}.


\bibitem[Nayak(2019)]%
        {nayak_understanding_2019}
\bibfield{author}{\bibinfo{person}{Pandu Nayak}.}
  \bibinfo{year}{2019}\natexlab{}.
\newblock \bibinfo{title}{Understanding searches better than ever before}.
\newblock
\newblock
\urldef\tempurl%
\url{https://blog.google/products/search/search-language-understanding-bert/}
\showURL{%
\tempurl}


\bibitem[Nissenbaum(1996)]%
        {nissenbaum_accountability_1996}
\bibfield{author}{\bibinfo{person}{Helen Nissenbaum}.}
  \bibinfo{year}{1996}\natexlab{}.
\newblock \showarticletitle{Accountability in a computerized society}.
\newblock \bibinfo{journal}{\emph{Science and Engineering Ethics}}
  \bibinfo{volume}{2}, \bibinfo{number}{1} (\bibinfo{date}{March}
  \bibinfo{year}{1996}), \bibinfo{pages}{25--42}.
\newblock
\showISSN{1471-5546}
\urldef\tempurl%
\url{https://doi.org/10.1007/BF02639315}
\showDOI{\tempurl}


\bibitem[Obermeyer and Mullainathan(2019)]%
        {obermeyer_dissecting_2019}
\bibfield{author}{\bibinfo{person}{Ziad Obermeyer} {and}
  \bibinfo{person}{Sendhil Mullainathan}.} \bibinfo{year}{2019}\natexlab{}.
\newblock \showarticletitle{Dissecting {Racial} {Bias} in an {Algorithm} that
  {Guides} {Health} {Decisions} for 70 {Million} {People}}. In
  \bibinfo{booktitle}{\emph{Proceedings of the {Conference} on {Fairness},
  {Accountability}, and {Transparency}}}. \bibinfo{publisher}{ACM}.
\newblock
\urldef\tempurl%
\url{https://doi.org/10.1145/3287560.3287593}
\showDOI{\tempurl}


\bibitem[Olsson et~al\mbox{.}(2022)]%
        {olsson_-context_2022}
\bibfield{author}{\bibinfo{person}{Catherine Olsson}, \bibinfo{person}{Nelson
  Elhage}, \bibinfo{person}{Neel Nanda}, \bibinfo{person}{Nicholas Joseph},
  \bibinfo{person}{Nova DasSarma}, \bibinfo{person}{Tom Henighan},
  \bibinfo{person}{Ben Mann}, \bibinfo{person}{Amanda Askell},
  \bibinfo{person}{Yuntao Bai}, \bibinfo{person}{Anna Chen},
  \bibinfo{person}{Tom Conerly}, \bibinfo{person}{Dawn Drain},
  \bibinfo{person}{Deep Ganguli}, \bibinfo{person}{Zac Hatfield-Dodds},
  \bibinfo{person}{Danny Hernandez}, \bibinfo{person}{Scott Johnston},
  \bibinfo{person}{Andy Jones}, \bibinfo{person}{Jackson Kernion},
  \bibinfo{person}{Liane Lovitt}, \bibinfo{person}{Kamal Ndousse},
  \bibinfo{person}{Dario Amodei}, \bibinfo{person}{Tom Brown},
  \bibinfo{person}{Jack Clark}, \bibinfo{person}{Jared Kaplan},
  \bibinfo{person}{Sam McCandlish}, {and} \bibinfo{person}{Chris Olah}.}
  \bibinfo{year}{2022}\natexlab{}.
\newblock \showarticletitle{In-context {Learning} and {Induction} {Heads}}.
\newblock \bibinfo{journal}{\emph{Transformer Circuits Thread}}
  (\bibinfo{year}{2022}).
\newblock


\bibitem[Omohundro(2008)]%
        {omohundro_basic_2008}
\bibfield{author}{\bibinfo{person}{Stephen~M Omohundro}.}
  \bibinfo{year}{2008}\natexlab{}.
\newblock \showarticletitle{The {Basic} {AI} {Drives}}. In
  \bibinfo{booktitle}{\emph{{AGI}}}, Vol.~\bibinfo{volume}{171}.
  \bibinfo{pages}{483--492}.
\newblock


\bibitem[OpenAI(2022)]%
        {openai_chatgpt_2022}
\bibfield{author}{\bibinfo{person}{OpenAI}.} \bibinfo{year}{2022}\natexlab{}.
\newblock \bibinfo{title}{{ChatGPT}: {Optimizing} {Language} {Models} for
  {Dialogue}}.
\newblock
\newblock
\urldef\tempurl%
\url{https://openai.com/blog/chatgpt/}
\showURL{%
\tempurl}


\bibitem[OpenAI(2023)]%
        {openai_chatgpt_2023}
\bibfield{author}{\bibinfo{person}{OpenAI}.} \bibinfo{year}{2023}\natexlab{}.
\newblock \bibinfo{title}{{ChatGPT} plugins}.
\newblock
\newblock
\urldef\tempurl%
\url{https://openai.com/blog/chatgpt-plugins}
\showURL{%
\tempurl}


\bibitem[Orseau et~al\mbox{.}(2018)]%
        {orseau_agents_2018}
\bibfield{author}{\bibinfo{person}{Laurent Orseau},
  \bibinfo{person}{Simon~McGregor McGill}, {and} \bibinfo{person}{Shane Legg}.}
  \bibinfo{year}{2018}\natexlab{}.
\newblock \bibinfo{title}{Agents and {Devices}: {A} {Relative} {Definition} of
  {Agency}}.
\newblock
\newblock
\urldef\tempurl%
\url{https://doi.org/10.48550/arXiv.1805.12387}
\showDOI{\tempurl}
\newblock
\shownote{arXiv:1805.12387 [cs, stat]}.


\bibitem[Pan et~al\mbox{.}(2022)]%
        {pan_effects_2022}
\bibfield{author}{\bibinfo{person}{Alexander Pan}, \bibinfo{person}{Kush
  Bhatia}, {and} \bibinfo{person}{Jacob Steinhardt}.}
  \bibinfo{year}{2022}\natexlab{}.
\newblock \showarticletitle{The {Effects} of {Reward} {Misspecification}:
  {Mapping} and {Mitigating} {Misaligned} {Models}}. In
  \bibinfo{booktitle}{\emph{International {Conference} on {Learning}
  {Representations}}}.
\newblock
\urldef\tempurl%
\url{https://openreview.net/forum?id=JYtwGwIL7ye}
\showURL{%
\tempurl}


\bibitem[Pan et~al\mbox{.}(2023)]%
        {pan_rewards_2023}
\bibfield{author}{\bibinfo{person}{Alexander Pan}, \bibinfo{person}{Chan~Jun
  Shern}, \bibinfo{person}{Andy Zou}, \bibinfo{person}{Nathaniel Li},
  \bibinfo{person}{Steven Basart}, \bibinfo{person}{Thomas Woodside},
  \bibinfo{person}{Jonathan Ng}, \bibinfo{person}{Hanlin Zhang},
  \bibinfo{person}{Scott Emmons}, {and} \bibinfo{person}{Dan Hendrycks}.}
  \bibinfo{year}{2023}\natexlab{}.
\newblock \bibinfo{title}{Do the {Rewards} {Justify} the {Means}? {Measuring}
  {Trade}-{Offs} {Between} {Rewards} and {Ethical} {Behavior} in the
  {MACHIAVELLI} {Benchmark}}.
\newblock
\newblock
\urldef\tempurl%
\url{https://doi.org/10.48550/arXiv.2304.03279}
\showDOI{\tempurl}
\newblock
\shownote{arXiv:2304.03279 [cs]}.


\bibitem[Park et~al\mbox{.}(2022)]%
        {park_social_2022}
\bibfield{author}{\bibinfo{person}{Joon~Sung Park}, \bibinfo{person}{Lindsay
  Popowski}, \bibinfo{person}{Carrie Cai}, \bibinfo{person}{Meredith~Ringel
  Morris}, \bibinfo{person}{Percy Liang}, {and} \bibinfo{person}{Michael~S.
  Bernstein}.} \bibinfo{year}{2022}\natexlab{}.
\newblock \showarticletitle{Social {Simulacra}: {Creating} {Populated}
  {Prototypes} for {Social} {Computing} {Systems}}. In
  \bibinfo{booktitle}{\emph{Proceedings of the 35th {Annual} {ACM} {Symposium}
  on {User} {Interface} {Software} and {Technology}}}
  \emph{(\bibinfo{series}{{UIST} '22})}. \bibinfo{publisher}{Association for
  Computing Machinery}, \bibinfo{address}{New York, NY, USA},
  \bibinfo{pages}{1--18}.
\newblock
\showISBNx{978-1-4503-9320-1}
\urldef\tempurl%
\url{https://doi.org/10.1145/3526113.3545616}
\showDOI{\tempurl}


\bibitem[Pasquinelli and Joler(2021)]%
        {pasquinelli_nooscope_2021}
\bibfield{author}{\bibinfo{person}{Matteo Pasquinelli} {and}
  \bibinfo{person}{Vladan Joler}.} \bibinfo{year}{2021}\natexlab{}.
\newblock \showarticletitle{The {Nooscope} manifested: {AI} as instrument of
  knowledge extractivism}.
\newblock \bibinfo{journal}{\emph{AI \& SOCIETY}} \bibinfo{volume}{36},
  \bibinfo{number}{4} (\bibinfo{date}{Dec.} \bibinfo{year}{2021}),
  \bibinfo{pages}{1263--1280}.
\newblock
\showISSN{0951-5666, 1435-5655}
\urldef\tempurl%
\url{https://doi.org/10.1007/s00146-020-01097-6}
\showDOI{\tempurl}


\bibitem[Perez et~al\mbox{.}(2022)]%
        {perez_discovering_2022}
\bibfield{author}{\bibinfo{person}{Ethan Perez}, \bibinfo{person}{Sam Ringer},
  \bibinfo{person}{Kamilė Lukošiūtė}, \bibinfo{person}{Karina Nguyen},
  \bibinfo{person}{Edwin Chen}, \bibinfo{person}{Scott Heiner},
  \bibinfo{person}{Craig Pettit}, \bibinfo{person}{Catherine Olsson},
  \bibinfo{person}{Sandipan Kundu}, \bibinfo{person}{Saurav Kadavath},
  \bibinfo{person}{Andy Jones}, \bibinfo{person}{Anna Chen},
  \bibinfo{person}{Ben Mann}, \bibinfo{person}{Brian Israel},
  \bibinfo{person}{Bryan Seethor}, \bibinfo{person}{Cameron McKinnon},
  \bibinfo{person}{Christopher Olah}, \bibinfo{person}{Da Yan},
  \bibinfo{person}{Daniela Amodei}, \bibinfo{person}{Dario Amodei},
  \bibinfo{person}{Dawn Drain}, \bibinfo{person}{Dustin Li},
  \bibinfo{person}{Eli Tran-Johnson}, \bibinfo{person}{Guro Khundadze},
  \bibinfo{person}{Jackson Kernion}, \bibinfo{person}{James Landis},
  \bibinfo{person}{Jamie Kerr}, \bibinfo{person}{Jared Mueller},
  \bibinfo{person}{Jeeyoon Hyun}, \bibinfo{person}{Joshua Landau},
  \bibinfo{person}{Kamal Ndousse}, \bibinfo{person}{Landon Goldberg},
  \bibinfo{person}{Liane Lovitt}, \bibinfo{person}{Martin Lucas},
  \bibinfo{person}{Michael Sellitto}, \bibinfo{person}{Miranda Zhang},
  \bibinfo{person}{Neerav Kingsland}, \bibinfo{person}{Nelson Elhage},
  \bibinfo{person}{Nicholas Joseph}, \bibinfo{person}{Noemí Mercado},
  \bibinfo{person}{Nova DasSarma}, \bibinfo{person}{Oliver Rausch},
  \bibinfo{person}{Robin Larson}, \bibinfo{person}{Sam McCandlish},
  \bibinfo{person}{Scott Johnston}, \bibinfo{person}{Shauna Kravec},
  \bibinfo{person}{Sheer~El Showk}, \bibinfo{person}{Tamera Lanham},
  \bibinfo{person}{Timothy Telleen-Lawton}, \bibinfo{person}{Tom Brown},
  \bibinfo{person}{Tom Henighan}, \bibinfo{person}{Tristan Hume},
  \bibinfo{person}{Yuntao Bai}, \bibinfo{person}{Zac Hatfield-Dodds},
  \bibinfo{person}{Jack Clark}, \bibinfo{person}{Samuel~R. Bowman},
  \bibinfo{person}{Amanda Askell}, \bibinfo{person}{Roger Grosse},
  \bibinfo{person}{Danny Hernandez}, \bibinfo{person}{Deep Ganguli},
  \bibinfo{person}{Evan Hubinger}, \bibinfo{person}{Nicholas Schiefer}, {and}
  \bibinfo{person}{Jared Kaplan}.} \bibinfo{year}{2022}\natexlab{}.
\newblock \bibinfo{title}{Discovering {Language} {Model} {Behaviors} with
  {Model}-{Written} {Evaluations}}.
\newblock
\newblock
\urldef\tempurl%
\url{https://doi.org/10.48550/arXiv.2212.09251}
\showDOI{\tempurl}
\newblock
\shownote{arXiv:2212.09251 [cs]}.


\bibitem[Perolat et~al\mbox{.}(2022)]%
        {perolat_mastering_2022}
\bibfield{author}{\bibinfo{person}{Julien Perolat}, \bibinfo{person}{Bart
  De~Vylder}, \bibinfo{person}{Daniel Hennes}, \bibinfo{person}{Eugene
  Tarassov}, \bibinfo{person}{Florian Strub}, \bibinfo{person}{Vincent de
  Boer}, \bibinfo{person}{Paul Muller}, \bibinfo{person}{Jerome~T. Connor},
  \bibinfo{person}{Neil Burch}, \bibinfo{person}{Thomas Anthony},
  \bibinfo{person}{Stephen McAleer}, \bibinfo{person}{Romuald Elie},
  \bibinfo{person}{Sarah~H. Cen}, \bibinfo{person}{Zhe Wang},
  \bibinfo{person}{Audrunas Gruslys}, \bibinfo{person}{Aleksandra Malysheva},
  \bibinfo{person}{Mina Khan}, \bibinfo{person}{Sherjil Ozair},
  \bibinfo{person}{Finbarr Timbers}, \bibinfo{person}{Toby Pohlen},
  \bibinfo{person}{Tom Eccles}, \bibinfo{person}{Mark Rowland},
  \bibinfo{person}{Marc Lanctot}, \bibinfo{person}{Jean-Baptiste Lespiau},
  \bibinfo{person}{Bilal Piot}, \bibinfo{person}{Shayegan Omidshafiei},
  \bibinfo{person}{Edward Lockhart}, \bibinfo{person}{Laurent Sifre},
  \bibinfo{person}{Nathalie Beauguerlange}, \bibinfo{person}{Remi Munos},
  \bibinfo{person}{David Silver}, \bibinfo{person}{Satinder Singh},
  \bibinfo{person}{Demis Hassabis}, {and} \bibinfo{person}{Karl Tuyls}.}
  \bibinfo{year}{2022}\natexlab{}.
\newblock \showarticletitle{Mastering the game of {Stratego} with model-free
  multiagent reinforcement learning}.
\newblock \bibinfo{journal}{\emph{Science}} \bibinfo{volume}{378},
  \bibinfo{number}{6623} (\bibinfo{date}{Dec.} \bibinfo{year}{2022}),
  \bibinfo{pages}{990--996}.
\newblock
\urldef\tempurl%
\url{https://doi.org/10.1126/science.add4679}
\showDOI{\tempurl}
\newblock
\shownote{Publisher: American Association for the Advancement of Science}.


\bibitem[Perrigo(2023)]%
        {perrigo_exclusive_2023}
\bibfield{author}{\bibinfo{person}{Billy Perrigo}.}
  \bibinfo{year}{2023}\natexlab{}.
\newblock \showarticletitle{Exclusive: {The} \$2 {Per} {Hour} {Workers} {Who}
  {Made} {ChatGPT} {Safer}}.
\newblock \bibinfo{journal}{\emph{Time}} (\bibinfo{date}{Jan.}
  \bibinfo{year}{2023}).
\newblock
\urldef\tempurl%
\url{https://time.com/6247678/openai-chatgpt-kenya-workers/}
\showURL{%
\tempurl}


\bibitem[Piantadosi and Hill(2022)]%
        {piantadosi_meaning_2022}
\bibfield{author}{\bibinfo{person}{Steven~T. Piantadosi} {and}
  \bibinfo{person}{Felix Hill}.} \bibinfo{year}{2022}\natexlab{}.
\newblock \bibinfo{title}{Meaning without reference in large language models}.
\newblock
\newblock
\urldef\tempurl%
\url{https://doi.org/10.48550/arXiv.2208.02957}
\showDOI{\tempurl}
\newblock
\shownote{arXiv:2208.02957 [cs]}.


\bibitem[Pratyusha(2020)]%
        {pratyusha_world_2020}
\bibfield{author}{\bibinfo{person}{By Pratyusha}.}
  \bibinfo{year}{2020}\natexlab{}.
\newblock \showarticletitle{World view}.
\newblock \bibinfo{journal}{\emph{Nature}}  \bibinfo{volume}{583}
  (\bibinfo{year}{2020}), \bibinfo{pages}{169}.
\newblock


\bibitem[Rae et~al\mbox{.}(2022)]%
        {rae_scaling_2022}
\bibfield{author}{\bibinfo{person}{Jack~W. Rae}, \bibinfo{person}{Sebastian
  Borgeaud}, \bibinfo{person}{Trevor Cai}, \bibinfo{person}{Katie Millican},
  \bibinfo{person}{Jordan Hoffmann}, \bibinfo{person}{Francis Song},
  \bibinfo{person}{John Aslanides}, \bibinfo{person}{Sarah Henderson},
  \bibinfo{person}{Roman Ring}, \bibinfo{person}{Susannah Young},
  \bibinfo{person}{Eliza Rutherford}, \bibinfo{person}{Tom Hennigan},
  \bibinfo{person}{Jacob Menick}, \bibinfo{person}{Albin Cassirer},
  \bibinfo{person}{Richard Powell}, \bibinfo{person}{George van~den Driessche},
  \bibinfo{person}{Lisa~Anne Hendricks}, \bibinfo{person}{Maribeth Rauh},
  \bibinfo{person}{Po-Sen Huang}, \bibinfo{person}{Amelia Glaese},
  \bibinfo{person}{Johannes Welbl}, \bibinfo{person}{Sumanth Dathathri},
  \bibinfo{person}{Saffron Huang}, \bibinfo{person}{Jonathan Uesato},
  \bibinfo{person}{John Mellor}, \bibinfo{person}{Irina Higgins},
  \bibinfo{person}{Antonia Creswell}, \bibinfo{person}{Nat McAleese},
  \bibinfo{person}{Amy Wu}, \bibinfo{person}{Erich Elsen},
  \bibinfo{person}{Siddhant Jayakumar}, \bibinfo{person}{Elena Buchatskaya},
  \bibinfo{person}{David Budden}, \bibinfo{person}{Esme Sutherland},
  \bibinfo{person}{Karen Simonyan}, \bibinfo{person}{Michela Paganini},
  \bibinfo{person}{Laurent Sifre}, \bibinfo{person}{Lena Martens},
  \bibinfo{person}{Xiang~Lorraine Li}, \bibinfo{person}{Adhiguna Kuncoro},
  \bibinfo{person}{Aida Nematzadeh}, \bibinfo{person}{Elena Gribovskaya},
  \bibinfo{person}{Domenic Donato}, \bibinfo{person}{Angeliki Lazaridou},
  \bibinfo{person}{Arthur Mensch}, \bibinfo{person}{Jean-Baptiste Lespiau},
  \bibinfo{person}{Maria Tsimpoukelli}, \bibinfo{person}{Nikolai Grigorev},
  \bibinfo{person}{Doug Fritz}, \bibinfo{person}{Thibault Sottiaux},
  \bibinfo{person}{Mantas Pajarskas}, \bibinfo{person}{Toby Pohlen},
  \bibinfo{person}{Zhitao Gong}, \bibinfo{person}{Daniel Toyama},
  \bibinfo{person}{Cyprien de~Masson d'Autume}, \bibinfo{person}{Yujia Li},
  \bibinfo{person}{Tayfun Terzi}, \bibinfo{person}{Vladimir Mikulik},
  \bibinfo{person}{Igor Babuschkin}, \bibinfo{person}{Aidan Clark},
  \bibinfo{person}{Diego de~Las Casas}, \bibinfo{person}{Aurelia Guy},
  \bibinfo{person}{Chris Jones}, \bibinfo{person}{James Bradbury},
  \bibinfo{person}{Matthew Johnson}, \bibinfo{person}{Blake Hechtman},
  \bibinfo{person}{Laura Weidinger}, \bibinfo{person}{Iason Gabriel},
  \bibinfo{person}{William Isaac}, \bibinfo{person}{Ed Lockhart},
  \bibinfo{person}{Simon Osindero}, \bibinfo{person}{Laura Rimell},
  \bibinfo{person}{Chris Dyer}, \bibinfo{person}{Oriol Vinyals},
  \bibinfo{person}{Kareem Ayoub}, \bibinfo{person}{Jeff Stanway},
  \bibinfo{person}{Lorrayne Bennett}, \bibinfo{person}{Demis Hassabis},
  \bibinfo{person}{Koray Kavukcuoglu}, {and} \bibinfo{person}{Geoffrey
  Irving}.} \bibinfo{year}{2022}\natexlab{}.
\newblock \bibinfo{title}{Scaling {Language} {Models}: {Methods}, {Analysis} \&
  {Insights} from {Training} {Gopher}}.
\newblock
\newblock
\urldef\tempurl%
\url{https://doi.org/10.48550/arXiv.2112.11446}
\showDOI{\tempurl}
\newblock
\shownote{arXiv:2112.11446 [cs]}.


\bibitem[Raji et~al\mbox{.}(2021)]%
        {raji_ai_2021}
\bibfield{author}{\bibinfo{person}{Deborah Raji}, \bibinfo{person}{Emily
  Denton}, \bibinfo{person}{Emily~M. Bender}, \bibinfo{person}{Alex Hanna},
  {and} \bibinfo{person}{Amandalynne Paullada}.}
  \bibinfo{year}{2021}\natexlab{}.
\newblock \showarticletitle{{AI} and the {Everything} in the {Whole} {Wide}
  {World} {Benchmark}}.
\newblock \bibinfo{journal}{\emph{Proceedings of the Neural Information
  Processing Systems Track on Datasets and Benchmarks}}  \bibinfo{volume}{1}
  (\bibinfo{date}{Dec.} \bibinfo{year}{2021}).
\newblock
\urldef\tempurl%
\url{https://datasets-benchmarks-proceedings.neurips.cc/paper/2021/hash/084b6fbb10729ed4da8c3d3f5a3ae7c9-Abstract-round2.html}
\showURL{%
\tempurl}


\bibitem[Raji and Buolamwini(2019)]%
        {raji_actionable_2019}
\bibfield{author}{\bibinfo{person}{Inioluwa~Deborah Raji} {and}
  \bibinfo{person}{Joy Buolamwini}.} \bibinfo{year}{2019}\natexlab{}.
\newblock \showarticletitle{Actionable {Auditing}}. In
  \bibinfo{booktitle}{\emph{Proceedings of the 2019 {AAAI}/{ACM} {Conference}
  on {AI}, {Ethics}, and {Society}}}. \bibinfo{publisher}{ACM}.
\newblock
\urldef\tempurl%
\url{https://doi.org/10.1145/3306618.3314244}
\showDOI{\tempurl}


\bibitem[Raji et~al\mbox{.}(2022)]%
        {raji_fallacy_2022}
\bibfield{author}{\bibinfo{person}{Inioluwa~Deborah Raji},
  \bibinfo{person}{I.~Elizabeth Kumar}, \bibinfo{person}{Aaron Horowitz}, {and}
  \bibinfo{person}{Andrew Selbst}.} \bibinfo{year}{2022}\natexlab{}.
\newblock \showarticletitle{The {Fallacy} of {AI} {Functionality}}. In
  \bibinfo{booktitle}{\emph{2022 {ACM} {Conference} on {Fairness},
  {Accountability}, and {Transparency}}}. \bibinfo{publisher}{ACM}.
\newblock
\urldef\tempurl%
\url{https://doi.org/10.1145/3531146.3533158}
\showDOI{\tempurl}


\bibitem[Raji et~al\mbox{.}(2020)]%
        {raji_closing_2020}
\bibfield{author}{\bibinfo{person}{Inioluwa~Deborah Raji},
  \bibinfo{person}{Andrew Smart}, \bibinfo{person}{Rebecca~N. White},
  \bibinfo{person}{Margaret Mitchell}, \bibinfo{person}{Timnit Gebru},
  \bibinfo{person}{Ben Hutchinson}, \bibinfo{person}{Jamila Smith-Loud},
  \bibinfo{person}{Daniel Theron}, {and} \bibinfo{person}{Parker Barnes}.}
  \bibinfo{year}{2020}\natexlab{}.
\newblock \showarticletitle{Closing the {AI} accountability gap}. In
  \bibinfo{booktitle}{\emph{Proceedings of the 2020 {Conference} on {Fairness},
  {Accountability}, and {Transparency}}}. \bibinfo{publisher}{ACM}.
\newblock
\urldef\tempurl%
\url{https://doi.org/10.1145/3351095.3372873}
\showDOI{\tempurl}


\bibitem[Reed et~al\mbox{.}(2022)]%
        {reed_generalist_2022}
\bibfield{author}{\bibinfo{person}{Scott Reed}, \bibinfo{person}{Konrad Zolna},
  \bibinfo{person}{Emilio Parisotto}, \bibinfo{person}{Sergio~Gómez
  Colmenarejo}, \bibinfo{person}{Alexander Novikov}, \bibinfo{person}{Gabriel
  Barth-maron}, \bibinfo{person}{Mai Giménez}, \bibinfo{person}{Yury Sulsky},
  \bibinfo{person}{Jackie Kay}, \bibinfo{person}{Jost~Tobias Springenberg},
  \bibinfo{person}{Tom Eccles}, \bibinfo{person}{Jake Bruce},
  \bibinfo{person}{Ali Razavi}, \bibinfo{person}{Ashley Edwards},
  \bibinfo{person}{Nicolas Heess}, \bibinfo{person}{Yutian Chen},
  \bibinfo{person}{Raia Hadsell}, \bibinfo{person}{Oriol Vinyals},
  \bibinfo{person}{Mahyar Bordbar}, {and} \bibinfo{person}{Nando~de Freitas}.}
  \bibinfo{year}{2022}\natexlab{}.
\newblock \showarticletitle{A {Generalist} {Agent}}.
\newblock \bibinfo{journal}{\emph{Transactions on Machine Learning Research}}
  (\bibinfo{year}{2022}).
\newblock
\urldef\tempurl%
\url{https://openreview.net/forum?id=1ikK0kHjvj}
\showURL{%
\tempurl}


\bibitem[Ribeiro et~al\mbox{.}(2020)]%
        {ribeiro_auditing_2020}
\bibfield{author}{\bibinfo{person}{Manoel~Horta Ribeiro},
  \bibinfo{person}{Raphael Ottoni}, \bibinfo{person}{Robert West},
  \bibinfo{person}{Virgílio A.~F. Almeida}, {and} \bibinfo{person}{Wagner
  Meira}.} \bibinfo{year}{2020}\natexlab{}.
\newblock \showarticletitle{Auditing radicalization pathways on {YouTube}}. In
  \bibinfo{booktitle}{\emph{Proceedings of the 2020 {Conference} on {Fairness},
  {Accountability}, and {Transparency}}}. \bibinfo{publisher}{ACM}.
\newblock
\urldef\tempurl%
\url{https://doi.org/10.1145/3351095.3372879}
\showDOI{\tempurl}


\bibitem[Robinette et~al\mbox{.}(2016)]%
        {robinette_overtrust_2016}
\bibfield{author}{\bibinfo{person}{Paul Robinette}, \bibinfo{person}{Wenchen
  Li}, \bibinfo{person}{Robert Allen}, \bibinfo{person}{Ayanna~M Howard}, {and}
  \bibinfo{person}{Alan~R Wagner}.} \bibinfo{year}{2016}\natexlab{}.
\newblock \showarticletitle{Overtrust of robots in emergency evacuation
  scenarios}. In \bibinfo{booktitle}{\emph{2016 11th {ACM}/{IEEE} international
  conference on human-robot interaction ({HRI})}}. \bibinfo{publisher}{IEEE},
  \bibinfo{pages}{101--108}.
\newblock


\bibitem[Russell and Norvig(2021)]%
        {russell_artificial_2021}
\bibfield{author}{\bibinfo{person}{Stuart~J. Russell} {and}
  \bibinfo{person}{Peter Norvig}.} \bibinfo{year}{2021}\natexlab{}.
\newblock \bibinfo{booktitle}{\emph{Artificial {Intelligence}: {A} {Modern}
  {Approach}} (\bibinfo{edition}{4} ed.)}.
\newblock


\bibitem[Schlosser(2019)]%
        {schlosser_agency_2019}
\bibfield{author}{\bibinfo{person}{Markus Schlosser}.}
  \bibinfo{year}{2019}\natexlab{}.
\newblock \showarticletitle{Agency}.
\newblock In \bibinfo{booktitle}{\emph{The {Stanford} {Encyclopedia} of
  {Philosophy}} (\bibinfo{edition}{winter 2019} ed.)},
  \bibfield{editor}{\bibinfo{person}{Edward~N. Zalta}} (Ed.).
  \bibinfo{publisher}{Metaphysics Research Lab, Stanford University}.
\newblock
\urldef\tempurl%
\url{https://plato.stanford.edu/archives/win2019/entries/agency/}
\showURL{%
\tempurl}


\bibitem[Schmelzer(2019)]%
        {schmelzer_amazon_2019}
\bibfield{author}{\bibinfo{person}{Ron Schmelzer}.}
  \bibinfo{year}{2019}\natexlab{}.
\newblock \showarticletitle{Amazon {Dives} {Deep} into {Reinforcement}
  {Learning}}.
\newblock \bibinfo{journal}{\emph{Forbes}} (\bibinfo{date}{June}
  \bibinfo{year}{2019}).
\newblock
\urldef\tempurl%
\url{https://www.forbes.com/sites/cognitiveworld/2019/06/14/amazon-dives-deep-into-reinforcement-learning/}
\showURL{%
\tempurl}


\bibitem[Schrittwieser et~al\mbox{.}(2020)]%
        {schrittwieser_mastering_2020}
\bibfield{author}{\bibinfo{person}{Julian Schrittwieser},
  \bibinfo{person}{Ioannis Antonoglou}, \bibinfo{person}{Thomas Hubert},
  \bibinfo{person}{Karen Simonyan}, \bibinfo{person}{Laurent Sifre},
  \bibinfo{person}{Simon Schmitt}, \bibinfo{person}{Arthur Guez},
  \bibinfo{person}{Edward Lockhart}, \bibinfo{person}{Demis Hassabis},
  \bibinfo{person}{Thore Graepel}, \bibinfo{person}{Timothy Lillicrap}, {and}
  \bibinfo{person}{David Silver}.} \bibinfo{year}{2020}\natexlab{}.
\newblock \showarticletitle{Mastering {Atari}, {Go}, chess and shogi by
  planning with a learned model}.
\newblock \bibinfo{journal}{\emph{Nature}} \bibinfo{volume}{588},
  \bibinfo{number}{7839} (\bibinfo{date}{Dec.} \bibinfo{year}{2020}),
  \bibinfo{pages}{604--609}.
\newblock
\showISSN{1476-4687}
\urldef\tempurl%
\url{https://doi.org/10.1038/s41586-020-03051-4}
\showDOI{\tempurl}
\newblock
\shownote{Number: 7839 Publisher: Nature Publishing Group}.


\bibitem[Schulz et~al\mbox{.}(2019)]%
        {schulz_industry_2019}
\bibfield{author}{\bibinfo{person}{Anne Schulz}, \bibinfo{person}{P Howard},
  {and} \bibinfo{person}{R Nielsen}.} \bibinfo{year}{2019}\natexlab{}.
\newblock \showarticletitle{Industry, {Experts}, or {Industry} {Experts}?
  {Academic} {Sourcing} in {News} {Coverage} of {AI}}.
\newblock  (\bibinfo{year}{2019}).
\newblock
\newblock
\shownote{Publisher: Reuters Institute for the Study of Journalism}.


\bibitem[Selbst et~al\mbox{.}(2019)]%
        {selbst_fairness_2019}
\bibfield{author}{\bibinfo{person}{Andrew~D. Selbst}, \bibinfo{person}{Danah
  Boyd}, \bibinfo{person}{Sorelle~A. Friedler}, \bibinfo{person}{Suresh
  Venkatasubramanian}, {and} \bibinfo{person}{Janet Vertesi}.}
  \bibinfo{year}{2019}\natexlab{}.
\newblock \showarticletitle{Fairness and {Abstraction} in {Sociotechnical}
  {Systems}}. In \bibinfo{booktitle}{\emph{Proceedings of the {Conference} on
  {Fairness}, {Accountability}, and {Transparency}}}. \bibinfo{publisher}{ACM}.
\newblock
\urldef\tempurl%
\url{https://doi.org/10.1145/3287560.3287598}
\showDOI{\tempurl}


\bibitem[Sendak et~al\mbox{.}(2020)]%
        {sendak_human_2020}
\bibfield{author}{\bibinfo{person}{Mark Sendak},
  \bibinfo{person}{Madeleine~Clare Elish}, \bibinfo{person}{Michael Gao},
  \bibinfo{person}{Joseph Futoma}, \bibinfo{person}{William Ratliff},
  \bibinfo{person}{Marshall Nichols}, \bibinfo{person}{Armando Bedoya},
  \bibinfo{person}{Suresh Balu}, {and} \bibinfo{person}{Cara O'Brien}.}
  \bibinfo{year}{2020}\natexlab{}.
\newblock \showarticletitle{"{The} human body is a black box"}. In
  \bibinfo{booktitle}{\emph{Proceedings of the 2020 {Conference} on {Fairness},
  {Accountability}, and {Transparency}}}. \bibinfo{publisher}{ACM}.
\newblock
\urldef\tempurl%
\url{https://doi.org/10.1145/3351095.3372827}
\showDOI{\tempurl}


\bibitem[Shah et~al\mbox{.}(2022)]%
        {shah_goal_2022}
\bibfield{author}{\bibinfo{person}{Rohin Shah}, \bibinfo{person}{Vikrant
  Varma}, \bibinfo{person}{Ramana Kumar}, \bibinfo{person}{Mary Phuong},
  \bibinfo{person}{Victoria Krakovna}, \bibinfo{person}{Jonathan Uesato}, {and}
  \bibinfo{person}{Zac Kenton}.} \bibinfo{year}{2022}\natexlab{}.
\newblock \bibinfo{title}{Goal {Misgeneralization}: {Why} {Correct}
  {Specifications} {Aren}'t {Enough} {For} {Correct} {Goals}}.
\newblock
\newblock
\urldef\tempurl%
\url{https://doi.org/10.48550/arXiv.2210.01790}
\showDOI{\tempurl}
\newblock
\shownote{arXiv:2210.01790 [cs]}.


\bibitem[Shavit(2023)]%
        {shavit_what_2023}
\bibfield{author}{\bibinfo{person}{Yonadav Shavit}.}
  \bibinfo{year}{2023}\natexlab{}.
\newblock \bibinfo{title}{What does it take to catch a {Chinchilla}?
  {Verifying} {Rules} on {Large}-{Scale} {Neural} {Network} {Training} via
  {Compute} {Monitoring}}.
\newblock
\newblock
\urldef\tempurl%
\url{https://doi.org/10.48550/arXiv.2303.11341}
\showDOI{\tempurl}
\newblock
\shownote{arXiv:2303.11341 [cs]}.


\bibitem[Shelby et~al\mbox{.}(2022)]%
        {shelby_sociotechnical_2022}
\bibfield{author}{\bibinfo{person}{Renee Shelby}, \bibinfo{person}{Shalaleh
  Rismani}, \bibinfo{person}{Kathryn Henne}, \bibinfo{person}{AJung Moon},
  \bibinfo{person}{Negar Rostamzadeh}, \bibinfo{person}{Paul Nicholas},
  \bibinfo{person}{N'Mah Yilla}, \bibinfo{person}{Jess Gallegos},
  \bibinfo{person}{Andrew Smart}, \bibinfo{person}{Emilio Garcia}, {and}
  \bibinfo{person}{Gurleen Virk}.} \bibinfo{year}{2022}\natexlab{}.
\newblock \bibinfo{title}{Sociotechnical {Harms}: {Scoping} a {Taxonomy} for
  {Harm} {Reduction}}.
\newblock
\newblock
\urldef\tempurl%
\url{https://doi.org/10.48550/arXiv.2210.05791}
\showDOI{\tempurl}
\newblock
\shownote{arXiv:2210.05791 [cs]}.


\bibitem[Silver et~al\mbox{.}(2016)]%
        {silver_mastering_2016}
\bibfield{author}{\bibinfo{person}{David Silver}, \bibinfo{person}{Aja Huang},
  \bibinfo{person}{Chris~J. Maddison}, \bibinfo{person}{Arthur Guez},
  \bibinfo{person}{Laurent Sifre}, \bibinfo{person}{George van~den Driessche},
  \bibinfo{person}{Julian Schrittwieser}, \bibinfo{person}{Ioannis Antonoglou},
  \bibinfo{person}{Veda Panneershelvam}, \bibinfo{person}{Marc Lanctot},
  \bibinfo{person}{Sander Dieleman}, \bibinfo{person}{Dominik Grewe},
  \bibinfo{person}{John Nham}, \bibinfo{person}{Nal Kalchbrenner},
  \bibinfo{person}{Ilya Sutskever}, \bibinfo{person}{Timothy Lillicrap},
  \bibinfo{person}{Madeleine Leach}, \bibinfo{person}{Koray Kavukcuoglu},
  \bibinfo{person}{Thore Graepel}, {and} \bibinfo{person}{Demis Hassabis}.}
  \bibinfo{year}{2016}\natexlab{}.
\newblock \showarticletitle{Mastering the game of {Go} with deep neural
  networks and tree search}.
\newblock \bibinfo{journal}{\emph{Nature}} \bibinfo{volume}{529},
  \bibinfo{number}{7587} (\bibinfo{date}{Jan.} \bibinfo{year}{2016}),
  \bibinfo{pages}{484--489}.
\newblock
\showISSN{1476-4687}
\urldef\tempurl%
\url{https://doi.org/10.1038/nature16961}
\showDOI{\tempurl}
\newblock
\shownote{Number: 7587 Publisher: Nature Publishing Group}.


\bibitem[Silver et~al\mbox{.}(2017)]%
        {silver_mastering_2017}
\bibfield{author}{\bibinfo{person}{David Silver}, \bibinfo{person}{Thomas
  Hubert}, \bibinfo{person}{Julian Schrittwieser}, \bibinfo{person}{Ioannis
  Antonoglou}, \bibinfo{person}{Matthew Lai}, \bibinfo{person}{Arthur Guez},
  \bibinfo{person}{Marc Lanctot}, \bibinfo{person}{Laurent Sifre},
  \bibinfo{person}{Dharshan Kumaran}, \bibinfo{person}{Thore Graepel},
  \bibinfo{person}{Timothy Lillicrap}, \bibinfo{person}{Karen Simonyan}, {and}
  \bibinfo{person}{Demis Hassabis}.} \bibinfo{year}{2017}\natexlab{}.
\newblock \bibinfo{title}{Mastering {Chess} and {Shogi} by {Self}-{Play} with a
  {General} {Reinforcement} {Learning} {Algorithm}}.
\newblock
\newblock
\urldef\tempurl%
\url{https://doi.org/10.48550/arXiv.1712.01815}
\showDOI{\tempurl}
\newblock
\shownote{arXiv:1712.01815 [cs]}.


\bibitem[Skalse et~al\mbox{.}(2022)]%
        {skalse_defining_2022}
\bibfield{author}{\bibinfo{person}{Joar Max~Viktor Skalse},
  \bibinfo{person}{Nikolaus H.~R. Howe}, \bibinfo{person}{Dmitrii
  Krasheninnikov}, {and} \bibinfo{person}{David Krueger}.}
  \bibinfo{year}{2022}\natexlab{}.
\newblock \showarticletitle{Defining and {Characterizing} {Reward} {Hacking}}.
  In \bibinfo{booktitle}{\emph{Advances in {Neural} {Information} {Processing}
  {Systems}}}, \bibfield{editor}{\bibinfo{person}{Alice~H. Oh},
  \bibinfo{person}{Alekh Agarwal}, \bibinfo{person}{Danielle Belgrave}, {and}
  \bibinfo{person}{Kyunghyun Cho}} (Eds.).
\newblock
\urldef\tempurl%
\url{https://arxiv.org/abs/2209.13085}
\showURL{%
\tempurl}


\bibitem[Spelke and Kinzler(2007)]%
        {spelke_core_2007}
\bibfield{author}{\bibinfo{person}{Elizabeth~S Spelke} {and}
  \bibinfo{person}{Katherine~D Kinzler}.} \bibinfo{year}{2007}\natexlab{}.
\newblock \showarticletitle{Core knowledge}.
\newblock \bibinfo{journal}{\emph{Developmental science}} \bibinfo{volume}{10},
  \bibinfo{number}{1} (\bibinfo{year}{2007}), \bibinfo{pages}{89--96}.
\newblock
\newblock
\shownote{Publisher: Wiley Online Library}.


\bibitem[Srivastava et~al\mbox{.}(2022)]%
        {srivastava_beyond_2022}
\bibfield{author}{\bibinfo{person}{Aarohi Srivastava}, \bibinfo{person}{Abhinav
  Rastogi}, \bibinfo{person}{Abhishek Rao}, \bibinfo{person}{Abu Awal~Md
  Shoeb}, \bibinfo{person}{Abubakar Abid}, \bibinfo{person}{Adam Fisch},
  \bibinfo{person}{Adam~R. Brown}, \bibinfo{person}{Adam Santoro},
  \bibinfo{person}{Aditya Gupta}, \bibinfo{person}{Adrià Garriga-Alonso},
  \bibinfo{person}{Agnieszka Kluska}, \bibinfo{person}{Aitor Lewkowycz},
  \bibinfo{person}{Akshat Agarwal}, \bibinfo{person}{Alethea Power},
  \bibinfo{person}{Alex Ray}, \bibinfo{person}{Alex Warstadt},
  \bibinfo{person}{Alexander~W. Kocurek}, \bibinfo{person}{Ali Safaya},
  \bibinfo{person}{Ali Tazarv}, \bibinfo{person}{Alice Xiang},
  \bibinfo{person}{Alicia Parrish}, \bibinfo{person}{Allen Nie},
  \bibinfo{person}{Aman Hussain}, \bibinfo{person}{Amanda Askell},
  \bibinfo{person}{Amanda Dsouza}, \bibinfo{person}{Ambrose Slone},
  \bibinfo{person}{Ameet Rahane}, \bibinfo{person}{Anantharaman~S. Iyer},
  \bibinfo{person}{Anders Andreassen}, \bibinfo{person}{Andrea Madotto},
  \bibinfo{person}{Andrea Santilli}, \bibinfo{person}{Andreas Stuhlmüller},
  \bibinfo{person}{Andrew Dai}, \bibinfo{person}{Andrew La},
  \bibinfo{person}{Andrew Lampinen}, \bibinfo{person}{Andy Zou},
  \bibinfo{person}{Angela Jiang}, \bibinfo{person}{Angelica Chen},
  \bibinfo{person}{Anh Vuong}, \bibinfo{person}{Animesh Gupta},
  \bibinfo{person}{Anna Gottardi}, \bibinfo{person}{Antonio Norelli},
  \bibinfo{person}{Anu Venkatesh}, \bibinfo{person}{Arash Gholamidavoodi},
  \bibinfo{person}{Arfa Tabassum}, \bibinfo{person}{Arul Menezes},
  \bibinfo{person}{Arun Kirubarajan}, \bibinfo{person}{Asher Mullokandov},
  \bibinfo{person}{Ashish Sabharwal}, \bibinfo{person}{Austin Herrick},
  \bibinfo{person}{Avia Efrat}, \bibinfo{person}{Aykut Erdem},
  \bibinfo{person}{Ayla Karakaş}, \bibinfo{person}{B.~Ryan Roberts},
  \bibinfo{person}{Bao~Sheng Loe}, \bibinfo{person}{Barret Zoph},
  \bibinfo{person}{Bartłomiej Bojanowski}, \bibinfo{person}{Batuhan Özyurt},
  \bibinfo{person}{Behnam Hedayatnia}, \bibinfo{person}{Behnam Neyshabur},
  \bibinfo{person}{Benjamin Inden}, \bibinfo{person}{Benno Stein},
  \bibinfo{person}{Berk Ekmekci}, \bibinfo{person}{Bill~Yuchen Lin},
  \bibinfo{person}{Blake Howald}, \bibinfo{person}{Cameron Diao},
  \bibinfo{person}{Cameron Dour}, \bibinfo{person}{Catherine Stinson},
  \bibinfo{person}{Cedrick Argueta}, \bibinfo{person}{César~Ferri Ramírez},
  \bibinfo{person}{Chandan Singh}, \bibinfo{person}{Charles Rathkopf},
  \bibinfo{person}{Chenlin Meng}, \bibinfo{person}{Chitta Baral},
  \bibinfo{person}{Chiyu Wu}, \bibinfo{person}{Chris Callison-Burch},
  \bibinfo{person}{Chris Waites}, \bibinfo{person}{Christian Voigt},
  \bibinfo{person}{Christopher~D. Manning}, \bibinfo{person}{Christopher
  Potts}, \bibinfo{person}{Cindy Ramirez}, \bibinfo{person}{Clara~E. Rivera},
  \bibinfo{person}{Clemencia Siro}, \bibinfo{person}{Colin Raffel},
  \bibinfo{person}{Courtney Ashcraft}, \bibinfo{person}{Cristina Garbacea},
  \bibinfo{person}{Damien Sileo}, \bibinfo{person}{Dan Garrette},
  \bibinfo{person}{Dan Hendrycks}, \bibinfo{person}{Dan Kilman},
  \bibinfo{person}{Dan Roth}, \bibinfo{person}{Daniel Freeman},
  \bibinfo{person}{Daniel Khashabi}, \bibinfo{person}{Daniel Levy},
  \bibinfo{person}{Daniel~Moseguí González}, \bibinfo{person}{Danielle
  Perszyk}, \bibinfo{person}{Danny Hernandez}, \bibinfo{person}{Danqi Chen},
  \bibinfo{person}{Daphne Ippolito}, \bibinfo{person}{Dar Gilboa},
  \bibinfo{person}{David Dohan}, \bibinfo{person}{David Drakard},
  \bibinfo{person}{David Jurgens}, \bibinfo{person}{Debajyoti Datta},
  \bibinfo{person}{Deep Ganguli}, \bibinfo{person}{Denis Emelin},
  \bibinfo{person}{Denis Kleyko}, \bibinfo{person}{Deniz Yuret},
  \bibinfo{person}{Derek Chen}, \bibinfo{person}{Derek Tam},
  \bibinfo{person}{Dieuwke Hupkes}, \bibinfo{person}{Diganta Misra},
  \bibinfo{person}{Dilyar Buzan}, \bibinfo{person}{Dimitri~Coelho Mollo},
  \bibinfo{person}{Diyi Yang}, \bibinfo{person}{Dong-Ho Lee},
  \bibinfo{person}{Ekaterina Shutova}, \bibinfo{person}{Ekin~Dogus Cubuk},
  \bibinfo{person}{Elad Segal}, \bibinfo{person}{Eleanor Hagerman},
  \bibinfo{person}{Elizabeth Barnes}, \bibinfo{person}{Elizabeth Donoway},
  \bibinfo{person}{Ellie Pavlick}, \bibinfo{person}{Emanuele Rodola},
  \bibinfo{person}{Emma Lam}, \bibinfo{person}{Eric Chu}, \bibinfo{person}{Eric
  Tang}, \bibinfo{person}{Erkut Erdem}, \bibinfo{person}{Ernie Chang},
  \bibinfo{person}{Ethan~A. Chi}, \bibinfo{person}{Ethan Dyer},
  \bibinfo{person}{Ethan Jerzak}, \bibinfo{person}{Ethan Kim},
  \bibinfo{person}{Eunice~Engefu Manyasi}, \bibinfo{person}{Evgenii
  Zheltonozhskii}, \bibinfo{person}{Fanyue Xia}, \bibinfo{person}{Fatemeh
  Siar}, \bibinfo{person}{Fernando Martínez-Plumed},
  \bibinfo{person}{Francesca Happé}, \bibinfo{person}{Francois Chollet},
  \bibinfo{person}{Frieda Rong}, \bibinfo{person}{Gaurav Mishra},
  \bibinfo{person}{Genta~Indra Winata}, \bibinfo{person}{Gerard de Melo},
  \bibinfo{person}{Germán Kruszewski}, \bibinfo{person}{Giambattista
  Parascandolo}, \bibinfo{person}{Giorgio Mariani}, \bibinfo{person}{Gloria
  Wang}, \bibinfo{person}{Gonzalo Jaimovitch-López}, \bibinfo{person}{Gregor
  Betz}, \bibinfo{person}{Guy Gur-Ari}, \bibinfo{person}{Hana Galijasevic},
  \bibinfo{person}{Hannah Kim}, \bibinfo{person}{Hannah Rashkin},
  \bibinfo{person}{Hannaneh Hajishirzi}, \bibinfo{person}{Harsh Mehta},
  \bibinfo{person}{Hayden Bogar}, \bibinfo{person}{Henry Shevlin},
  \bibinfo{person}{Hinrich Schütze}, \bibinfo{person}{Hiromu Yakura},
  \bibinfo{person}{Hongming Zhang}, \bibinfo{person}{Hugh~Mee Wong},
  \bibinfo{person}{Ian Ng}, \bibinfo{person}{Isaac Noble},
  \bibinfo{person}{Jaap Jumelet}, \bibinfo{person}{Jack Geissinger},
  \bibinfo{person}{Jackson Kernion}, \bibinfo{person}{Jacob Hilton},
  \bibinfo{person}{Jaehoon Lee}, \bibinfo{person}{Jaime~Fernández Fisac},
  \bibinfo{person}{James~B. Simon}, \bibinfo{person}{James Koppel},
  \bibinfo{person}{James Zheng}, \bibinfo{person}{James Zou},
  \bibinfo{person}{Jan Kocoń}, \bibinfo{person}{Jana Thompson},
  \bibinfo{person}{Jared Kaplan}, \bibinfo{person}{Jarema Radom},
  \bibinfo{person}{Jascha Sohl-Dickstein}, \bibinfo{person}{Jason Phang},
  \bibinfo{person}{Jason Wei}, \bibinfo{person}{Jason Yosinski},
  \bibinfo{person}{Jekaterina Novikova}, \bibinfo{person}{Jelle Bosscher},
  \bibinfo{person}{Jennifer Marsh}, \bibinfo{person}{Jeremy Kim},
  \bibinfo{person}{Jeroen Taal}, \bibinfo{person}{Jesse Engel},
  \bibinfo{person}{Jesujoba Alabi}, \bibinfo{person}{Jiacheng Xu},
  \bibinfo{person}{Jiaming Song}, \bibinfo{person}{Jillian Tang},
  \bibinfo{person}{Joan Waweru}, \bibinfo{person}{John Burden},
  \bibinfo{person}{John Miller}, \bibinfo{person}{John~U. Balis},
  \bibinfo{person}{Jonathan Berant}, \bibinfo{person}{Jörg Frohberg},
  \bibinfo{person}{Jos Rozen}, \bibinfo{person}{Jose Hernandez-Orallo},
  \bibinfo{person}{Joseph Boudeman}, \bibinfo{person}{Joseph Jones},
  \bibinfo{person}{Joshua~B. Tenenbaum}, \bibinfo{person}{Joshua~S. Rule},
  \bibinfo{person}{Joyce Chua}, \bibinfo{person}{Kamil Kanclerz},
  \bibinfo{person}{Karen Livescu}, \bibinfo{person}{Karl Krauth},
  \bibinfo{person}{Karthik Gopalakrishnan}, \bibinfo{person}{Katerina
  Ignatyeva}, \bibinfo{person}{Katja Markert}, \bibinfo{person}{Kaustubh~D.
  Dhole}, \bibinfo{person}{Kevin Gimpel}, \bibinfo{person}{Kevin Omondi},
  \bibinfo{person}{Kory Mathewson}, \bibinfo{person}{Kristen Chiafullo},
  \bibinfo{person}{Ksenia Shkaruta}, \bibinfo{person}{Kumar Shridhar},
  \bibinfo{person}{Kyle McDonell}, \bibinfo{person}{Kyle Richardson},
  \bibinfo{person}{Laria Reynolds}, \bibinfo{person}{Leo Gao},
  \bibinfo{person}{Li Zhang}, \bibinfo{person}{Liam Dugan},
  \bibinfo{person}{Lianhui Qin}, \bibinfo{person}{Lidia Contreras-Ochando},
  \bibinfo{person}{Louis-Philippe Morency}, \bibinfo{person}{Luca Moschella},
  \bibinfo{person}{Lucas Lam}, \bibinfo{person}{Lucy Noble},
  \bibinfo{person}{Ludwig Schmidt}, \bibinfo{person}{Luheng He},
  \bibinfo{person}{Luis~Oliveros Colón}, \bibinfo{person}{Luke Metz},
  \bibinfo{person}{Lütfi~Kerem Şenel}, \bibinfo{person}{Maarten Bosma},
  \bibinfo{person}{Maarten Sap}, \bibinfo{person}{Maartje ter Hoeve},
  \bibinfo{person}{Maheen Farooqi}, \bibinfo{person}{Manaal Faruqui},
  \bibinfo{person}{Mantas Mazeika}, \bibinfo{person}{Marco Baturan},
  \bibinfo{person}{Marco Marelli}, \bibinfo{person}{Marco Maru},
  \bibinfo{person}{Maria Jose~Ramírez Quintana}, \bibinfo{person}{Marie
  Tolkiehn}, \bibinfo{person}{Mario Giulianelli}, \bibinfo{person}{Martha
  Lewis}, \bibinfo{person}{Martin Potthast}, \bibinfo{person}{Matthew~L.
  Leavitt}, \bibinfo{person}{Matthias Hagen}, \bibinfo{person}{Mátyás
  Schubert}, \bibinfo{person}{Medina~Orduna Baitemirova},
  \bibinfo{person}{Melody Arnaud}, \bibinfo{person}{Melvin McElrath},
  \bibinfo{person}{Michael~A. Yee}, \bibinfo{person}{Michael Cohen},
  \bibinfo{person}{Michael Gu}, \bibinfo{person}{Michael Ivanitskiy},
  \bibinfo{person}{Michael Starritt}, \bibinfo{person}{Michael Strube},
  \bibinfo{person}{Michał Swędrowski}, \bibinfo{person}{Michele Bevilacqua},
  \bibinfo{person}{Michihiro Yasunaga}, \bibinfo{person}{Mihir Kale},
  \bibinfo{person}{Mike Cain}, \bibinfo{person}{Mimee Xu},
  \bibinfo{person}{Mirac Suzgun}, \bibinfo{person}{Mo Tiwari},
  \bibinfo{person}{Mohit Bansal}, \bibinfo{person}{Moin Aminnaseri},
  \bibinfo{person}{Mor Geva}, \bibinfo{person}{Mozhdeh Gheini},
  \bibinfo{person}{Mukund~Varma T}, \bibinfo{person}{Nanyun Peng},
  \bibinfo{person}{Nathan Chi}, \bibinfo{person}{Nayeon Lee},
  \bibinfo{person}{Neta Gur-Ari Krakover}, \bibinfo{person}{Nicholas Cameron},
  \bibinfo{person}{Nicholas Roberts}, \bibinfo{person}{Nick Doiron},
  \bibinfo{person}{Nikita Nangia}, \bibinfo{person}{Niklas Deckers},
  \bibinfo{person}{Niklas Muennighoff}, \bibinfo{person}{Nitish~Shirish
  Keskar}, \bibinfo{person}{Niveditha~S. Iyer}, \bibinfo{person}{Noah
  Constant}, \bibinfo{person}{Noah Fiedel}, \bibinfo{person}{Nuan Wen},
  \bibinfo{person}{Oliver Zhang}, \bibinfo{person}{Omar Agha},
  \bibinfo{person}{Omar Elbaghdadi}, \bibinfo{person}{Omer Levy},
  \bibinfo{person}{Owain Evans}, \bibinfo{person}{Pablo Antonio~Moreno
  Casares}, \bibinfo{person}{Parth Doshi}, \bibinfo{person}{Pascale Fung},
  \bibinfo{person}{Paul~Pu Liang}, \bibinfo{person}{Paul Vicol},
  \bibinfo{person}{Pegah Alipoormolabashi}, \bibinfo{person}{Peiyuan Liao},
  \bibinfo{person}{Percy Liang}, \bibinfo{person}{Peter Chang},
  \bibinfo{person}{Peter Eckersley}, \bibinfo{person}{Phu~Mon Htut},
  \bibinfo{person}{Pinyu Hwang}, \bibinfo{person}{Piotr Miłkowski},
  \bibinfo{person}{Piyush Patil}, \bibinfo{person}{Pouya Pezeshkpour},
  \bibinfo{person}{Priti Oli}, \bibinfo{person}{Qiaozhu Mei},
  \bibinfo{person}{Qing Lyu}, \bibinfo{person}{Qinlang Chen},
  \bibinfo{person}{Rabin Banjade}, \bibinfo{person}{Rachel~Etta Rudolph},
  \bibinfo{person}{Raefer Gabriel}, \bibinfo{person}{Rahel Habacker},
  \bibinfo{person}{Ramón~Risco Delgado}, \bibinfo{person}{Raphaël Millière},
  \bibinfo{person}{Rhythm Garg}, \bibinfo{person}{Richard Barnes},
  \bibinfo{person}{Rif~A. Saurous}, \bibinfo{person}{Riku Arakawa},
  \bibinfo{person}{Robbe Raymaekers}, \bibinfo{person}{Robert Frank},
  \bibinfo{person}{Rohan Sikand}, \bibinfo{person}{Roman Novak},
  \bibinfo{person}{Roman Sitelew}, \bibinfo{person}{Ronan LeBras},
  \bibinfo{person}{Rosanne Liu}, \bibinfo{person}{Rowan Jacobs},
  \bibinfo{person}{Rui Zhang}, \bibinfo{person}{Ruslan Salakhutdinov},
  \bibinfo{person}{Ryan Chi}, \bibinfo{person}{Ryan Lee}, \bibinfo{person}{Ryan
  Stovall}, \bibinfo{person}{Ryan Teehan}, \bibinfo{person}{Rylan Yang},
  \bibinfo{person}{Sahib Singh}, \bibinfo{person}{Saif~M. Mohammad},
  \bibinfo{person}{Sajant Anand}, \bibinfo{person}{Sam Dillavou},
  \bibinfo{person}{Sam Shleifer}, \bibinfo{person}{Sam Wiseman},
  \bibinfo{person}{Samuel Gruetter}, \bibinfo{person}{Samuel~R. Bowman},
  \bibinfo{person}{Samuel~S. Schoenholz}, \bibinfo{person}{Sanghyun Han},
  \bibinfo{person}{Sanjeev Kwatra}, \bibinfo{person}{Sarah~A. Rous},
  \bibinfo{person}{Sarik Ghazarian}, \bibinfo{person}{Sayan Ghosh},
  \bibinfo{person}{Sean Casey}, \bibinfo{person}{Sebastian Bischoff},
  \bibinfo{person}{Sebastian Gehrmann}, \bibinfo{person}{Sebastian Schuster},
  \bibinfo{person}{Sepideh Sadeghi}, \bibinfo{person}{Shadi Hamdan},
  \bibinfo{person}{Sharon Zhou}, \bibinfo{person}{Shashank Srivastava},
  \bibinfo{person}{Sherry Shi}, \bibinfo{person}{Shikhar Singh},
  \bibinfo{person}{Shima Asaadi}, \bibinfo{person}{Shixiang~Shane Gu},
  \bibinfo{person}{Shubh Pachchigar}, \bibinfo{person}{Shubham Toshniwal},
  \bibinfo{person}{Shyam Upadhyay}, \bibinfo{person}{Shyamolima},
  \bibinfo{person}{Debnath}, \bibinfo{person}{Siamak Shakeri},
  \bibinfo{person}{Simon Thormeyer}, \bibinfo{person}{Simone Melzi},
  \bibinfo{person}{Siva Reddy}, \bibinfo{person}{Sneha~Priscilla Makini},
  \bibinfo{person}{Soo-Hwan Lee}, \bibinfo{person}{Spencer Torene},
  \bibinfo{person}{Sriharsha Hatwar}, \bibinfo{person}{Stanislas Dehaene},
  \bibinfo{person}{Stefan Divic}, \bibinfo{person}{Stefano Ermon},
  \bibinfo{person}{Stella Biderman}, \bibinfo{person}{Stephanie Lin},
  \bibinfo{person}{Stephen Prasad}, \bibinfo{person}{Steven~T. Piantadosi},
  \bibinfo{person}{Stuart~M. Shieber}, \bibinfo{person}{Summer Misherghi},
  \bibinfo{person}{Svetlana Kiritchenko}, \bibinfo{person}{Swaroop Mishra},
  \bibinfo{person}{Tal Linzen}, \bibinfo{person}{Tal Schuster},
  \bibinfo{person}{Tao Li}, \bibinfo{person}{Tao Yu}, \bibinfo{person}{Tariq
  Ali}, \bibinfo{person}{Tatsu Hashimoto}, \bibinfo{person}{Te-Lin Wu},
  \bibinfo{person}{Théo Desbordes}, \bibinfo{person}{Theodore Rothschild},
  \bibinfo{person}{Thomas Phan}, \bibinfo{person}{Tianle Wang},
  \bibinfo{person}{Tiberius Nkinyili}, \bibinfo{person}{Timo Schick},
  \bibinfo{person}{Timofei Kornev}, \bibinfo{person}{Timothy Telleen-Lawton},
  \bibinfo{person}{Titus Tunduny}, \bibinfo{person}{Tobias Gerstenberg},
  \bibinfo{person}{Trenton Chang}, \bibinfo{person}{Trishala Neeraj},
  \bibinfo{person}{Tushar Khot}, \bibinfo{person}{Tyler Shultz},
  \bibinfo{person}{Uri Shaham}, \bibinfo{person}{Vedant Misra},
  \bibinfo{person}{Vera Demberg}, \bibinfo{person}{Victoria Nyamai},
  \bibinfo{person}{Vikas Raunak}, \bibinfo{person}{Vinay Ramasesh},
  \bibinfo{person}{Vinay~Uday Prabhu}, \bibinfo{person}{Vishakh Padmakumar},
  \bibinfo{person}{Vivek Srikumar}, \bibinfo{person}{William Fedus},
  \bibinfo{person}{William Saunders}, \bibinfo{person}{William Zhang},
  \bibinfo{person}{Wout Vossen}, \bibinfo{person}{Xiang Ren},
  \bibinfo{person}{Xiaoyu Tong}, \bibinfo{person}{Xinran Zhao},
  \bibinfo{person}{Xinyi Wu}, \bibinfo{person}{Xudong Shen},
  \bibinfo{person}{Yadollah Yaghoobzadeh}, \bibinfo{person}{Yair Lakretz},
  \bibinfo{person}{Yangqiu Song}, \bibinfo{person}{Yasaman Bahri},
  \bibinfo{person}{Yejin Choi}, \bibinfo{person}{Yichi Yang},
  \bibinfo{person}{Yiding Hao}, \bibinfo{person}{Yifu Chen},
  \bibinfo{person}{Yonatan Belinkov}, \bibinfo{person}{Yu Hou},
  \bibinfo{person}{Yufang Hou}, \bibinfo{person}{Yuntao Bai},
  \bibinfo{person}{Zachary Seid}, \bibinfo{person}{Zhuoye Zhao},
  \bibinfo{person}{Zijian Wang}, \bibinfo{person}{Zijie~J. Wang},
  \bibinfo{person}{Zirui Wang}, {and} \bibinfo{person}{Ziyi Wu}.}
  \bibinfo{year}{2022}\natexlab{}.
\newblock \bibinfo{title}{Beyond the {Imitation} {Game}: {Quantifying} and
  extrapolating the capabilities of language models}.
\newblock
\newblock
\urldef\tempurl%
\url{https://doi.org/10.48550/arXiv.2206.04615}
\showDOI{\tempurl}
\newblock
\shownote{arXiv:2206.04615 [cs, stat]}.


\bibitem[Stapleton et~al\mbox{.}(2022)]%
        {stapleton_imagining_2022}
\bibfield{author}{\bibinfo{person}{Logan Stapleton}, \bibinfo{person}{Min~Hun
  Lee}, \bibinfo{person}{Diana Qing}, \bibinfo{person}{Marya Wright},
  \bibinfo{person}{Alexandra Chouldechova}, \bibinfo{person}{Ken Holstein},
  \bibinfo{person}{Zhiwei~Steven Wu}, {and} \bibinfo{person}{Haiyi Zhu}.}
  \bibinfo{year}{2022}\natexlab{}.
\newblock \showarticletitle{Imagining new futures beyond predictive systems in
  child welfare: {A} qualitative study with impacted stakeholders}. In
  \bibinfo{booktitle}{\emph{2022 {ACM} {Conference} on {Fairness},
  {Accountability}, and {Transparency}}}. \bibinfo{publisher}{ACM}.
\newblock
\urldef\tempurl%
\url{https://doi.org/10.1145/3531146.3533177}
\showDOI{\tempurl}


\bibitem[{steven t. piantadosi [@spiantado]}(2022)]%
        {steven_t_piantadosi_spiantado_yes_2022}
\bibfield{author}{\bibinfo{person}{{steven t. piantadosi [@spiantado]}}.}
  \bibinfo{year}{2022}\natexlab{}.
\newblock \bibinfo{title}{Yes, {ChatGPT} is amazing and impressive. {No},
  @{OpenAI} has not come close to addressing the problem of bias. {Filters}
  appear to be bypassed with simple tricks, and superficially masked. {And}
  what is lurking inside is egregious. @{Abebab} @sama tw racism, sexism.
  https://t.co/{V4fw1fY9dY}}.
\newblock
\newblock
\urldef\tempurl%
\url{https://twitter.com/spiantado/status/1599462375887114240}
\showURL{%
\tempurl}


\bibitem[Stokes et~al\mbox{.}(2020)]%
        {stokes_deep_2020}
\bibfield{author}{\bibinfo{person}{Jonathan~M. Stokes}, \bibinfo{person}{Kevin
  Yang}, \bibinfo{person}{Kyle Swanson}, \bibinfo{person}{Wengong Jin},
  \bibinfo{person}{Andres Cubillos-Ruiz}, \bibinfo{person}{Nina~M. Donghia},
  \bibinfo{person}{Craig~R. MacNair}, \bibinfo{person}{Shawn French},
  \bibinfo{person}{Lindsey~A. Carfrae}, \bibinfo{person}{Zohar
  Bloom-Ackermann}, \bibinfo{person}{Victoria~M. Tran}, \bibinfo{person}{Anush
  Chiappino-Pepe}, \bibinfo{person}{Ahmed~H. Badran}, \bibinfo{person}{Ian~W.
  Andrews}, \bibinfo{person}{Emma~J. Chory}, \bibinfo{person}{George~M.
  Church}, \bibinfo{person}{Eric~D. Brown}, \bibinfo{person}{Tommi~S.
  Jaakkola}, \bibinfo{person}{Regina Barzilay}, {and} \bibinfo{person}{James~J.
  Collins}.} \bibinfo{year}{2020}\natexlab{}.
\newblock \showarticletitle{A {Deep} {Learning} {Approach} to {Antibiotic}
  {Discovery}}.
\newblock \bibinfo{journal}{\emph{Cell}} \bibinfo{volume}{180},
  \bibinfo{number}{4} (\bibinfo{date}{Feb.} \bibinfo{year}{2020}),
  \bibinfo{pages}{688--702.e13}.
\newblock
\showISSN{0092-8674}
\urldef\tempurl%
\url{https://doi.org/10.1016/j.cell.2020.01.021}
\showDOI{\tempurl}


\bibitem[Sullivan and Fosso~Wamba(2022)]%
        {sullivan_moral_2022}
\bibfield{author}{\bibinfo{person}{Yulia~W. Sullivan} {and}
  \bibinfo{person}{Samuel Fosso~Wamba}.} \bibinfo{year}{2022}\natexlab{}.
\newblock \showarticletitle{Moral {Judgments} in the {Age} of {Artificial}
  {Intelligence}}.
\newblock \bibinfo{journal}{\emph{Journal of Business Ethics}}
  \bibinfo{volume}{178}, \bibinfo{number}{4} (\bibinfo{date}{July}
  \bibinfo{year}{2022}), \bibinfo{pages}{917--943}.
\newblock
\showISSN{1573-0697}
\urldef\tempurl%
\url{https://doi.org/10.1007/s10551-022-05053-w}
\showDOI{\tempurl}


\bibitem[Sutton(2022)]%
        {sutton_notitle_2022}
\bibfield{author}{\bibinfo{person}{Richard Sutton}.}
  \bibinfo{year}{2022}\natexlab{}.
\newblock
\newblock
\urldef\tempurl%
\url{https://twitter.com/richardssutton/status/1575619651563708418}
\showURL{%
\tempurl}


\bibitem[Sutton and Barto(2018)]%
        {sutton_reinforcement_2018}
\bibfield{author}{\bibinfo{person}{Richard~S Sutton} {and}
  \bibinfo{person}{Andrew~G Barto}.} \bibinfo{year}{2018}\natexlab{}.
\newblock \bibinfo{booktitle}{\emph{Reinforcement learning: {An}
  introduction}}.
\newblock \bibinfo{publisher}{MIT press}.
\newblock


\bibitem[Sutton et~al\mbox{.}(2022)]%
        {sutton_alberta_2022}
\bibfield{author}{\bibinfo{person}{Richard~S. Sutton}, \bibinfo{person}{Michael
  Bowling}, {and} \bibinfo{person}{Patrick~M. Pilarski}.}
  \bibinfo{year}{2022}\natexlab{}.
\newblock \bibinfo{title}{The {Alberta} {Plan} for {AI} {Research}}.
\newblock
\newblock
\urldef\tempurl%
\url{https://doi.org/10.48550/arXiv.2208.11173}
\showDOI{\tempurl}
\newblock
\shownote{arXiv:2208.11173 [cs]}.


\bibitem[Sühr et~al\mbox{.}(2021)]%
        {suhr_does_2021}
\bibfield{author}{\bibinfo{person}{Tom Sühr}, \bibinfo{person}{Sophie
  Hilgard}, {and} \bibinfo{person}{Himabindu Lakkaraju}.}
  \bibinfo{year}{2021}\natexlab{}.
\newblock \showarticletitle{Does {Fair} {Ranking} {Improve} {Minority}
  {Outcomes}? {Understanding} the {Interplay} of {Human} and {Algorithmic}
  {Biases} in {Online} {Hiring}}. In \bibinfo{booktitle}{\emph{Proceedings of
  the 2021 {AAAI}/{ACM} {Conference} on {AI}, {Ethics}, and {Society}}}.
  \bibinfo{publisher}{ACM}.
\newblock
\urldef\tempurl%
\url{https://doi.org/10.1145/3461702.3462602}
\showDOI{\tempurl}


\bibitem[Team et~al\mbox{.}(2023)]%
        {adaptive_agent_team_human-timescale_2023}
\bibfield{author}{\bibinfo{person}{Adaptive~Agent Team}, \bibinfo{person}{Jakob
  Bauer}, \bibinfo{person}{Kate Baumli}, \bibinfo{person}{Satinder Baveja},
  \bibinfo{person}{Feryal Behbahani}, \bibinfo{person}{Avishkar Bhoopchand},
  \bibinfo{person}{Nathalie Bradley-Schmieg}, \bibinfo{person}{Michael Chang},
  \bibinfo{person}{Natalie Clay}, \bibinfo{person}{Adrian Collister},
  \bibinfo{person}{Vibhavari Dasagi}, \bibinfo{person}{Lucy Gonzalez},
  \bibinfo{person}{Karol Gregor}, \bibinfo{person}{Edward Hughes},
  \bibinfo{person}{Sheleem Kashem}, \bibinfo{person}{Maria Loks-Thompson},
  \bibinfo{person}{Hannah Openshaw}, \bibinfo{person}{Jack Parker-Holder},
  \bibinfo{person}{Shreya Pathak}, \bibinfo{person}{Nicolas Perez-Nieves},
  \bibinfo{person}{Nemanja Rakicevic}, \bibinfo{person}{Tim Rocktäschel},
  \bibinfo{person}{Yannick Schroecker}, \bibinfo{person}{Jakub Sygnowski},
  \bibinfo{person}{Karl Tuyls}, \bibinfo{person}{Sarah York},
  \bibinfo{person}{Alexander Zacherl}, {and} \bibinfo{person}{Lei Zhang}.}
  \bibinfo{year}{2023}\natexlab{}.
\newblock \bibinfo{title}{Human-{Timescale} {Adaptation} in an {Open}-{Ended}
  {Task} {Space}}.
\newblock
\newblock
\urldef\tempurl%
\url{https://doi.org/10.48550/arXiv.2301.07608}
\showDOI{\tempurl}
\newblock
\shownote{arXiv:2301.07608 [cs]}.


\bibitem[Tetlock and Gardner(2016)]%
        {tetlock_superforecasting_2016}
\bibfield{author}{\bibinfo{person}{Philip~E Tetlock} {and} \bibinfo{person}{Dan
  Gardner}.} \bibinfo{year}{2016}\natexlab{}.
\newblock \bibinfo{booktitle}{\emph{Superforecasting: {The} {Art} and {Science}
  of {Prediction}}}.
\newblock \bibinfo{publisher}{Random House}.
\newblock


\bibitem[Trager(2022)]%
        {trager_deliberating_2022}
\bibfield{author}{\bibinfo{person}{Robert Trager}.}
  \bibinfo{year}{2022}\natexlab{}.
\newblock \showarticletitle{Deliberating {Autonomous} {Weapons}}.
\newblock \bibinfo{journal}{\emph{Issues in Science and Technology}}
  \bibinfo{volume}{XXXVIII}, \bibinfo{number}{4} (\bibinfo{year}{2022}).
\newblock
\urldef\tempurl%
\url{https://issues.org/autonomous-weapons-russell-forum/}
\showURL{%
\tempurl}


\bibitem[Tutt(2017)]%
        {tutt_fda_2017}
\bibfield{author}{\bibinfo{person}{Andrew Tutt}.}
  \bibinfo{year}{2017}\natexlab{}.
\newblock \showarticletitle{An {FDA} for {Algorithms}}.
\newblock \bibinfo{journal}{\emph{Administrative Law Review}}
  \bibinfo{volume}{69}, \bibinfo{number}{1} (\bibinfo{year}{2017}),
  \bibinfo{pages}{83--124}.
\newblock
\urldef\tempurl%
\url{https://heinonline.org/HOL/P?h=hein.journals/admin69&i=95}
\showURL{%
\tempurl}


\bibitem[Valmeekam et~al\mbox{.}(2022)]%
        {valmeekam_large_2022}
\bibfield{author}{\bibinfo{person}{Karthik Valmeekam}, \bibinfo{person}{Alberto
  Olmo}, \bibinfo{person}{Sarath Sreedharan}, {and} \bibinfo{person}{Subbarao
  Kambhampati}.} \bibinfo{year}{2022}\natexlab{}.
\newblock \bibinfo{title}{Large {Language} {Models} {Still} {Can}'t {Plan} ({A}
  {Benchmark} for {LLMs} on {Planning} and {Reasoning} about {Change})}.
\newblock
\newblock
\urldef\tempurl%
\url{https://doi.org/10.48550/arXiv.2206.10498}
\showDOI{\tempurl}
\newblock
\shownote{arXiv:2206.10498 [cs]}.


\bibitem[van~der Loeff et~al\mbox{.}(2019)]%
        {van_der_loeff_ai_2019}
\bibfield{author}{\bibinfo{person}{Agnes~Schim van~der Loeff},
  \bibinfo{person}{Iggy Bassi}, \bibinfo{person}{Sachin Kapila}, {and}
  \bibinfo{person}{Jevgenij Gamper}.} \bibinfo{year}{2019}\natexlab{}.
\newblock \bibinfo{title}{{AI} {Ethics} for {Systemic} {Issues}: {A}
  {Structural} {Approach}}.
\newblock
\newblock
\urldef\tempurl%
\url{https://doi.org/10.48550/arXiv.1911.03216}
\showDOI{\tempurl}
\newblock
\shownote{arXiv:1911.03216 [cs]}.


\bibitem[Vinsel(2021)]%
        {vinsel_youre_2021}
\bibfield{author}{\bibinfo{person}{Lee Vinsel}.}
  \bibinfo{year}{2021}\natexlab{}.
\newblock \bibinfo{title}{You’re {Doing} {It} {Wrong}: {Notes} on {Criticism}
  and {Technology} {Hype}}.
\newblock
\newblock
\urldef\tempurl%
\url{https://sts-news.medium.com/youre-doing-it-wrong-notes-on-criticism-and-technology-hype-18b08b4307e5}
\showURL{%
\tempurl}


\bibitem[Vogell et~al\mbox{.}(2022)]%
        {vogell_how_2022}
\bibfield{author}{\bibinfo{person}{Heather Vogell}, \bibinfo{person}{Haru
  Coryne}, {and} \bibinfo{person}{Ryan Little}.}
  \bibinfo{year}{2022}\natexlab{}.
\newblock \bibinfo{title}{How a secret rent algorithm pushes rents higher}.
\newblock
\newblock
\urldef\tempurl%
\url{https://www.propublica.org/article/yieldstar-rent-increase-realpage-rent}
\showURL{%
\tempurl}
\newblock
\shownote{Publication Title: ProPublica}.


\bibitem[Volkery and Ribeiro(2009)]%
        {volkery_scenario_2009}
\bibfield{author}{\bibinfo{person}{Axel Volkery} {and} \bibinfo{person}{Teresa
  Ribeiro}.} \bibinfo{year}{2009}\natexlab{}.
\newblock \showarticletitle{Scenario planning in public policy: {Understanding}
  use, impacts and the role of institutional context factors}.
\newblock \bibinfo{journal}{\emph{Technological forecasting and social change}}
  \bibinfo{volume}{76}, \bibinfo{number}{9} (\bibinfo{year}{2009}),
  \bibinfo{pages}{1198--1207}.
\newblock
\newblock
\shownote{Publisher: Elsevier}.


\bibitem[Wallace et~al\mbox{.}(2019)]%
        {wallace_universal_2019}
\bibfield{author}{\bibinfo{person}{Eric Wallace}, \bibinfo{person}{Shi Feng},
  \bibinfo{person}{Nikhil Kandpal}, \bibinfo{person}{Matt Gardner}, {and}
  \bibinfo{person}{Sameer Singh}.} \bibinfo{year}{2019}\natexlab{}.
\newblock \showarticletitle{Universal {Adversarial} {Triggers} for {Attacking}
  and {Analyzing} {NLP}}. In \bibinfo{booktitle}{\emph{Proceedings of the 2019
  {Conference} on {Empirical} {Methods} in {Natural} {Language} {Processing}
  and the 9th {International} {Joint} {Conference} on {Natural} {Language}
  {Processing} ({EMNLP}-{IJCNLP})}}. \bibinfo{publisher}{Association for
  Computational Linguistics}, \bibinfo{address}{Hong Kong, China},
  \bibinfo{pages}{2153--2162}.
\newblock
\urldef\tempurl%
\url{https://doi.org/10.18653/v1/D19-1221}
\showDOI{\tempurl}


\bibitem[Wei et~al\mbox{.}(2022)]%
        {wei_emergent_2022}
\bibfield{author}{\bibinfo{person}{Jason Wei}, \bibinfo{person}{Yi Tay},
  \bibinfo{person}{Rishi Bommasani}, \bibinfo{person}{Colin Raffel},
  \bibinfo{person}{Barret Zoph}, \bibinfo{person}{Sebastian Borgeaud},
  \bibinfo{person}{Dani Yogatama}, \bibinfo{person}{Maarten Bosma},
  \bibinfo{person}{Denny Zhou}, \bibinfo{person}{Donald Metzler},
  \bibinfo{person}{Ed~H. Chi}, \bibinfo{person}{Tatsunori Hashimoto},
  \bibinfo{person}{Oriol Vinyals}, \bibinfo{person}{Percy Liang},
  \bibinfo{person}{Jeff Dean}, {and} \bibinfo{person}{William Fedus}.}
  \bibinfo{year}{2022}\natexlab{}.
\newblock \showarticletitle{Emergent {Abilities} of {Large} {Language}
  {Models}}.
\newblock \bibinfo{journal}{\emph{Transactions on Machine Learning Research}}
  (\bibinfo{year}{2022}).
\newblock
\urldef\tempurl%
\url{https://openreview.net/forum?id=yzkSU5zdwD}
\showURL{%
\tempurl}


\bibitem[Wei et~al\mbox{.}(2023)]%
        {wei_chain--thought_2023}
\bibfield{author}{\bibinfo{person}{Jason Wei}, \bibinfo{person}{Xuezhi Wang},
  \bibinfo{person}{Dale Schuurmans}, \bibinfo{person}{Maarten Bosma},
  \bibinfo{person}{Brian Ichter}, \bibinfo{person}{Fei Xia},
  \bibinfo{person}{Ed Chi}, \bibinfo{person}{Quoc Le}, {and}
  \bibinfo{person}{Denny Zhou}.} \bibinfo{year}{2023}\natexlab{}.
\newblock \bibinfo{title}{Chain-of-{Thought} {Prompting} {Elicits} {Reasoning}
  in {Large} {Language} {Models}}.
\newblock
\newblock
\urldef\tempurl%
\url{https://doi.org/10.48550/arXiv.2201.11903}
\showDOI{\tempurl}
\newblock
\shownote{arXiv:2201.11903 [cs]}.


\bibitem[Weidinger et~al\mbox{.}(2022)]%
        {weidinger_taxonomy_2022}
\bibfield{author}{\bibinfo{person}{Laura Weidinger}, \bibinfo{person}{Jonathan
  Uesato}, \bibinfo{person}{Maribeth Rauh}, \bibinfo{person}{Conor Griffin},
  \bibinfo{person}{Po-Sen Huang}, \bibinfo{person}{John Mellor},
  \bibinfo{person}{Amelia Glaese}, \bibinfo{person}{Myra Cheng},
  \bibinfo{person}{Borja Balle}, \bibinfo{person}{Atoosa Kasirzadeh},
  \bibinfo{person}{Courtney Biles}, \bibinfo{person}{Sasha Brown},
  \bibinfo{person}{Zac Kenton}, \bibinfo{person}{Will Hawkins},
  \bibinfo{person}{Tom Stepleton}, \bibinfo{person}{Abeba Birhane},
  \bibinfo{person}{Lisa~Anne Hendricks}, \bibinfo{person}{Laura Rimell},
  \bibinfo{person}{William Isaac}, \bibinfo{person}{Julia Haas},
  \bibinfo{person}{Sean Legassick}, \bibinfo{person}{Geoffrey Irving}, {and}
  \bibinfo{person}{Iason Gabriel}.} \bibinfo{year}{2022}\natexlab{}.
\newblock \showarticletitle{Taxonomy of {Risks} posed by {Language} {Models}}.
  In \bibinfo{booktitle}{\emph{2022 {ACM} {Conference} on {Fairness},
  {Accountability}, and {Transparency}}}. \bibinfo{publisher}{ACM}.
\newblock
\urldef\tempurl%
\url{https://doi.org/10.1145/3531146.3533088}
\showDOI{\tempurl}


\bibitem[Welsh(2019)]%
        {welsh_regulating_2019}
\bibfield{author}{\bibinfo{person}{Sean Welsh}.}
  \bibinfo{year}{2019}\natexlab{}.
\newblock \showarticletitle{Regulating {Lethal} and {Harmful} {Autonomy}}. In
  \bibinfo{booktitle}{\emph{Proceedings of the 2019 {AAAI}/{ACM} {Conference}
  on {AI}, {Ethics}, and {Society}}}. \bibinfo{publisher}{ACM}.
\newblock
\urldef\tempurl%
\url{https://doi.org/10.1145/3306618.3314295}
\showDOI{\tempurl}


\bibitem[Whittaker et~al\mbox{.}(2021)]%
        {whittaker_recommender_2021}
\bibfield{author}{\bibinfo{person}{Joe Whittaker}, \bibinfo{person}{Seán
  Looney}, \bibinfo{person}{Alastair Reed}, {and} \bibinfo{person}{Fabio
  Votta}.} \bibinfo{year}{2021}\natexlab{}.
\newblock \showarticletitle{Recommender systems and the amplification of
  extremist content}.
\newblock \bibinfo{journal}{\emph{Internet Policy Review}}
  \bibinfo{volume}{10}, \bibinfo{number}{2} (\bibinfo{date}{June}
  \bibinfo{year}{2021}).
\newblock
\showISSN{2197-6775}
\urldef\tempurl%
\url{https://policyreview.info/articles/analysis/recommender-systems-and-amplification-extremist-content}
\showURL{%
\tempurl}


\bibitem[Wieringa(2020)]%
        {wieringa_what_2020}
\bibfield{author}{\bibinfo{person}{Maranke Wieringa}.}
  \bibinfo{year}{2020}\natexlab{}.
\newblock \showarticletitle{What to account for when accounting for
  algorithms}. In \bibinfo{booktitle}{\emph{Proceedings of the 2020
  {Conference} on {Fairness}, {Accountability}, and {Transparency}}}.
  \bibinfo{publisher}{ACM}.
\newblock
\urldef\tempurl%
\url{https://doi.org/10.1145/3351095.3372833}
\showDOI{\tempurl}


\bibitem[Wolfe and Caliskan(2022)]%
        {wolfe_american_2022}
\bibfield{author}{\bibinfo{person}{Robert Wolfe} {and} \bibinfo{person}{Aylin
  Caliskan}.} \bibinfo{year}{2022}\natexlab{}.
\newblock \showarticletitle{American == {White} in {Multimodal}
  {Language}-and-{Image} {AI}}. In \bibinfo{booktitle}{\emph{Proceedings of the
  2022 {AAAI}/{ACM} {Conference} on {AI}, {Ethics}, and {Society}}}.
  \bibinfo{publisher}{ACM}.
\newblock
\urldef\tempurl%
\url{https://doi.org/10.1145/3514094.3534136}
\showDOI{\tempurl}


\bibitem[Wyatt(2008)]%
        {wyatt_technological_2008}
\bibfield{author}{\bibinfo{person}{S. Wyatt}.} \bibinfo{year}{2008}\natexlab{}.
\newblock \showarticletitle{Technological determinism is dead; {Long} live
  technological determinism.}
\newblock In \bibinfo{booktitle}{\emph{Handbook of {Science} and {Technology}
  {Studies}}}, \bibfield{editor}{\bibinfo{person}{E.~Hackett},
  \bibinfo{person}{O.~Amsterdamska}, \bibinfo{person}{M.~Lynch}, {and}
  \bibinfo{person}{J.~Wajcman}} (Eds.). \bibinfo{publisher}{MIT Press},
  \bibinfo{address}{Cambridge}, \bibinfo{pages}{165--180}.
\newblock


\bibitem[Ye et~al\mbox{.}(2021)]%
        {ye_mastering_2021}
\bibfield{author}{\bibinfo{person}{Weirui Ye}, \bibinfo{person}{Shaohuai Liu},
  \bibinfo{person}{Thanard Kurutach}, \bibinfo{person}{Pieter Abbeel}, {and}
  \bibinfo{person}{Yang Gao}.} \bibinfo{year}{2021}\natexlab{}.
\newblock \showarticletitle{Mastering {Atari} {Games} with {Limited} {Data}}.
  In \bibinfo{booktitle}{\emph{Advances in {Neural} {Information} {Processing}
  {Systems}}}, Vol.~\bibinfo{volume}{34}. \bibinfo{publisher}{Curran
  Associates, Inc.}, \bibinfo{pages}{25476--25488}.
\newblock
\urldef\tempurl%
\url{https://proceedings.neurips.cc/paper/2021/hash/d5eca8dc3820cad9fe56a3bafda65ca1-Abstract.html}
\showURL{%
\tempurl}


\bibitem[Yoon et~al\mbox{.}(2019)]%
        {yoon_is_2019}
\bibfield{author}{\bibinfo{person}{Sunkyung Yoon}, \bibinfo{person}{Mary
  Kleinman}, \bibinfo{person}{Jessica Mertz}, {and} \bibinfo{person}{Michael
  Brannick}.} \bibinfo{year}{2019}\natexlab{}.
\newblock \showarticletitle{Is social network site usage related to depression?
  {A} meta-analysis of {Facebook}–depression relations}.
\newblock \bibinfo{journal}{\emph{Journal of Affective Disorders}}
  \bibinfo{volume}{248} (\bibinfo{date}{April} \bibinfo{year}{2019}),
  \bibinfo{pages}{65--72}.
\newblock
\showISSN{0165-0327}
\urldef\tempurl%
\url{https://doi.org/10.1016/j.jad.2019.01.026}
\showDOI{\tempurl}


\bibitem[Young et~al\mbox{.}(2022)]%
        {young_confronting_2022}
\bibfield{author}{\bibinfo{person}{Meg Young}, \bibinfo{person}{Michael
  Katell}, {and} \bibinfo{person}{P.~M. Krafft}.}
  \bibinfo{year}{2022}\natexlab{}.
\newblock \showarticletitle{Confronting {Power} and {Corporate} {Capture} at
  the {FAccT} {Conference}}. In \bibinfo{booktitle}{\emph{2022 {ACM}
  {Conference} on {Fairness}, {Accountability}, and {Transparency}}}.
  \bibinfo{publisher}{ACM}.
\newblock
\urldef\tempurl%
\url{https://doi.org/10.1145/3531146.3533194}
\showDOI{\tempurl}


\bibitem[Zeng et~al\mbox{.}(2022)]%
        {zeng_socratic_2022}
\bibfield{author}{\bibinfo{person}{Andy Zeng}, \bibinfo{person}{Maria
  Attarian}, \bibinfo{person}{Brian Ichter}, \bibinfo{person}{Krzysztof
  Choromanski}, \bibinfo{person}{Adrian Wong}, \bibinfo{person}{Stefan Welker},
  \bibinfo{person}{Federico Tombari}, \bibinfo{person}{Aveek Purohit},
  \bibinfo{person}{Michael Ryoo}, \bibinfo{person}{Vikas Sindhwani},
  \bibinfo{person}{Johnny Lee}, \bibinfo{person}{Vincent Vanhoucke}, {and}
  \bibinfo{person}{Pete Florence}.} \bibinfo{year}{2022}\natexlab{}.
\newblock \bibinfo{title}{Socratic {Models}: {Composing} {Zero}-{Shot}
  {Multimodal} {Reasoning} with {Language}}.
\newblock
\newblock
\urldef\tempurl%
\url{https://doi.org/10.48550/arXiv.2204.00598}
\showDOI{\tempurl}
\newblock
\shownote{arXiv:2204.00598 [cs]}.


\bibitem[Zhang and Liu(2020)]%
        {zhang_fairness_2020}
\bibfield{author}{\bibinfo{person}{Xueru Zhang} {and} \bibinfo{person}{M.
  Liu}.} \bibinfo{year}{2020}\natexlab{}.
\newblock \showarticletitle{Fairness in {Learning}-{Based} {Sequential}
  {Decision} {Algorithms}: {A} {Survey}}.
\newblock \bibinfo{journal}{\emph{ArXiv}}  \bibinfo{volume}{abs/2001.04861}
  (\bibinfo{year}{2020}).
\newblock


\bibitem[Zheng et~al\mbox{.}(2021)]%
        {zheng_ai_2021}
\bibfield{author}{\bibinfo{person}{Stephan Zheng}, \bibinfo{person}{Alexander
  Trott}, \bibinfo{person}{Sunil Srinivasa}, \bibinfo{person}{David~C. Parkes},
  {and} \bibinfo{person}{Richard Socher}.} \bibinfo{year}{2021}\natexlab{}.
\newblock \bibinfo{title}{The {AI} {Economist}: {Optimal} {Economic} {Policy}
  {Design} via {Two}-level {Deep} {Reinforcement} {Learning}}.
\newblock
\newblock
\urldef\tempurl%
\url{https://doi.org/10.48550/arXiv.2108.02755}
\showDOI{\tempurl}
\newblock
\shownote{arXiv:2108.02755 [cs, econ, q-fin]}.


\bibitem[Zilka et~al\mbox{.}(2022)]%
        {zilka_transparency_2022}
\bibfield{author}{\bibinfo{person}{Miri Zilka}, \bibinfo{person}{Holli
  Sargeant}, {and} \bibinfo{person}{Adrian Weller}.}
  \bibinfo{year}{2022}\natexlab{}.
\newblock \showarticletitle{Transparency, {Governance} and {Regulation} of
  {Algorithmic} {Tools} {Deployed} in the {Criminal} {Justice} {System}: a {UK}
  {Case} {Study}}. In \bibinfo{booktitle}{\emph{Proceedings of the 2022
  {AAAI}/{ACM} {Conference} on {AI}, {Ethics}, and {Society}}}.
  \bibinfo{publisher}{ACM}.
\newblock
\urldef\tempurl%
\url{https://doi.org/10.1145/3514094.3534200}
\showDOI{\tempurl}


\bibitem[Zwetsloot and Dafoe(2019)]%
        {zwetsloot_thinking_2019}
\bibfield{author}{\bibinfo{person}{Remco Zwetsloot} {and}
  \bibinfo{person}{Allan Dafoe}.} \bibinfo{year}{2019}\natexlab{}.
\newblock \showarticletitle{Thinking about risks from {AI}: {Accidents}, misuse
  and structure}.
\newblock \bibinfo{journal}{\emph{Lawfare. February}}  \bibinfo{volume}{11}
  (\bibinfo{year}{2019}), \bibinfo{pages}{2019}.
\newblock


\end{thebibliography}

\clearpage

\appendix

\end{document}